%% file: sn-article.tex
\documentclass[sn-basic, linktocpage, pdflatex]{sn-jnl}

\usepackage{hyperref}
\hypersetup{pdfproducer={}, pdfcreator={Me}}
\usepackage{graphicx}%
\usepackage{multirow}%
\usepackage{amsmath,amssymb,amsfonts}%
\usepackage{mathrsfs}%
\usepackage[title]{appendix}%
\usepackage{xcolor}%
\usepackage{textcomp}%
\usepackage{manyfoot}%
\usepackage{booktabs}%
\usepackage{algorithm}%
\usepackage{algorithmicx}%
\usepackage{algpseudocode}%
\usepackage{listings}%
\usepackage{siunitx}%
\usepackage{array, makecell}%
\usepackage{comment}%
\usepackage{tabularx}%
\usepackage{anyfontsize}%
\usepackage{orcidlink}

\usepackage{lineno}

\setcitestyle{authoryear,open={(},close={)}}

\def\kms{\hbox{km$\;$s$^{-1}$}} 

\begin{document}

\title[Properties of Magnetic Switchbacks in the Near-Sun Solar Wind]{Properties of Magnetic Switchbacks in the Near-Sun Solar Wind}

\include{authors.tex}


\abstract{Magnetic switchbacks are fluctuations in the solar wind in which the interplanetary magnetic field sharply deflects away from its background direction so as to create folds in magnetic field lines while remaining of roughly constant magnitude. The magnetic field and velocity fluctuations are extremely well correlated in a way corresponding to Alfv\'enic fluctuations propagating away from the Sun. For a background field which is nearly radial this causes an outwardly propagating jet to form. Switchbacks and their characteristic velocity jets have recently been observed to be nearly ubiquitous by Parker Solar Probe with \textit{in situ} measurements in the inner heliosphere within 0.3~AU. Their prevalence, substantial energy content, and potentially fundamental role in the dynamics of the outer corona and solar wind motivate the significant research efforts into their understanding. Here we review the \textit{in situ} measurements of these structures (primarily by Parker Solar Probe). We discuss how they are identified and measured, and present an overview of the primary observational properties of these structures, both in terms of individual switchbacks and their collective arrangement into ``patches''. We identify both properties for which there is a strong consensus and those that have limited or qualified support and require further investigation. We identify and collate several open questions and recommendations for future studies. 
}



\maketitle
\tableofcontents

\section{On the Striking Nature of Switchbacks }\label{sec: 1_Intro}
\subsection{First Observations in the Young Solar Wind} \label{subsec: 1_first_obs}
    \begin{figure}
        \centering
        \includegraphics[width=\textwidth]{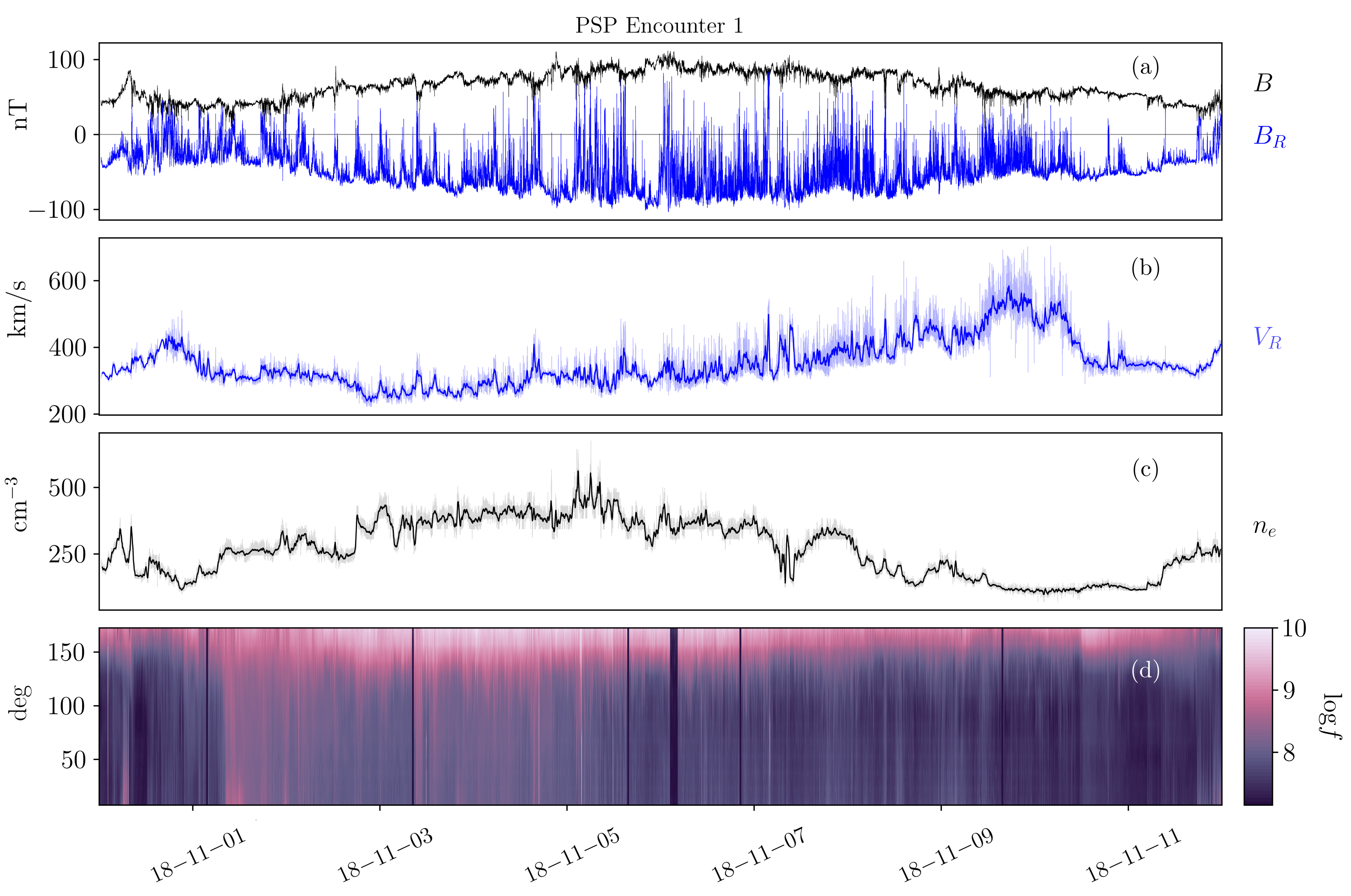}
        \caption{Parker measurements of the young solar wind during the mission's first orbit, showing (a) the magnetic field amplitude and radial component as the spacecraft goes from 60 to 35~$R_{\odot}$ on November 6; (b) the radial solar wind speed together with a 20-minute average; (c) the electron density inferred using quasi-thermal noise spectroscopy (see Sect.~ \ref{subsec: 1_instru}) with a 20-minute average; (d) the pitch angle distribution of suprathermal electrons (314~eV). }
        \label{fig: 1_ubiquitous_SB}
    \end{figure}

    From the very first orbit of the Parker Solar Probe \citep[Parker; ][]{Fox2016,Raouafi2023a} mission with the Sun, \textit{in situ} data revealed surprising and notable features: 
    the solar wind exhibited frequent magnetic deflections, accompanied by velocity enhancements (``spikes") and significant changes in the radial magnetic field component \citep{Bale2019, Kasper2019}. 
    These structures, of Alfv\'enic nature – i.e., implying high correlation between magnetic field and velocity variations – are commonly referred to as \textit{magnetic switchbacks} as the magnetic field fluctuations, especially in the early Parker orbits, lead to local reversals of the radial magnetic field component. A more comprehensive description of the properties defining switchbacks is provided in Sect. \ref{sec: 2_SB_definition}.
    
    Figure~\ref{fig: 1_ubiquitous_SB} highlights the prevalence of magnetic switchbacks in the young solar wind.
    As Parker approaches perihelion (November 6, 2018), its radial distance from the Sun decreasing from  60 solar radii (R$_\odot$) down to 35~R$_\odot$, rapid reversals in the radial component of the magnetic field $B_R$ are observed to pervade the solar wind (top panel). 
    Though the mean radial field is negative, innumerable rapid oscillations to positive are observed while the pitch angle distribution (PAD) of suprathermal electrons (bottom panel) remains concentrated around 180$^{\circ}$. This means that the faster electrons, that are channelled along the field, are actually propagating backwards during the reversals \citep[see figure 10 of][]{Owen2013}, implying that the reversals are only local folds in magnetic field lines \citep{Balogh1999, Kasper2019, Bale2019}, while the spacecraft remains magnetically connected to the same polarity region on the Sun, here a region of inward field.
    
    Although switchbacks are commonly depicted as 2D kinks in  magnetic field lines (Fig.~\ref{fig: 1_SBs_First_Obs_1966}), their internal structure is probably more complex \citep[see e.g.,][]{Shi2024}. 

\subsection{A Brief Historical Overview} \label{subsec: 1_history}
    \begin{figure}
        \centering
        \includegraphics[width=\textwidth]{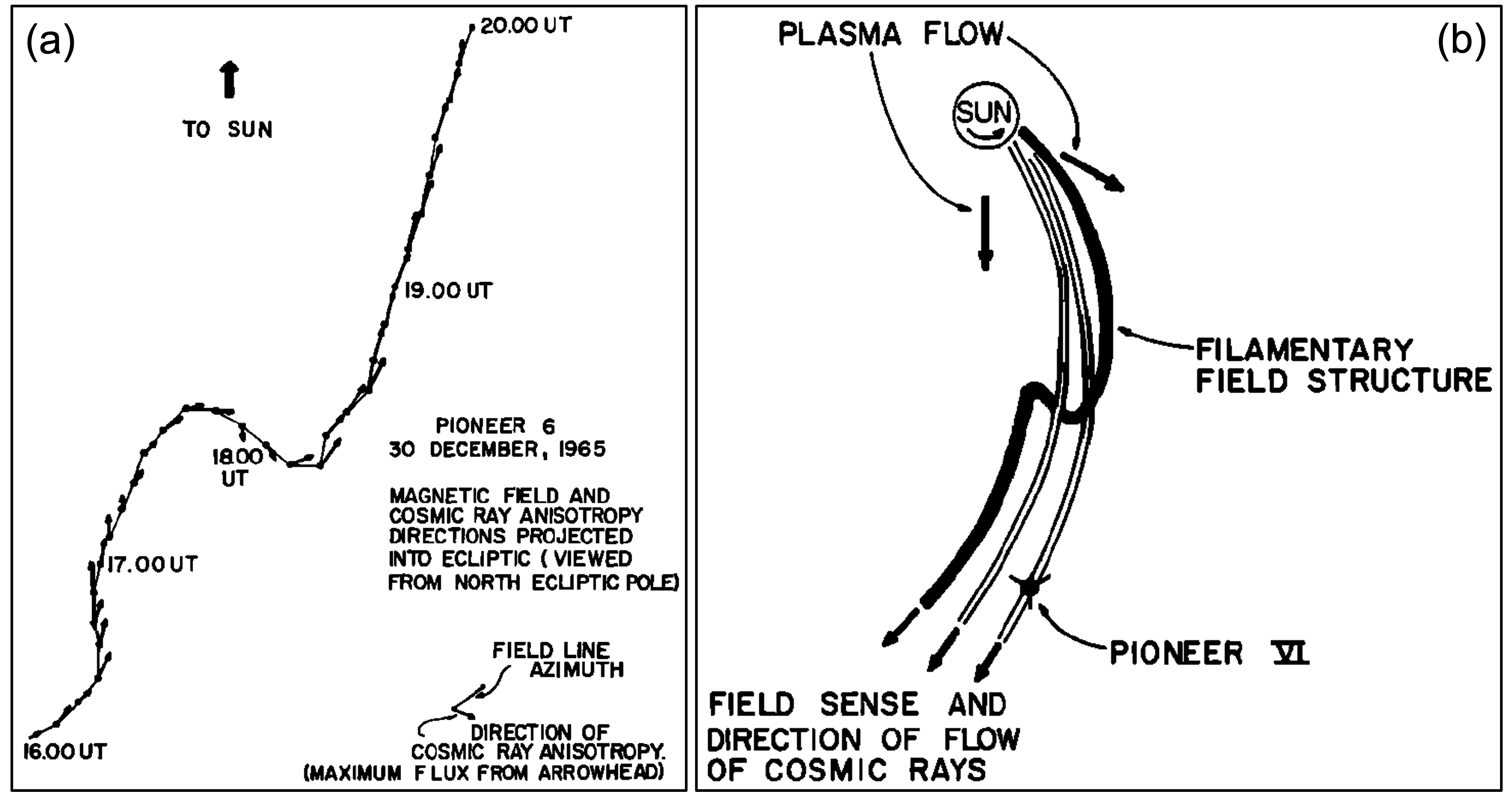}
        \caption{Observation from Mariner-II showcasing abrupt changes in the direction of the interplanetary magnetic field. Left panel: magnetic field and cosmic ray anisotropy; right panel: derived field line and (solar) cosmic ray flow. Reproduced from \citet{McCracken1966}.}
        \label{fig: 1_SBs_First_Obs_1966}
    \end{figure}

    Measurements from various spacecraft have reported the occurrence of abrupt changes in the interplanetary magnetic field direction. 
    In historical observations conducted at distances beyond 0.3~AU, such occurrences were relatively infrequent. 
    These events were subject to different interpretations, only some of which align with our current understanding of magnetic switchbacks.      
    The initial indication of the potential existence of magnetic field switchbacks was reported by \citet{McCracken1966} through the analysis of magnetic field and cosmic ray measurements obtained from Mariner-II. 
    Their interpretation, based on the close alignment between cosmic ray anisotropy and magnetic field lines, led them to conclude that filamentary structures existed within the interplanetary magnetic field (Fig.~\ref{fig: 1_SBs_First_Obs_1966}). 
    \citet{Michel1967} provided one of the first suggestions that a velocity modulation (jet-like enhancement) is associated with the folding of the interplanetary field in switchbacks, although we now understand that this is due to their Alfvénic nature, rather than background velocity shears \citep{Matteini2014}.
    \citet{Horbury2018} established clear evidence for the occurrence of these plasma jets in measurements obtained from the Helios solar wind, particularly in proximity to 0.3 AU.        
    Observations made by the Ulysses mission beyond 1 AU further contributed to our understanding, revealing magnetic field rotations exceeding 90$^{\circ}$ in relation to the Parker spiral \citep[as documented by][]{Balogh1999, Yamauchi2004}. 
    Additional evidence supporting the existence of magnetic switchbacks was identified through observations made by the ISEE-3 mission \citep{Kahler1996} and the ACE mission \citep{Gosling2009, Li2016}. 
    A more detailed historical overview is available in \cite{Velli_this_issue}.

\subsection{Parker Solar Probe Instrumentation} \label{subsec: 1_instru}
    In this review, we primarily examine magnetic switchback properties as measured by the Parker mission \citep{Fox2016}. 
    As some discussions require an understanding of the instrumentation limitations on board, we briefly discuss the \textit{in situ} instrumentation on board the Parker spacecraft. 
    
    Parker carries three \textit{in situ} instrument suites whose data are shown throughout this work: \begin{itemize}
        \item The Electromagnetic Fields Investigation suite \citep[FIELDS;][]{Bale2016, Malaspina2016, Pulupa2017} measures AC and DC electric and magnetic fields, radio waves, and quasi-thermal noise spectroscopy (QTN) plasma electron diagnostics.
        \item The Solar Wind Alphas, Electrons and Protons suite \citep[SWEAP;][]{Kasper2016, Case2020,Whittlesey2020,Livi2022} measures particle velocity distribution functions up to 30~keV and derives bulk properties of electrons, protons, and alpha particles.
        \item The Integrated Science Investigation of the Sun suite \citep[IS$\odot$IS;][]{McComas2016} measures the direction-resolved fluxes and energy distributions of energetic electrons, protons, and heavy ions (from 25~keV to 6~MeV for electrons, and 20~keV/nucleon to 200~MeV/nucleon for ions). 
    \end{itemize} 

    The key \textit{in situ} signatures of switchbacks are shown in Fig.~\ref{fig: 1_ubiquitous_SB} during E1, i.e., the first encounter of the Parker mission with the Sun (in the remainder of the paper, E$x$ stands for Encounter number $x$ and where encounter formally refers to continuous intervals where Parker was within 0.25~au of the Sun (although it can be used somewhat more loosely in the literature to index perihelia).    
    Data is shown in the RTN frame of reference, where $\mathbf{R}$ (radial) is the Sun to spacecraft unit vector, $\mathbf{T}$  (tangential) is the cross product between the Sun's spin axis and $\mathbf{R}$, and $\mathbf{N}$ (normal) completes the direct orthogonal frame.
    
    The FIELDS DC magnetic field 3D vector (whose radial component is shown in panel \ref{fig: 1_ubiquitous_SB}a) provides one of the primary signatures of switchbacks \citep{Bale2019} is measured by a pair of fluxgate magnetometers located on a boom trailing the spacecraft. 
    FIELDS also provides a robust measurement of the electron density (Fig.~\ref{fig: 1_ubiquitous_SB}c) via QTN from the FIELDS Radio Frequency Spectrometer \citep[FIELDS/RFS;][]{Pulupa2017, Moncuquet2020}, but the measurement is only available when the spacecraft is sufficiently close to the Sun for the effective length of the antennae to be longer than the Debye length and therefore for the plasma frequency to be well inside the frequency range of the instrument (typically within a heliocentric distance of around 70 $R_\odot$).
    
    The proton bulk properties (including velocity, Fig. \ref{fig: 1_ubiquitous_SB}b) are measured by the Solar Probe Cup \citep[SPC;][]{Case2020} and the Solar Probe Analyzer for ions \citep[SPAN-i;][]{Livi2022}. 
    SPC is a Faraday cup with a narrow field of view oriented towards the Sun. SPAN-i is an electrostatic analyzer with a wider field of view oriented on the ram side of the spacecraft. Both SPC and Span-i give partial coverage of the velocity distribution functions (VDFs). SPAN-i further has a time-of-flight chamber which allows it to discriminate different mass per charge ratios and therefore provide separate data products for protons and alpha particles.
    No matter the species, the quality of the determination, particularly in density and temperature, depends on the location of the peak of the VDF in these two fields of view. This can change strongly during different parts of the orbit, due to the combined effects of aberration (changing speed of the plasma relative to the spacecraft throughout the orbit) as well as to intrinsic variations in solar wind speed (which occur throughout the orbit but also especially during switchbacks). 

    The last panel of Fig.~\ref{fig: 1_ubiquitous_SB} shows the pitch angle distributions (PAD) of suprathermal electrons as measured by the Solar Probe Analyzer for electrons \citep[SPAN-e;][]{Whittlesey2020}, which is an electrostatic analyzer with two sensors oriented on the spacecraft on the ram and anti-ram sides (thereby providing nearly 4$\pi$ steradians coverage of electron VDFs). 
    The pitch angles show the relative orientation between the flow direction of suprathermal electrons (or strahl) and the magnetic field. 

\subsection{Key Questions Associated with Switchbacks} \label{subsec: 1_key_questions}

    The ubiquity of switchbacks in Parker measurements below 0.3~AU \citep[e.g.,][]{Bale2019,Kasper2019}, their substantial energy content \citep[e.g.,][]{Halekas2023,Rivera2024a}, and their diagnostic potential for coronal processes and solar wind basic physics, have made their understanding a focal point  of heliophysics research. They have been widely studied in the early phases of the Parker mission, and formed a large segment of early results reviewed previously in   \citet{Raouafi2023a}.
    Their existence raises the following key questions:
    \begin{enumerate}
        \item {\bf Are switchbacks dynamically important in the initial acceleration of the solar wind and potentially in magnetized astrophysical winds in general?} Are switchbacks actively contributing to the acceleration of the flow in the corona and below the Alfvén radius? 
        \item {\bf Are switchbacks a significant source of heating for non-adiabatic expansion of the plasma in interplanetary space?} Switchbacks carry away a considerable extra amount of Poynting flux and bulk kinetic energy from the Corona; this energy excess can be converted into thermal energy during expansion and could be a potential main contribution to solar wind heating farther from the Sun. Recent works suggest they are significant for both the acceleration and heating of the fast wind beyond the Alfv\'en point \citep{Rivera2024a}, injecting energy directly into the turbulent cascade \citep{Hernandez2021} but with decreasing importance for slower wind types \citep{Halekas2023}.
        \item {\bf Are switchbacks direct signatures of the processes responsible for solar wind acceleration in the lower corona?} Even if switchbacks are not directly responsible for the acceleration of the plasma, they could be produced by the same mechanisms that cause the plasma acceleration, e.g., interchange reconnection. Therefore, understanding switchbacks may shed light on the acceleration processes. Different models for switchback generation and their link to solar dynamics and sources are reviewed in \citet{Tripathi_this_issue} and \citet{Wyper_this_issue}. 
        \item {\bf Are switchbacks passive tracers of solar dynamics?}
        Switchbacks are nearly ubiquitous in the near-Sun solar wind. This presence makes them an indirect probe of solar dynamics imprinted in the solar wind. By studying switchback modulation and properties, we can probe properties of source regions that cannot be explored \textit{in situ}. An example of this is discussed in Sect.~\ref{sec: 5_sw_structure} of this paper.
    \end{enumerate}

\subsection{Outline}

This review is structured as follows: 

\begin{itemize}
    \item In Sect.~\ref{sec: 2_SB_definition}, we present the defining features of magnetic switchbacks and illustrate these features with real examples. \citep[See also section 4.1 of ][]{Raouafi2023a}
    \item In Sect.~\ref{sec: 3_methodology}, we discuss the diverse methodologies used throughout the literature to detect switchbacks and build statistics on them.
    \item In Sect.~\ref{sec: 4_SB_properties} we review investigations into different types of properties of individual switchback spikes, focusing on their plasma populations (Sect.~\ref{subsec: 4.1_inside_SB}), their geometry and boundary properties (Sect.~\ref{subsec: 4.2_geometry_boundaries}) and their relationship to more general solar wind turbulence and electromagnetic wave activity (Sect.~\ref{subsec: 4.3_turbulence}). \citep[See also section 4.2 of ][which this section updates and expands on]{Raouafi2023a}
    \item In Sect.~\ref{sec: 5_sw_structure}, we zoom out to examine the collective behavior of switchbacks, including their arrangement into patches and differences in their properties in different types of solar wind streams.
    \item Finally, in Sect.~\ref{sec: 6_open_questions}, we close by summarizing the main elements of this review and presenting some key open questions for which further study is required to reach definitive conclusions.
\end{itemize}

\section{Switchback Definition} \label{sec: 2_SB_definition} 

To study magnetic switchback properties in the solar wind, it is particularly important that the scientific community agrees on a common definition for these structures (see Sect.~\ref{sec: 3_methodology} for a review of how definitions and methodologies may impact statistical studies).
Here, we propose a consensus definition as a list of expected switchback features.

\begin{itemize}
    \item A switchback is a sharp deflection of the magnetic field vector away from the ambient direction and back; ``sharp" meaning that their boundaries have a short timescale compared to the switchback duration.
    The deviation should be significant with respect to the local level of fluctuations, typically at least a few tens of degrees. 
    While most switchbacks deflect less than 90$^{\circ}$, some do lead to changes in the polarity of the magnetic field. We refer to these more than 90$^{\circ}$ deflections as local ``polarity reversals" throughout the paper.
    \item Switchbacks have an approximately constant magnetic field magnitude $B$, meaning that, geometrically, the magnetic field vector $\mathbf{B}$ evolves on a sphere throughout the structure.
    The switchback boundaries are thus 1D arcs on the sphere surface.
    \item Switchbacks are local folds in the magnetic field and are not associated with a change of polarity at the source. They thus exhibit the same electron strahl pitch angle direction throughout the switchback. 
    \item Switchbacks are Alfvénic, displaying the high correlation between magnetic field and velocity variations that corresponds to fluctuations propagating away from the Sun in the background field. As a consequence, kinks in the magnetic field, leading to changes in the radial magnetic field ($B_R$), are usually associated with enhancements in the radial bulk proton velocity ($V_R$),  with the peak in velocity occurring at the maximum field deflection. We note that this spike in $V_R$ is in fact a 1D projection of the velocity vector also moving on a sphere centered on the reference frame in which the motional electric field goes to zero \citep{Matteini2015_dHT}, sometimes referred to as the DeHoffman Teller frame \citep[][see also Sect.~\ref{subsec: 2_spikes}]{Horbury2020}.
\end{itemize}   

It is important to note that while our definition of switchbacks allows for a spectrum of deflection angles relative to some choice of background field, there are distinct lower thresholds applied to this spectrum throughout the literature (see Sect.~\ref{sec: 3_methodology}).    
The most restrictive of these is a 90$^o$ threshold from which the term ``switchback'' originates. This additional criterion is often applied to assess whether models are sufficient to explain these largest switchbacks \citep[see][for a review]{Wyper_this_issue}. 

Finally, note that, because of their typical MHD scale (duration of minutes/tens of seconds), switchbacks can be considered at first order an ideal-MHD solution, therefore their magnetic-velocity structure is solely supported by the self-consistent motional electric field $\mathbf{E}=-\mathbf{V}\times \mathbf{B}$ \citep[see e.g.,][]{Matteini2015_dHT}. However, if deflections are particularly sharp, non-ideal effect can play a role at boundaries, see Sect.~\ref{subsubsec: 4.2_SB_boundaries}.

\subsection{Prototypical Example} \label{subsec: 2_SB_example} 

    \begin{figure}
        \centering
        \includegraphics[width=\textwidth]{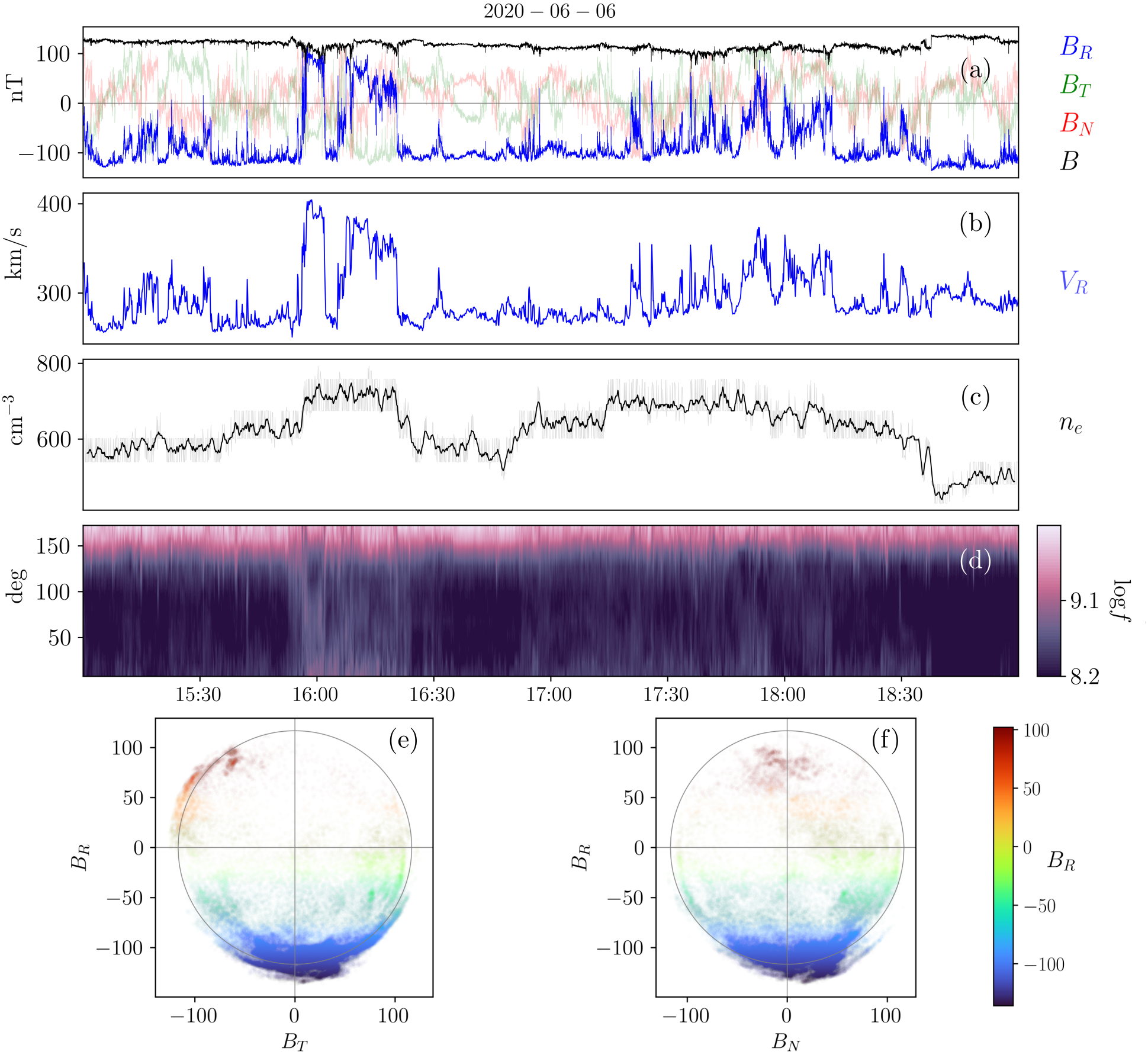}
        \caption{Magnetic switchbacks observed by Parker at 28~R$_{\odot}$ during E5. The top panels display timeseries of (a) the magnetic field vector, (b) the radial solar wind velocity from SPAN, (c) the 1-minute average of the electron density from QTN, and (d) the PAD of suprathermal electrons (314~eV). The bottom panels show scatter plots of $B_R$ vs $B_T$ (e) and $B_R$ vs $B_N$ (f) colored by $B_R$ during the considered time interval.}
        \label{fig: 2_SB_example}
    \end{figure}

   We illustrate these defining features in a realistic context in Fig.~\ref{fig: 2_SB_example}, where we show observations from Parker during its 5$^{th}$ perihelion (June, 2020). 
   During the selected time interval, several typical signatures of magnetic switchbacks are observed. 
   Sharp deflections from the background field are present at different scales, including two impressive large, close-to 180$^{\circ}$ reversals in the first half of the interval (between 15:50 and 16:20~UT) and multiple smaller deflections that sometimes reverse the radial magnetic field in the second half between (17:20 and 18:40~UT).
   The solar wind is mostly Alfvénic during this interval, with the radial velocity $V_R$ positively correlated to $B_R$, exhibiting radial velocity spikes coincidental with the switchbacks.
   The electron strahl orientation with respect to the magnetic field remains constant around 180$^{\circ}$ throughout the interval, meaning that at maximum deflection, the strahl is briefly moving sunward.
   The density fluctuates between 450 and 800~cm$^{-3}$, with some compression potentially associated with the switchbacks considered here. Note that, as switchbacks observed in the inner Heliosphere are typically embedded in low-beta plasma, the relative jump in density is often larger than in $B$, as expected for pressure balance.
   The magnetic field deflections occur at roughly constant $B$, as further illustrated by the hodograms of $\mathbf{B}$ shown in the bottom panels (\ref{fig: 2_SB_example}e, \ref{fig: 2_SB_example}f). 
   There, the two large deflections (15:55--16:20~UT) are associated with red points ($B_R \in [50, 100]$~nT, $B_T$ negative) and closely follow the constant $B = 117$~nT sphere.

   This time interval also illustrates that the exact distinction between switchbacks and background solar wind turbulence is not easily captured. 
   Often, potential switchback structures are ambiguous in the solar wind, due to smaller deflection angles (15:10--15:45~UT), internal fluctuations inside a larger deflection (18:00--18:10~UT), or imperfect $B$ conservation, for instance.
   The choice of whether to label such structures as switchbacks varies between studies, and some features like constant $B$ or Alfvénicity are discussed in Sect.~\ref{sec: 4_SB_properties}.

\subsection{Link to Velocity Spikes} \label{subsec: 2_spikes} 
    Fig.~\ref{fig: 2_SB_example} shows that velocity spikes associated with switchbacks are always positive. 
An explanation for this apparent one-sidedness of solar wind velocity fluctuations, first noticed by \cite{Gosling2009}, was provided by \cite{Matteini2014}. Large amplitude Alfvénic fluctuations in the MHD regime display the following correlation between magnetic field and velocity fluctuation vectors:
    
     \begin{equation}\label{eq_AW}
        \frac{\delta \mathbf{V}}{V_A}=\pm\frac{\delta \mathbf{B}}{B}\,,
    \end{equation}
    where $V_A = B/\sqrt{\mu_0 \rho}$~is the Alfvén speed, with $\rho$ the mass density, and $\mu_0$ the vacuum permeability constant. 
    The positive (negative) correlation corresponds to waves propagating along the negative (positive) direction of the magnetic field.
    As a consequence, the expected radial velocity change in a switchback is:
        \begin{equation}\label{eq_dvr_sb}
            |\delta V_R|=\frac{|\delta B_R|}{B}V_{ph}\,,
        \end{equation} 
        where we have introduced an effective ``phase speed'' $V_{ph}$
     that is different and smaller than $V_A$ because the ratio of kinetic to magnetic energies $r_A$ in the fluctuations is always somewhat smaller than 1. In other words, the effective velocity is $V_{ph} = V_A$ when there is equal content in magnetic and kinetic energy fluctuations (no residual energy), i.e. in intervals of high correlation between $\delta \mathbf{B}-\delta \mathbf{V}$ (high cross-helicity -- see Sect.~\ref{subsubsec: 4.3_alfv_turb} for definitions), while typically the relation $V_{ph}<V_A$ is observed in intervals with lower cross-helicity. Alfvénicity in switchbacks and observed deviations are discussed in Sect.~\ref{subsubsec: 4.2_SB_boundaries}.
    
    Since fluctuations in solar wind streams with high Alfvénicity correspond to waves that predominantly propagate away from the Sun, the positive or negative sign in Eq.~(\ref{eq_AW}) holds for the correlation in a background magnetic field pointing back towards the Sun or away from the Sun, respectively. In a switchback that changes the sign of the field, the result will always be an outward jet, $\delta V_R>0$, as long as the background field is predominantly radial \citep{Matteini2014}. 
    Assuming a radial solar wind flow $V_0$, it is then possible to express variations of the local speed $V$ associated with switchbacks in the inner Heliosphere, as:
    \begin{equation}\label{eq_v_in_sb}
    V= V_0+ V_{ph}\left[1- \cos(\theta_{BR})\right],
    \end{equation}
    where $V_0$ is the minimum speed of the background, associated here with a positive radial magnetic field \citep{Matteini2014}.
    Under the assumption of little magnetic field compression, the magnitude of the magnetic field ($|\mathbf{B}|$) is approximately constant, implying that the maximum magnetic field variation is $|\delta \mathbf{B}|=2B$. The maximum speed enhancement in a switchback is therefore $|\delta \mathbf{V}|=2V_A$, associated with a full reversal of the background magnetic field.
    For switchbacks at $90^\circ$, $V\sim V_0+V_A$.   
    
\section{Review of Methodologies}\label{sec: 3_methodology}      

The definition of switchbacks provided in Sect.~\ref{sec: 2_SB_definition} is abstracted from observations collected by spacecraft \textit{in situ} over a great range of distances and throughout the solar cycle. 
However, as exhibited in Fig.~\ref{fig: 2_SB_example}, any individual event observed \textit{in situ} more often than not presents departures in some features from the ideal definition.
Some switchback characteristics are easier to quantify and detect algorithmically using thresholds in some quantity (such as the deflection angle in the magnetic field), whereas others present subjective aspects or are more difficult to quantify (such as rotation abruptness or required Alfv\'enicity).
For example, although measuring the deflection of the magnetic field vector does not automatically test for Alfv\'enicity or reject current sheet crossings, it allows for comprehensive statistical event collection. 
Stricter methods tend to require manual input, leading to more subjectivity and smaller event numbers in statistics.  
In this section, we review the range of different methodologies used to identify switchbacks in the present literature and discuss their relative assumptions. 
It is important to keep track of the method chosen to identify switchbacks, as it may affect the inferred properties of switchbacks discussed in the later sections of the paper.

\subsection{Identification using Deflection from a Background Field} \label{subsec: 3_SB_methods}

    Since switchbacks are partly defined as sudden magnetic field rotations (Sect.~\ref{sec: 2_SB_definition}), it is common to study them as a simple deflection from a background field.
    Such a treatment requires two parts: first, a choice of background orientation from which the switchback deflects, and second, a method to characterize and separate the deflection relative to the background. 
    Here, we first highlight the impact the choice of the background field may have on switchback analyses and results (Sect.~\ref{subsubsec: 3_SB_background}).
    Next, we examine the effects of the varying deflection thresholds used by various authors and show that no clear consensus exists on \textit{how small} a deflection should be categorized as a switchback (Sect.~\ref{subsubsec: 3_threshold}).
        
    \subsubsection{Impact of the Magnetic Field Background Choice} \label{subsubsec: 3_SB_background}
        Various background magnetic field definitions have been used to identify switchbacks, and two kinds of approaches are typically used. 
        One is to compute the background field from the data, using different statistical parameters like the mean, median \citep{DudokdeWit2020} or mode values \citep{Bale2019}. 
        The other consists of modeling the expected background field independently, either assuming a radial nominal magnetic field \citep[e.g.,][]{Horbury2018, Larosa2021, AkhavanTafti2021, Woolley2020} or using the Parker spiral model \citep[e.g.,][]{Horbury2020, Laker2021, Fargette2021}.
        The time interval over which the background is computed should exceed the timescale of magnetic switchbacks, and is typically chosen to be a few hours.
        All methods have certain limitations: mean, median, and mode values might be biased by the switchbacks themselves \citep[see e.g.,][]{Badman2021}, while model accuracy may vary.

        \begin{figure}
            \centering
            \includegraphics[width=\textwidth]{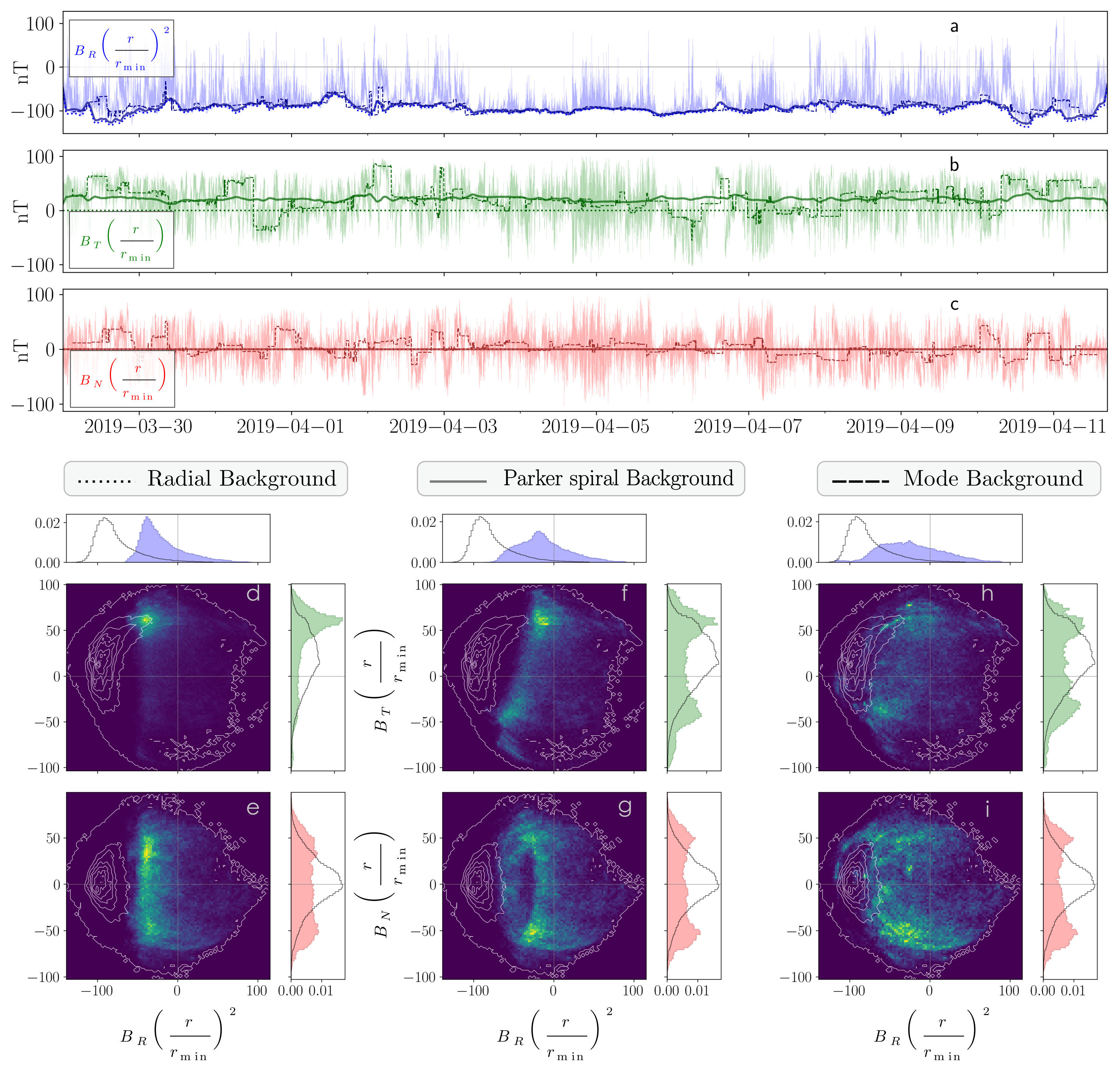}
            \caption[Impact of switchback definition]{Impact of switchback definition. In panels $a$ to $c$, we show the magnetic field components in the RTN frame during E2, normalized by the closest approach radial distance of 35~R$_{\odot}$. We overplot the radial field (dotted lines), the Parker spiral field (full lines), and a 6h-mode field (dashed lines). In panels $d$ to $i$, we plot in white the 2D distribution contours of the normalized magnetic field components ($B_R$, $B_T$ in panels $d$, $f$, $h$, and $B_R$, $B_N$ in panels ($e$, $g$, $i$). Superimposed in color are the 2D histograms of the points that are located more than 60$^o$ away from the computed background fields, i.e., the radial direction ($d$, $e$), the Parker spiral ($f$, $g$), and the 6h-mode vector ($h$, $i$). We also add the normalized projected distributions on the side, as a black line for the full 2D distribution and color-shaded for the ``more than 60$^o$ away'' points.
            Reproduced from \citet{fargette_phd}, copyright by the author}
            \label{fig: 3_SB_method}
        \end{figure}
    
        In Fig.~\ref{fig: 3_SB_method}, we illustrate how different background choices will affect a given analysis.  
        Three different background fields computed over 6h are compared: a purely radial magnetic field, the Parker spiral field, and a 6h-mode magnetic field. 
        By definition, both the $B_N$ component of the Parker spiral in the ecliptic and the $B_T$ and $B_N$ components of the radial field are equal to zero throughout the interval. 
        While all methods converge to a similar $B_R$ background, the differences in the $B_T$ and $B_N$ backgrounds are particularly striking.
        In the bottom panels, we illustrate how these different backgrounds lead to the selection of different solar wind structures.
        The 2D histograms show the distribution of points deflected more than 60$^{\circ}$ away from each background field, while the white contour tracks the nominal distribution of the magnetic field components.
        These distributions of the deflected magnetic field differ significantly between the three methods.
        The radial background field neglects the $B_T$ component of the expected Parker spiral and, consequently, the detected deviations are strongly biased toward a positive $B_T$. 
        By contrast, the distributions 60$^{\circ}$ away from both the Parker spiral model and the sliding mode are more isotropic in $B_T$.
        All distributions of $B_R$ are different.
        Overall, structures labeled as ``switchbacks deflected by more than 60$^{\circ}$'' will be different across the three methods.
        
        From this example, it seems clear that the switchback definition cannot be solely based on the radial direction, as the tangential component of the Parker spiral remains significant in portions of Parker's orbit. This is especially important in continuing to identify these structures in other spacecraft datasets farther from the Sun, 
        where, because the background magnetic field may be at large angles to the radial, the velocity signature is no longer apparent as a one-sided spike \citep{Gosling2009, Bourouaine2022}.
        The Parker spiral model or the ambient background field (median or mode) produces more accurate references to define magnetic switchbacks.
        While each will call for a different interpretation of the results (deflection from an expected physical model or an ambient field), we advise choosing one of these two methods in future statistical studies.

        \begin{figure}
            \centering
            \includegraphics[width=\textwidth]{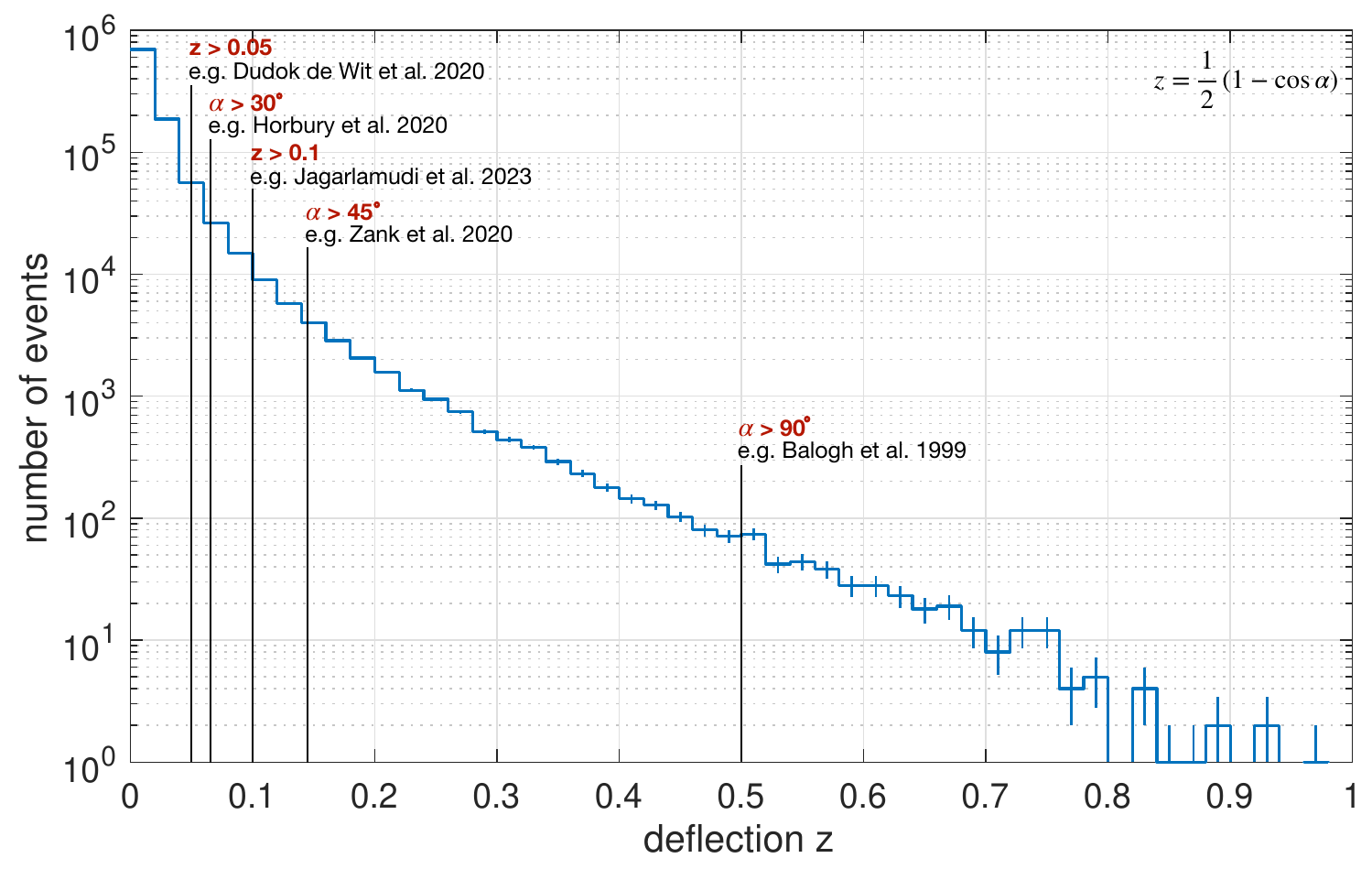}
            \caption{Histogram showing the distribution of \textit{normalized deflection} $z$, as defined by \citet{DudokdeWit2020}, for individual switchback events during the E1--E13. Some common choices of thresholds in magnetic field deflections are indicated with a mention of the first study that made that choice to define a switchback. 
            }
            \label{fig: 3_z_angle_annotated}
        \end{figure}
    
    \subsubsection{Deflection Angle Thresholds} \label{subsubsec: 3_threshold}

        Figure~\ref{fig: 3_z_angle_annotated} shows a histogram of observed switchback deflection angles in terms of the normalized deflection parameter $z = (1 - \cos{\alpha})/2$, where $\alpha$ is the deviation angle from the identified background field \citep[a median over 6 hour intervals,][]{DudokdeWit2020}.
        These switchbacks, compiled from Parker E1 through E13, show a largely featureless distribution in $z$, confirming the initial result of \citet{DudokdeWit2020}. 
        This hints towards a continuum of nearly self-affine deflections, preventing a quantitative threshold for deflection angle that categorizes a switchback. 
        Demarcating some common choices of $z$ values in Fig.~\ref{fig: 3_z_angle_annotated} reveals that studies that do not impose a strong reversal in the magnetic field direction $(\alpha < 90^{\circ})$ have all demanded a rather low $z$ threshold. 
        
        The radial magnetic field reversal as a  requirement to identify switchbacks has been used in the analysis of both Ulysses \citep{Balogh1999} and Parker data \citep{Macneil2020, Mozer2021, Tenerani2021,Pecora2022}. 
        These authors use the $\alpha > 90^{\circ} (z>0.5)$ criterion on the basis that switchbacks were first reported as field reversals in Parker data \citep{Bale2019}. 
        However, such fluctuations lie on a continuum of deflection angles, and there is no clear departure from self-similarity at $z=0.5$, although the statistical significance of larger fluctuations, especially for near-one z values is quite poor. 
        
        
        Several studies have relaxed the deflection threshold to intermediate angles, such as $25^{\circ}$ \citep[$z=0.05$,][]{DudokdeWit2020},
        $30^{\circ}$ \citep{Horbury2020}, $37^{\circ}$ \citep{Jagarlamudi2023} and $45^{\circ}$ \citep{Zank2020, Laker2021, Woolley2020, Laker2022}. Such choices, highlighted in Fig.~\ref{fig: 3_z_angle_annotated}, also appear somewhat arbitrary in the context of the featureless spectrum of deflection angles, although for very low deflection angles, such fluctuations are not meaningfully distinct from general stochastic variation in the \textit{in situ} data (see Sect.~\ref{subsubsec: 4.2_occurrence}).  Further, it has been shown that quiet solar wind intervals (devoid of switchbacks) present a standard deflection of around 15$^{\circ}$ around the Parker spiral, while larger amplitudes tend to have a systematic offset compared to this direction \citep{Fargette2022}. Therefore, this value may be viewed as a reasonable minimum threshold value.

        The choice of lower deflection Alfv\'enic fluctuations to qualify as switchbacks is further bolstered by the distribution of deflection angle $\alpha$ with regards to the Alfv\'en Mach number $M_A$ (defined as $v_{sw}/v_A$, where $v_{sw}$ is the magnitude of the proton velocity vector defined by plasma moments).
        It displays a ``herringbone" structure where individual striations (conjectured to arise from flows of similar origin) show a systematic increase in deflection angle with an increase in $M_A$ \citep{Bandyopadhyay2022,Liu2023_SB}. 
        Since an increase in $M_A$ is expected due to solar wind acceleration, switchback deflection angles are thus expected to increase with distance from the Sun.  
        With an abundance of deflections being below $90^{\circ}$, the renaming of ``switchbacks'' to ``Alfv\'enic deflections'' has been suggested, but has not yet been adopted \citep{Liu2023_SB}. The connection between switchbacks and sub-Alfv\'enic wind is discussed further in Sect.~\ref{subsec: 5_low_MA}.
    
    \begin{table}
    \begin{tabular}{p{1.7cm}p{4.2cm} p{4.5cm} p{4cm}}     
        \hline
        \textbf{Method} & \textbf{References} & \textbf{Method Summary} & \textbf{Assumptions} \\ 
        \hline
        & & & \\
        Deflection thresholds & 
        \begin{tabular}[t]{@{}l@{}} 
            $\boldsymbol\alpha \geq \mathbf{25^{\circ}}$ \\ 
            \cite{DudokdeWit2020}$^{\rm{E}1}$ \\ 
            \cite{Tatum2024}$^{\rm{E}1,6,8,11}$ \\
            $\boldsymbol\alpha \geq \mathbf{30^{\circ}}$ \\ 
            \cite{Horbury2020}$^{\rm{E}1}$\\ 
            $\boldsymbol\alpha \geq \mathbf{37^{\circ}}$ \\ 
            \cite{Jagarlamudi2023}$^{\rm{E}1,2,4-10}$\\ 
            $\boldsymbol\alpha \geq \mathbf{45^{\circ}}$ \\  
            \cite{Zank2020}$^{\rm{E}1}$ \\ 
            \cite{Woolley2020}$^{\rm{E}1,2}$ \\ 
            \cite{Laker2021}$^{\rm{E}1,2}$ \\ 
            \cite{Laker2022}$^{\rm{E}1-8}$\\ 
            $\boldsymbol\alpha \geq \mathbf{90^{\circ}}$ \\ 
            \cite{Balogh1999}$^{\rm{U}}$\\ 
            \cite{Macneil2020}$^{\rm{H}}$ \\ 
            \cite{Badman2021}$^{\rm{E1-5}}$\\
            \cite{Mozer2021}$^{\rm{E}3-7}$\\ 
            \cite{Tenerani2021}$^{\rm{H,U,E1-6}}$ \\ \cite{Pecora2022}$^{\rm{E1-8}}$ 
        \end{tabular} &         
        Local, instantaneous deflections ($z$-angle) of the magnetic field are computed with respect to a chosen background model. Time periods with $z$-angle larger than a prescribed threshold are labeled as a switchback. & 
        Ad-hoc choices of $z$-angles that are not consistent across studies. Generally, the method does not impose a lower limit of window length for switchback patches and/or strength of magnetic field deflections. \\     
        & & & \\
        \hline  
        & & & \\
        \multirow{2}{=}{Manual ``by-eye"}  & 
        \begin{tabular}[t]{@{}l@{}}\cite{Burlaga69}$^{\rm{P}}$\\ 
            \cite{Tsurutani79}$^{\rm{P}}$ \\ \cite{Horbury2001}$^{\rm{G,W,I}}$\\ \cite{Knetter2003}$^{\rm{C}}$ \\ \cite{Knetter2004}$^{\rm{C}}$ \\ \cite{Borovsky2016}$^{\rm{H,A,W,U}}$ \\ \cite{Kasper2019}$^{\rm{E}1,2}$\\ \cite{Woolley2020}$^{\rm{E}1,2}$\\ \cite{Martinovic2021}$^{\rm{E}1,2}$ \\ \cite{Hernandez2021}$^{\rm{E}1}$ \\ \cite{Larosa2021}$^{\rm{E}1}$ \\ \cite{Fedorov2021}$^{\rm{S}}$\\ \cite{McManus2022}$^{\rm{E}3,4}$ \\ \cite{HuangJ2023a}$^{\rm{E}1,2,4-8}$\\ \cite{HuangJ2023b}$^{\rm{E}1,2,4-8}$\\
            \cite{AkhavanTafti2024}$^{\rm{E}1-14}$
        \end{tabular} &
        Generally involves a two-step process with an initial automated timeseries picking followed by a manual or visual screening of events. Step 1 may be based on criteria such as a minimum $z$-angle or strength of magnetic field deflection or a minimum number of data points constituting the spike. &
        Short-lived switchback patches and/or weaker amplitude field deflections might be overlooked due to subjective bias. Choice of Step 1 criteria may lead to variable switchback identification across different studies.  \\ 
        & & & \\
        \hline  
        & & & \\
        Velocity spike & 
        \cite{Horbury2018}$^{\rm{H}}$   &         
        Velocity threshold is used to pick out Alfv\'enic deflections over a smoothed background velocity obtained from a running average using a boxcar function. & 
        No threshold for magnetic field deflection. Mandates a specific ad-hoc choice for velocity spike threshold and length of background averaging window. \\ 
        & & & \\
        \hline
        &&& \\ 
        \multirow{2}{=}{Velocity spike and deflection combination} & 
        \begin{tabular}[t]{@{}l@{}}
        \cite{Farrell2020}$^{\rm{E1}}$ \\ 
        \cite{Farrell2021}$^{\rm{E1}}$ \\ 
        \cite{Rasca2021}$^{\rm{E1,2}}$ \\ 
        \cite{Rasca2022}$^{\rm{E1,2}}$ \\
        \cite{Rasca2023}$^{\rm{E8,12}}$
        \end{tabular} &         
        ``Switchback index'' defined as the product of velocity and magnetic fluctuations, $\delta B_r \cdot \delta V_r$. & 
        The approach prioritizes identifying sharp boundaries. Assumes an averaging timescale (usually 30~s) to define fluctuation quantities.  \\ 
        & & & \\
        \hline  
        & & & \\
        \multirow{2}{=}{MCMC optimization of Gaussian populations} & 
        \cite{Fargette2022}$^{\rm{E}1,2,4-9}$  &
        Continuous timeseries of magnetic field is converted from RTN coordinates to spherical polar coordinates aligned with the Parker spiral to find remnant deflections. Involves Bayesian fitting of these deflections under two Gaussian populations --- quiet solar wind and SBs. &
        Only one dominant population of SBs is assumed to be present in the solar wind. No lower limit for the minimum angle of deflection. \\ 
        & & & \\
        \hline  
        & & & \\
        Individual events &      
        \cite{Telloni2022}$^{\rm{Sm}}$ &
        Isolated switchback structures analyzed with specialized tools on a case-by-case basis. &
        Observed an S-shaped kink structure in white light coronagraph observations, requiring a very large structure with a density enhancement. \\
        & & & \\
        \hline  
    \end{tabular}\label{tab:SB_definitions} 
    \caption{Summary of a few major methods of switchback identification. Superscripts on the references indicate `A' for ACE, `C' for Cluster, `EX' for the X$^{\rm{th}}$ Parker encounter, `G' for Geotail, `H' for Helios, `P' for Pioneer, `S' for Solar Orbiter, `Sm' for Metis coronagraph onboard Solar Orbiter, `U' for Ulysses, and `W' for Wind.}
    \end{table}
    
\subsection{Manual Identification of Switchbacks} \label{subsec: 3_manual_id_of_sb}
    In several studies, the selection of switchbacks to perform statistical studies has been the result of a manual identification by the authors. 
    A two-step process is usually involved, where the raw time-series of observations is first processed using some criteria from Sect.~\ref{sec: 2_SB_definition} definition (field deflection and Alfv\'enic bulk velocity enhancements) and is followed by data visual inspection to build a catalog of switchbacks. 
    While visual inspection necessarily introduces a bias in switchback selection and requires a large workload, it remains an efficient and robust way of identifying switchbacks. 
    Here, we review how some switchback catalogs were built, based partly or totally on visual inspection. 
    
    Visual inspection has been extensively used on past mission data to identify solar wind discontinuities.
    Deflections of the interplanetary magnetic field were visually identified and studied with Pioneer \citep{Tsurutani79, Burlaga69}, the Geotail, Wind, and IMP 8 satellites at 1~AU \citep{Horbury2001}, as well as the Cluster mission \citep{Knetter2003, Knetter2004}. 
    Strong deviations from the Parker spiral direction were also manually identified consistently across four satellites at different radial distances \citep[Helios 1 at 0.3~AU, Wind and ACE at 1~AU, and Ulysses at 2.3~AU,][]{Borovsky2016}. 
    More recently, several papers released switchback catalogs in Parker data based on visual inspection, examples of which follow:        
    A list of 1074 events from E1 and E2 was assembled by \cite{Martinovic2021}. Through multiple independent visual inspections of the magnetic field rotations coincident with bulk velocity enhancements, they further confirmed 921 events with five regions for each switchback: (1) the leading quiet region (LQR) with stable velocities and magnetic fields before the switchback; (2) the leading transition region (LTR), where the magnetic field rotates from LQR toward its switchback orientation; (3) the switchback itself with stable field orientation; (4) the trailing transition region (TTR); and (5) the trailing quiet region (TQR). 
    This method was later used to identify 92 additional switchbacks from E3 and E4 \citep{McManus2022}.
    In parallel, 70 switchbacks were identified during the first encounter by \cite{Larosa2021} based on several signatures, namely a deflection of the magnetic field, an increase in magnetic fluctuation accompanied by an increase in proton bulk velocity and radial temperature.
    Another catalog of 1748 switchback candidates was produced by \cite{HuangJ2023a} and \cite{HuangJ2023b} using data spanning E1--E8 (excluding E3).
    The thresholds were first set in terms of $B_R / B$ and a minimum number of points constituting the switchback spike. This was followed by manual screening of better switchback candidates based on the requirements that suprathermal electrons do not change their main distributions, and that the magnetic and velocity fluctuations are dominantly Alfv\'enic.
    Finally, visual inspection has also been used to identify periods of quiet solar wind and periods of high switchback occurrence \citep{Hernandez2021, Fargette2022}.
    
    All these catalogs have been used to perform statistical analysis of the properties of switchbacks, such as 
    kinetic properties \citep{Martinovic2021, Larosa2021} and variation of plasma parameters inside and outside switchbacks \citep{McManus2022, HuangJ2023b}. 
    These different lists of switchbacks may not be mutually consistent, as the criteria used to identify each structure vary between references. One may also note from table \ref{tab:SB_definitions} that many ``by-eye'' catalogs in the Parker era only examined the first 1-2 encounters. The extent to which existing catalogs are consistent remains to be systematically examined and separated by identification method. However, as will be shown in table \ref{tab: 4_SB_prop}, it is already apparent that the differing identification methods result in very different number statistics, with more strict or laborious definitions necessarily resulting in sparser numbers of events.

\subsection{Alternative Methods and Takeaways} \label{subsec: 3_takeaways}
    In addition to magnetic field deflections and visual identification of individual structures, other methods and criteria have been used to detect switchbacks. For example, some studies have implemented thresholds on the amplitude of ``velocity spikes" \citep{Horbury2018, Kasper2019}.
    The amplitude of the velocity increase, however, depends on the background Alfvén speed, so this velocity threshold is bound to change with radial distance for similar deflection angles.       

    A more probabilistic approach has also been proposed by \cite{Fargette2022}: rather than setting arbitrary thresholds on the magnetic field deflection angle, they 
    assume that two populations of solar wind co-exist and calculate the probability for a given data point to belong to a quiet solar wind interval or to a switchback.  
       
    Finally, \cite{Telloni2022} reported the first possible observation of a switchback in the solar corona. 
    They reported a single large propagating S-shaped 
    structure observed by the Metis coronagraph onboard Solar Orbiter, possibly the signature of a large solar jet. 
    They argued that the switchback structure was generated via interchange reconnection between the coronal loops above an active region and nearby open-field regions. 
    However, by virtue of being observed in white light (implying a global density enhancement of the structure), such a switchback may not be typical since most \textit{in situ} events are not typically elevated in plasma density (see Sect.~\ref{sec: 4_SB_properties}). Further, this remote sensing method necessarily involves line of sight integration which may permit non-unique solutions to the apparent magnetic topology consistent with the observation. More robust inferences would require more information than apparent morphology such as using multiple independent lines of sight or polarization-based 3D localization.
    
    We have seen in this section that several methods exist in the literature to define and identify magnetic switchbacks, each with its own assets and drawbacks.
    Scientific results on switchbacks may vary significantly depending on the choice of background solar wind, the choice of threshold, and the choice of switchback definition. 
    We therefore advise caution when interpreting results, depending on the method used.
    Additionally, the criteria for switchback identification may change as they evolve in the solar wind, and we refer to \citet{Mallet_this_issue} for a deeper understanding of switchback evolution.
    In the remainder of this review, we present results obtained studying the properties of magnetic switchbacks, highlighting where the methodologies used can explain differences in results between studies. 

\section{Switchback Properties and Structure}\label{sec: 4_SB_properties}


We move now to a discussion of the major properties inferred about switchbacks and their implication in terms of origin and impact. We separate these studies into three broad categories: 1) Investigations pertaining to whether switchbacks contain distinct steady plasma populations from the general solar wind (Sect.~\ref{subsec: 4.1_inside_SB}); 2) Studies on switchback geometry and boundaries (Sect.~\ref{subsec: 4.2_geometry_boundaries} and 3) Studies that determine whether switchbacks contain distinct fluctuation characteristics compared to general Alfv\'enic turbulence in the solar wind (Sect.~\ref{subsec: 4.3_turbulence}).

\subsection{Plasma Population Inside Switchbacks} \label{subsec: 4.1_inside_SB}
    
    Exploring the plasma characteristics inside a switchback is important to the question of switchback formation. 
    The processes leading to the origin of switchbacks are still a matter of open debate \citep[see][]{Wyper_this_issue}. 
    Three broad categories include the generation and injection at coronal heights, e.g. by interchange reconnection \citep{Fisk2020, Drake2021, Bale2021}, large-scale velocity shear due to coronal jets or motion of the magnetic footpoint between source regions of fast and slow streams \citep{Landi2006, Schwadron2021, Toth2023arXiv}, and the continuous \textit{in situ} formation by expansion and nonlinear processes, e.g. turbulence \citep{Squire2020,Ruffolo2020,Mallet2021, Shoda2021, Johnston2022, Matteini2024}. 
        
    The presence of a distinct plasma population inside switchbacks, different from the surrounding solar wind, would be a clear indication that switchbacks are the result of direct injection via impulsive processes. If the switchback plasma is indistinguishable from its background, it suggests either switchbacks are formed in the accelerating solar wind, or that the switchback plasma has been able to mix significantly with the background during its transport from the sun. Another reason for the absence of specific plasma signatures might be that an originating impulsive event does not involve direct injection of plasma, for example, a pulse in the magnetic field only. Here, we review investigations pertaining to this overarching question. In the following subsections, two different types of investigations are discussed: studies that compare individual switchback spikes to their immediate surroundings, and studies that consider entire macroscopic streams in which switchbacks occur, and compare them with streams that lack significant switchback signatures, i.e., are ``switchback quiet''. The results often differ between these two classes of studies but in instances when the results agree, the results may be considered more reliable. 
    
    \subsubsection{Steady Plasma Properties Overview} \label{subsubsec: 4.1_overview}

        We begin by collating recent results that compare typical steady plasma properties inside and outside switchbacks. These are presented in Table~\ref{tab: 4_SB_prop}, where quantities such as magnetic field magnitude, the proton and alpha density, proton velocity, plasma $\beta$, pressure and electron strahl intensity and anisotropy inside and outside switchbacks are contrasted.. The parameters are drawn from several statistical studies, and the methodology used to identify switchbacks is noted in the table. For each entry, the property inside individual switchbacks is compared to its typical value immediately outside. As shown in the third column, these studies draw from different encounters (and phases of encounters therein) and so it should be noted that differing solar wind conditions, distances from the sun and direction of motion of the spacecraft are generally not differentiated in such studies.
        
        \begin{table}
        \begin{tabular}{p{0.1\textwidth}|p{0.17\textwidth}|p{0.12\textwidth}|p{0.18\textwidth}|p{0.3\textwidth}}
            \hline
            \textbf{Parameter} & \textbf{Interior/Exterior} & \textbf{Encounters} & \textbf{Switchback Definition and Number} & \textbf{Reference} \\ \hline
            $|B|$    & \begin{tabular}[c]{@{}c@{}} $\simeq 1$ \\ Median=0.98 \end{tabular}    & Six days in E1   &  Manual “by-eye”, 70   &  \cite{Larosa2021} \\ 
                     & $\simeq 1$   & E1-E2   &  Manual “by-eye”, 1074   &  \cite{Martinovic2021} \\ 
                     & \begin{tabular}[c]{@{}c@{}} Median = 1 \\  \end{tabular}    & E1-E2, E4-E10   &  Thresholding   &  Fig.~\ref{fig:plasma ratios Vamsee} and \cite{Jagarlamudi2023}\\ \hline
            $N_e (QTN)$    & \begin{tabular}[c]{@{}c@{}} $\simeq 1$ \\ Median=0.98 \end{tabular} & E1-E2, E4-E10   &  Thresholding   &  Fig.~\ref{fig:plasma ratios Vamsee} and \cite{Jagarlamudi2023} \\ \hline
            $N_p$    & \begin{tabular}[c]{@{}c@{}} $\simeq 1$ \\ Median=1.00 \end{tabular}    & Six days in E1   &  Manual “by-eye”, 70   &  \cite{Larosa2021} \\ 
                     & $\simeq 1$   & E1-E2   &  Manual “by-eye”, 1074   &  \cite{Martinovic2021} \\ \hline
            $N_{\alpha}/N_p$    & $\simeq 1$  &  E3-E4   &  Manual “by-eye”, 92   &  \cite{McManus2022} \\ 
                                & Similar distribution  &  E1-E8 (no E3)   &  Manual “by-eye”, 1748   & \cite{HuangJ2023b} \\ \hline
            $V_p$    & \begin{tabular}[c]{@{}c@{}} $> 1$ \\ Median=1.18 \end{tabular}    & Six days in E1   & Manual “by-eye”, 70  &  \cite{Larosa2021} \\ 
                     & increment of $V_A/2$   & E1-E2   &  Manual “by-eye”, 1074   &  \cite{Martinovic2021} \\ 
                     & \begin{tabular}[c]{@{}c@{}} $> 1$ \\ Median=1.1 \end{tabular}    & E1-E2 $\&$ E4-E10   &  Thresholding   &  Fig.~\ref{fig:plasma ratios Vamsee} and \cite{Jagarlamudi2023}\\  \hline
            $V_{\alpha p}/V_A$  & Similar distribution  &  E1-E8 (no E3)   &  Manual “by-eye”, 1748   & \cite{HuangJ2023b} \\ \hline
            $\beta$  & \begin{tabular}[c]{@{}c@{}} $> 1$ \\ Median=1.45 \end{tabular}    & Six days in E1   &  Manual “by-eye”, 70   &  \cite{Larosa2021} \\ \hline
            $P_K$           & \begin{tabular}[c]{@{}c@{}} Slightly larger $\geq 1$ \\ Median=1.12$\pm$0.21 \end{tabular}  &  E1-E8 (no E3)   &  Manual “by-eye”, 1748   & \cite{HuangJ2023a}  \\ \hline
            $P_B$           &  \begin{tabular}[c]{@{}c@{}} Slightly smaller $\leq 1$ \\ Median=0.90$\pm$0.73 \end{tabular}  &  E1-E8 (no E3)   &  Manual “by-eye”, 1748   & \cite{HuangJ2023a}  \\ \hline
            $P_{total}$     &  \begin{tabular}[c]{@{}c@{}} $\simeq 1$ \\ Median=1.05$\pm$0.12 \end{tabular} &  E1-E8 (no E3)   &  Manual “by-eye”, 1748   & \cite{HuangJ2023a}  \\ \hline
            e-PADs  $A_E$   &  \begin{tabular}[c]{@{}c@{}} $> 1$ \\ Median=1.30$\pm$0.45 \end{tabular} &  E1-E8 (no E3)   &  Manual “by-eye”, 1748   & \cite{HuangJ2023a}  \\ \hline
            e-PADs  $F_E$   &  \begin{tabular}[c]{@{}c@{}} $\simeq 1$ \\ Median=1.00$\pm$0.01 \end{tabular} &  E1-E8 (no E3)   &  Manual “by-eye”, 1748   & \cite{HuangJ2023a}  \\ \hline
        \end{tabular}
        \caption{Comparison of general plasma properties between the interior and exterior of switchbacks. From left to right, the columns present the parameters, Inside-to-Outside comparisons, encounters of the data, the switchback definition in Table~\ref{tab:SB_definitions}, switchback numbers, and the references. From top to bottom, the parameters include magnetic field strength $|B|$, electron number density $N_e$, proton number density $N_p$, alpha-to-proton abundance $N_{\alpha}/N_p$, proton bulk speed $V_p$, alpha-to-proton differential speed normalized by local Alfvén speed $V_{\alpha p}/V_A$, plasma beta $\beta$, kinetic pressure $P_K$, magnetic pressure $P_B$, total pressure $P_{total}$, anisotropy ($A_E$) and integral intensity ($F_E$) of suprathermal electron pitch angle distributions (e-PADs). }\label{tab: 4_SB_prop}
        \end{table}
        
        Surveying this table, almost all plasma properties remain unchanged. This includes the magnetic field magnitude, electron and proton number density, alpha-to-proton abundance ratio, and differential alpha streaming (normalized by the Alfv\'en speed).
        This implies switchbacks are mostly Alfv\'enic fluctuations for which protons and alpha velocities rotate in rigid rotation in a frame co-moving with the switchback \citep{Neugebauer2013, Matteini2015_dHT, McManus2022}. 
        The latter paper points out that, while a lack of compositional difference implies \textit{in situ} generation, a clear compositional signature could be hidden by differential velocity between the switchback wave speed and background solar wind.
        
        The sum of kinetic (proton, alpha, and electron) pressure and magnetic pressure also remains constant during typical switchback fluctuations, implying the structures are in pressure balance with their surroundings. 
        Lastly, while the suprathermal electron pitch angle distribution remains strongly peaked in a consistent direction either parallel or antiparallel to the magnetic field throughout each switchback \citep[a key marker identifying switchbacks as topological folds rather than current sheets or loops,][]{Kasper2019}, the total distribution does exhibit signs of isotropization \citep[the magnitude of the suprathermal electron pitch angle anisotropy, $A_E$, decreases,][]{HuangJ2023b} suggesting switchbacks can scatter electrons to some extent. This finding, which comes from Parker/SPAN-e data, should be revisited to assess the potential impact of switchback-related Doppler shifts: the higher energy tail of the thermal population could appear in suprathermal energy ranges during the large velocity spikes comprising switchbacks.

\begin{figure}
            \centering
            \includegraphics[width=\textwidth]{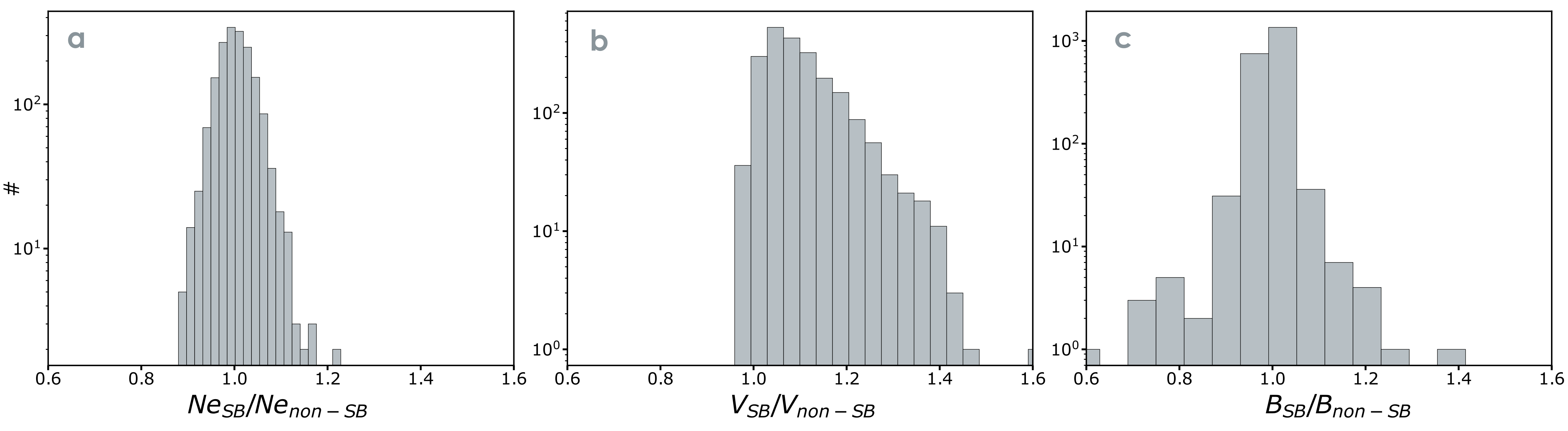}%
            \caption{Histogram of plasma parameter ratios between the switchback to non-switchback regions. Panel (a) electron density ratios, Panel (b) Bulk velocity ratios, and Panel (c) magnetic field magnitude ratios. \citep[selected regions from][]{Jagarlamudi2023}}
            \label{fig:plasma ratios Vamsee}
        \end{figure}

        The above results imply \textit{individual} switchback's plasma properties are largely not distinct from those of their surroundings. A different way to analyze these properties is to compare the plasma in intervals with many consecutive switchbacks with adjacent ``quiet'' intervals. Using this method, one finds that the plasma properties populate a finite width distribution whose peak generally matches the results presented in Table~\ref{tab: 4_SB_prop} for individual spikes. In Fig.~\ref{fig:plasma ratios Vamsee}, we present statistical distributions of the electron density, velocity, and magnetic field magnitude ratios between switchback and non-switchback regions \citep[selected regions from][]{Jagarlamudi2023}. Non-switchback intervals are defined by requiring that the deflection parameter (see Sect.~\ref{subsubsec: 3_threshold}) satisfy $z<0.1$ at least $60\%$ of the time or the mean of $z$ values inside the region is less than 0.1. The minimum duration of a non-switchback interval used is 20s. Intervals are chosen so that switchback and non-switchback intervals are adjacent to each other in time to avoid comparison of plasmas with very different background solar wind conditions.
        
        We observe that the in-out ratio of magnetic field magnitudes and densities is nearly one for the majority of the intervals. However, there are a non-negligible number of intervals for which the ratios are sometimes higher and sometimes lower \citep[see also][]{Larosa2021}. 

        Moreover, while usually the magnitude field magnitude, pressure, and energy density are approximately constant in, i.e., fractional changes are much less than 1 (as shown above), it is often the case that switchbacks cause more of a disruption to magnetic pressure balance (and conservation of $|B|$) than the surrounding plasma. \citet{Ruffolo2021} examined domains of nearly constant $|B|$ using Parker observations, and concluded that many (but not all) transitions between different regions of uniform $|B|$ were associated with switchbacks. In this sense, these constant-$|B|$ domains often represent regions of space in between the switchbacks or in between the boundaries of switchback patches, as will be discussed further in Sect.~\ref{sec: 5_sw_structure}. As for the bulk velocity, a clear one-sided distribution in velocity is observed for switchbacks \citep{Gosling2009}. This is consistent with their identification as Alfv\'enic velocity spikes and readily explained in terms of outwardly propagating Alfv\'enic fluctuations (see Sect.~\ref{sec: 2_SB_definition}).

    \subsubsection{Switchback Proton and Alpha Temperatures}  \label{subsubsec: 4.1_proton_alpha}
        
        The previous subsection did not discuss any thermodynamic properties of the plasma and in particular proton and alpha particle temperatures and anisotropies. The reason is that obtaining such quantities is complicated by instrumental effects arising from the finite portion of velocity space observable at a given instant by the SWEAP instrument (see Sect.~\ref{subsec: 1_instru}). The observed velocity distribution functions (VDFs) can systematically move out of the instrument field of view during a switchback due to the velocity perturbation of the structure, distorting the measured temperature. The way such instrumental effects are modeled can easily affect the trends reported in the Parker Solar Probe literature. Nevertheless, it is an area of high interest, since it determines whether the switchback plasma may be considered to be in thermal equilibrium with its surroundings. The comparison of proton and alpha particle parallel and perpendicular temperatures and their ratios inside and outside switchbacks, together with the methodology used to obtain them, are summarized in Table~\ref{table:sb-temp-Huang}. 
        
        \begin{table}
        \caption{Comparison of various plasma temperatures between inside and outside switchbacks where the subscripts ($p,\alpha,\perp$ and $\parallel$) refer respectively to protons, alphas, and perpendicular or parallel components of the temperature tensor with respect to the orientation of the magnetic field. From left to right, the columns present the parameters, Inside-to-Outside results, encounters of the data, the selection methods of the switchbacks, and the related notes and references.}
        \begin{tabular}{c|c|c|p{0.25\textwidth}|c}
        \hline
            \textbf{Parameters}   & \textbf{Inside/Outside}   & \textbf{Encounters} & \textbf{Switchback Definition and Number Statistics} &  \textbf{Reference} \\ \hline
     
            $T_{p}$          	& $\simeq 1$   & E1-E2   &   Manual “by-eye”, 1074   &  \cite{Martinovic2021} \\
                                & \begin{tabular}[c]{@{}c@{}} $> 1$ \\ Median=1.12$\pm$0.20 \end{tabular} &  E1-E8 (no E3)   &   Manual “by-eye”, 1748   & \cite{HuangJ2023a}\\ 
                                & $> 1$   & E1-E2   &   ``Switchback Index'', 25   &  \cite{Rasca2021} \\ \hline
            $T_{\parallel p}$	& $\simeq 1$  &  E1-E2   &  Deflection thresholds, 5   & \cite{Woolley2020} \\
                                & $> 1$   	&  six hours in E2   &   Manual “by-eye”, 6 patches   & \cite{Woodham2021} \\
                                & \begin{tabular}[c]{@{}c@{}} $> 1$ \\ Median=1.26$\pm$0.50 \end{tabular} &  E1-E8 (no E3)   &  Manual “by-eye”,  1748   & \cite{HuangJ2023a}\\
                                & $> 1$   	&  E5-E11 (no E9)   &   \begin{tabular}[c]{@{}c@{}} Deflection thresholds \\ 2-4 days in each encounter \end{tabular} & \cite{Laker2024} \\ \hline
            $T_{\perp p}$    	& $\simeq 1$   &  six hours in E2  &   Manual “by-eye”, 6 patches   & \cite{Woodham2021} \\
                                & \begin{tabular}[c]{@{}c@{}} $> 1$ \\ Median=1.06$\pm$0.18 \end{tabular} &  E1-E8 (no E3)   &   Manual “by-eye”, 1748   & \cite{HuangJ2023a}\\
                                & $> 1$   	&  E5-E11 (no E9)   &  \begin{tabular}[c]{@{}c@{}} Deflection thresholds \\ 2-4 days in each encounter \end{tabular}  & \cite{Laker2024} \\ \hline
            $T_{\alpha}$                 	& \begin{tabular}[c]{@{}c@{}} $< 1$ \\ Median=0.86$\pm$0.44 \end{tabular} &  E1-E8 (no E3)   &   Manual “by-eye”, 1748   & \cite{HuangJ2023a}\\ \hline
            $T_{\parallel \alpha}$       	& \begin{tabular}[c]{@{}c@{}} $< 1$ \\ Median=0.82$\pm$0.48 \end{tabular} &  E1-E8 (no E3)   &   Manual “by-eye”, 1748   & \cite{HuangJ2023a}\\ \hline
            $T_{\perp \alpha}$           	& \begin{tabular}[c]{@{}c@{}} $< 1$ \\ Median=0.85$\pm$0.53 \end{tabular} &  E1-E8 (no E3)   &   Manual “by-eye”, 1748   & \cite{HuangJ2023a}\\ \hline
            $T_{\alpha}/T_{p}$                     	& \begin{tabular}[c]{@{}c@{}} $< 1$ \\ Median=0.75$\pm$0.41 \end{tabular} &  E1-E8 (no E3)   &   Manual “by-eye”, 1748   & \cite{HuangJ2023a}\\ \hline
            $T_{\parallel \alpha}/T_{\parallel p}$         	& \begin{tabular}[c]{@{}c@{}} $< 1$ \\ Median=0.56$\pm$0.40 \end{tabular} &  E1-E8 (no E3)   &   Manual “by-eye”, 1748   & \cite{HuangJ2023a}\\ \hline
            $T_{\perp \alpha}/T_{\perp p}$ 	& \begin{tabular}[c]{@{}c@{}} $< 1$ \\ Median=0.79$\pm$0.48 \end{tabular} &  E1-E8 (no E3)   &   Manual “by-eye”, 1748   & \cite{HuangJ2023a}\\ \hline
         
        \end{tabular}
       \label{table:sb-temp-Huang}
        \end{table}
     
        As shown in the table, the results are generally mixed, but with a preponderance of results suggesting the proton temperature inside switchbacks is generally higher than in their surroundings. Results are also mixed when considering temperature anisotropy, with some studies concluding that the parallel and perpendicular temperatures are individually enhanced or unchanged. No studies report examples of lower proton temperature (parallel, perpendicular, or scalar). 
        
        On the other hand, Table~\ref{table:sb-temp-Huang} shows that the alpha temperature in switchbacks is typically lower with respect to their surroundings, both in absolute terms and with respect to the ratio to proton temperatures. These results remain somewhat controversial, and future studies are needed that carefully follow the systematic evolution of the VDFs during switchbacks. Close to the Sun, such studies when possible would ideally use both plasma instruments together on Parker (SPAN-i and SPC) to reduce field of view issues, while \textbf{farther} from the Sun, the excellent angular coverage of the Proton-Alpha sensor in the Solar Orbiter's Solar Wind Analyser suite \citep[SWA/PAS;][]{Owen2020} instrument should be capitalized on. To highlight the importance of instrumental considerations, Table~\ref{tab:sbhot} presents a few studies that use temperature measurements by Parker/SWEAP in differing ways and from different periods of the mission with careful attention to systematic measurement effects. We go into some detail to highlight that this is a non-trivial task.
        
         \begin{table}
            \caption{Comparison of results on proton temperature evolution during switchbacks (SBs) as observed on Parker Solar Probe. For each study, the dataset, the switchback definition used, the instrumental geometry effects considered, and the ultimate takeaways are listed.}
            \begin{tabular}{cccc}
            \hline &\citet{Woolley2020} & \citet{Woodham2021} & \cite{Laker2024} 
            \\ \hline Data  
            & \begin{tabular}[c]{@{}c@{}} 
            SPC Proton Core Fits (1D)\\ E01,E02
            \end{tabular} 
            & \begin{tabular}[c]{@{}c@{}} 
            SPAN-i Proton Core Fits (3D)\\ E02
            \end{tabular} 
            & \begin{tabular} [c]{@{}c@{}}
            SPAN-i Proton Core Fits (3D)\\ E05-E11
            \end{tabular}
            \\ \hline SB Definition 
            &  Full reversal in $B_R$ 
            & \begin{tabular} [c]{@{}c@{}}
            Patches selected by eye.
            \end{tabular} 
            & \begin{tabular}[c]{@{}c@{}} 
            No threshold deflection set, \\ 
            general quasi-periodic deflections 
            \end{tabular} 
            \\ \hline 
            \begin{tabular}[c]{@{}c@{}} 
            Geometrical \\ Considerations 
            \end{tabular}
            & \begin{tabular}[c]{@{}c@{}} 
            Measured deviation from \\
            theoretical instrument \\ response.
            \end{tabular}
            & \begin{tabular}[c]{@{}c@{}} 
            Instances of large B field deflections \\ ($ > 10^o$)  occurring faster than the \\ SPAN-i integration time are \\ rejected.
            \end{tabular}
            & \begin{tabular}[c]{@{}c@{}} 
            Data during large positive \\ 
            s/c frame tangential flow angle \\ deflections 
            ($< -5^\circ $)
            rejected.
            \end{tabular}
            \\ \hline Conclusions: 
            &  Individual SBs unchanged
            &  SB patches hotter
            & \begin{tabular}[c]{@{}c@{}} 
            Groups of neighboring\\ SBs hotter
            \end{tabular}
            \\ \hline
            \end{tabular}
            \label{tab:sbhot}
        \end{table}

        \cite{Woolley2020} used proton core fits from SPC, which captures the 1D reduced VDF best along the radial direction, meaning only switchbacks with a near-full $B_R$ reversal were considered. SPC data show clear temperature enhancements in lockstep with the magnetic field vector \citep[see also][]{Rasca2021}. However, these were compared to the artificial dependence of the temperature on magnetic field orientation dependence common to Faraday Cups \citep{Kasper2002,Huang2020_temperature}. Plotting the radial component of the proton temperature tensor as a function of the instantaneous angle between the magnetic field and radial direction \citep[see figure~4 of][]{Woolley2020} showed little deviation from this geometrical prediction. \citet{Woolley2020} therefore concludes that switchbacks do not appear hotter in SPC measurements, although they caution about the instrumental ambiguity and that SPC may therefore not be able to measure a temperature enhancement during large deflections.
        On the other hand, \cite{Woodham2021} reported different conclusions but using SPAN-I 3D bi-Maxwellian proton core fits and expanding the switchback definition to general 3D field deflections superimposed on a larger scale ``patch'' (see Sect.~\ref{sec: 5_sw_structure}), evoking a picture of individual switchbacks embedded in a larger 3D structure with systematic flow properties. They considered intervals with minimal artificial temperature broadening effects shared by electrostatic analyzers (ESAs) \citep{Verscharen:2011} by rejecting time intervals where the field deflected significantly (as compared to the detector angular resolution) during intervals faster than a SPAN-ion integration time (see Table~\ref{tab:sbhot}). They observed  $T_\parallel$ enhancements over large-scale patches of many switchbacks (see also Sect.~\ref{sec: 5_sw_structure}) and argued similar enhancements could be observed in individual spikes. They interpreted this as evidence that patches of switchbacks may have a common solar origin. The competing conclusion found in \cite{Woolley2020} was reconciled with a suggestion that the specific relationship between $T_\parallel$ and the field orientation could be the same as the artificial SPC response ($T_\parallel \propto \sin\Theta_{RB}$ where $\Theta_{RB}$ is the angle between the magnetic field and the radial direction) such that an actual temperature enhancement signature may be indistinguishable from the systematic instrumental signature observed with SPC data in \citet{Woolley2020}. They suggested that future work with later orbits, where SPAN-i field of view issues were less severe, would be necessary to resolve this.
    
        \cite{Laker2024} expanded upon the results of \cite{Woodham2021} for later encounters E05-E11, using the same proton core bi-Maxwellian fitting methodology.  They highlighted that in later Parker encounters, the systematic azimuthal motion of the spacecraft in the solar wind frame means VDFs are more completely sampled by the SPAN instrument. They loosened the switchback definition by not defining a threshold for deflection angle. By doing so, the switchback was then generalized to include quasi-periodic deflections (on a minute timescale) of the full magnetic field vector. They interpreted these Alfv\'enic quasi-periodic patterns as torsional Alfvén waves propagating from interchange reconnection sites in the corona. Associated with these quasi-periodic deflections, they found quasi-periodic variation in the proton core temperature (both $T_\perp$ and $T_\parallel$). They additionally found a positive correlation between the change in perpendicular and parallel temperatures during the switchbacks and $\theta_{RB}$, and argued that this does not rule out an \textit{in situ} switchback generation mechanism. To ensure reliability of the SPAN-I measurements, they avoided intervals of large $V_T$ flows that swung the proton core out of the SPAN-I field of view, defining a threshold angle that ensures the angle between the flow and spacecraft frame velocity does not deviate from radial in the +T direction. 
    
        By not applying a deflection threshold to the switchback definition, a less extreme underlying structure was identified in \cite{Laker2024}, defining individual switchbacks as sub-structures. This slightly differs from \cite{Woodham2021}, who defined the switchback based on deflection angle, and concluded increases of $T_\parallel$ were linked to switchback patches. \cite{Laker2024} instead identifies the deflection and $T_\parallel$ increase as the bigger structure, with an embedded sub-structure of constant $T_\parallel$. To reconcile the conclusions in \cite{Woolley2020}, \cite{Laker2021} also points out that in these later encounters, the switchbacks do not undergo a full deflection, and there is no SPC vs SPAN comparison available (due to data gaps and anti-coincident reliability). \\

    \subsubsection{Steady Plasma Properties Takeaways} \label{subsubsec: 4.1_takeaway}

        Overall, switchback steady plasma properties are mostly consistent with that of steepened large amplitude Alfv\'en waves in which the magnetic field and velocity vector instantaneously vary together while most other plasma properties stay close to consistent with the surrounding wind. That is, they are largely incompressible in terms of the magnitude of the magnetic field and the plasma density (see Figure \ref{fig:plasma ratios Vamsee}). 

        The notable exception to this picture is the temperature for which there are indications of potential enhancements and also modulation in anisotropy. Instrumental considerations and limited statistics mean that further work is needed to establish this concretely. Similarly, limited by instrumental capabilities, the presence of proton and alpha beams, alpha abundance and any evidence of heavier species modulation in and outside switchbacks remains an open question. Strategies including detailed velocity distribution function reconstruction at Parker, full VDF measurements at Solar Orbiter \textbf{farther} from the Sun, and combining multi-spacecraft instruments during conjunctions, are needed to make progress here. Further, the sources for these findings generally only span up to E11 and need updating with more recent encounters which have taken Parker even closer to the Sun, potentially swinging the VDF more favorably into the SPAN-I field of view.
            
\subsection{Switchback Occurrence, Geometry, and Boundaries} \label{subsec: 4.2_geometry_boundaries}  
    Switchback structure, occurrence rate, boundary evolution may provide fundamental clues both into the origin of switchbacks, their growth and decay, as well as to the question of how switchbacks interact and exchange plasma and energy with the background solar wind. 
    One major feature of switchbacks relevant to these points is that they have ``sharp'' boundaries, and their magnetic and velocity vectors rotate on a sphere in a specific way. Here we collate results that investigate switchback geometry and the properties of switchback boundaries, to shed light on the nature of these fluctuations. We first introduce occurrence statistics based on the classification of these structures as sharp deflections (Sect.~\ref{subsubsec: 4.2_occurrence}).
    Second, we discuss their spatial scales and geometrical aspects in Sect.~\ref{subsubsec: 4.2_SB_geometry}. Lastly, we present a series of discussions on the nature of the switchback boundaries: their non-Alfv\'enic characteristics (Sect.~\ref{subsubsec: 4.2_SB_boundaries}), their classification as discontinuities (Sect.~\ref{subsubsec: 4.2_discontinuities}), their relation to magnetic reconnection (Sect.~\ref{subsubsec: 4.2_reconnection}), and lastly their relationship to electromagnetic wave activity (Sect.~\ref{subsubsec: 4.2_waves_boundaries}). 
    
    \subsubsection{Deflection Angle and Duration Statistics} \label{subsubsec: 4.2_occurrence} 

        Establishing the distribution of switchback sizes, occurrence rates, and lifetimes provides powerful diagnostic information regarding their generation, decay, and physical nature. Automated identification and characterization of switchbacks (see Sect.~\ref{sec: 3_methodology}) enable in-depth studies of these statistical properties. 
        Here, we highlight some results based on a recent reanalysis of Parker data from E1 to E17, identifying switchbacks as sharp deflections from the Parker spiral as characterized by changes in the normalized deflection parameter, $z$ \citep{DudokdeWit2020}. 
  
\begin{figure}
            \includegraphics[width=\textwidth]{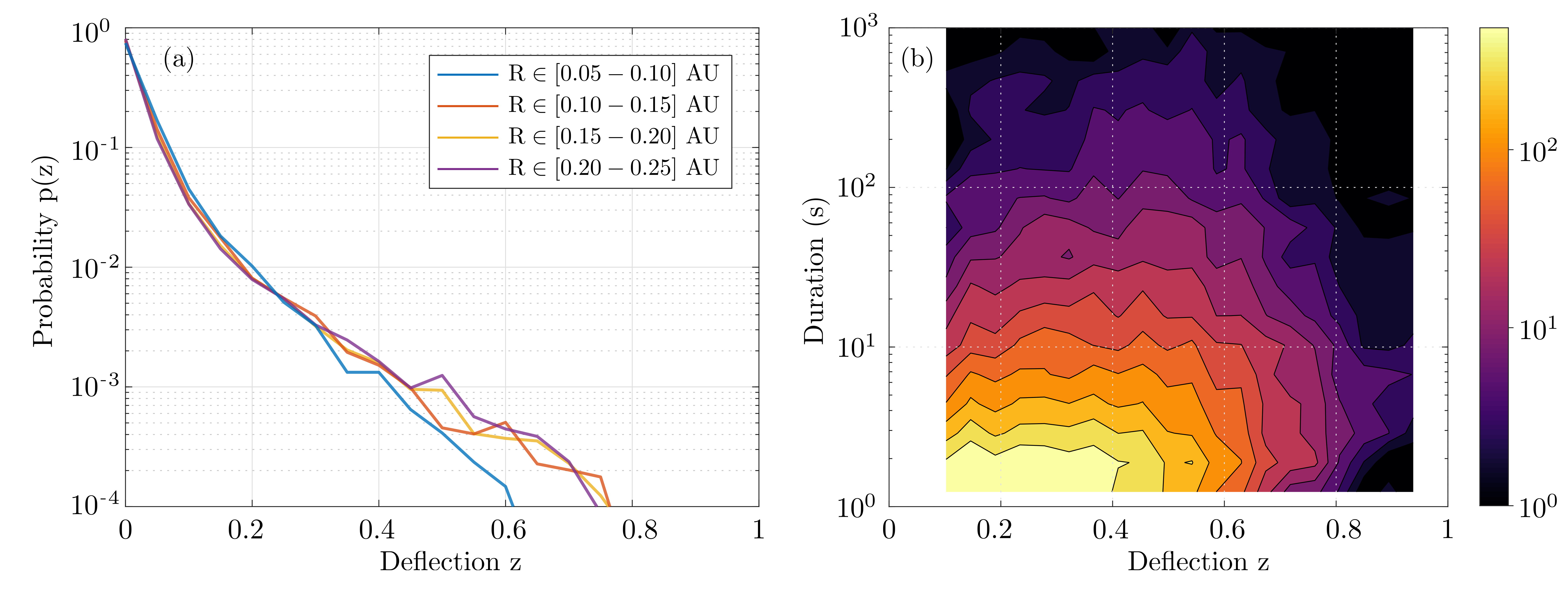}
            \caption{Left panel (a): Probability distribution of the normalized deflection $z$ for Parker data from E1 to E17, binned by distance from the Sun in AU. Right panel (b): 2D histogram versus the duration and the normalized deflection $z$ for E1--E17, using the same dataset.}
            \label{fig: 4.2_deflection_duration}
        \end{figure}

        Figure~\ref{fig: 4.2_deflection_duration} disaggregates what was previously shown in Fig.~\ref{fig: 3_z_angle_annotated}. 
        In the left panel, the distribution of switchback deflections is shown for different distances from the Sun.         
        These distributions are close to exponential except for an excess of very small deflections with $z<0.1$. 
        As mentioned before in Sect.~\ref{sec: 3_methodology}, such small deflections are increasingly difficult to distinguish from stochastic fluctuations, and for that reason, ought to be discarded, even if this excludes some which technically meet the definition presented in Sect.~\ref{sec: 2_SB_definition}. In what follows, we exclude these smallest deflections ($z<0.1$) to avoid ambiguity with background solar wind. 
        
        From Fig.~(\ref{fig: 4.2_deflection_duration}), we first observe a continuum of magnitudes with an abundance of smaller ones. 
        Out of these, less than 2\% can be called full-reversal switchbacks because they have a deflection of more than \ang{90}. 
        Numerous other statistical indicators \citep[e.g.,][]{DudokdeWit2020} suggest that small and large deflections belong to the same statistical distribution. 
        This stresses the importance of considering all deflections as relevant, and not focusing only on full-reversal switchbacks of more than \ang{90}.   
        Secondly, Fig.~\ref{fig: 4.2_deflection_duration}a shows that the probability distribution remains remarkably constant when approaching the Sun, except for a growing deficit of large deflections within 0.1 AU \citep[see also][]{Bandyopadhyay2022, Larosa2024}. 
        The distribution does somewhat vary with solar wind conditions, i.e., velocity and other plasma parameters.
        However because all data are included in Fig.~\ref{fig: 4.2_deflection_duration} one gets the (false) impression that the distribution is largely invariant with distance.
        Establishing what causes the distribution of $z$ to vary will be crucial for constraining the generation mechanisms of switchbacks. 
        This and other disambiguation of switchback occurrence by solar wind type currently remain largely understudied (see Sect.~\ref{subsec: 5_streams}).

        The same statistics allow us to investigate the connection between switchback lifetime and deflection. We show in Fig.~\ref{fig: 4.2_deflection_duration}b the joint distribution of these key parameters, which shows they are poorly correlated. Very short switchbacks with a duration of typically less than 2~s predominantly correspond to small deflections of much less than \ang{90}. This may be due to the finite thickness of their boundaries (see Sect.~\ref{subsubsec: 4.2_SB_boundaries}), 
        which may translate into a duration in the spacecraft frame which is greater $\ge 2$~s \citep{Larosa2021,Bizien2023} thus limiting the magnitude of the deflection for very short switchbacks. 
        Though one sees that with increasing duration the contours of equal counts extend to greater deflections, the two parameters are essentially disconnected beyond a few seconds duration. 
        This suggests that deflection and duration are most likely driven by different physical processes.        
        These simple occurrence statistics, therefore, evidently contain powerful diagnostic information but are currently underutilized and underinterpreted.
    
    \subsubsection{Size, Shape, and Orientation} \label{subsubsec: 4.2_SB_geometry}

        The size, shape, and orientation of switchback structures are important for understanding their origins, evolution, and impact on the solar wind. 
        While many thousands of switchbacks have been measured, determining their characteristics is challenging: the structures are very variable in amplitude and duration; they contain considerable substructure; 
        and as we have seen, different researchers have used differing criteria to define switchbacks so there is no observational standard.
        Nevertheless, it is possible to estimate the properties of switchbacks, at least in a statistical sense, if one assumes that these are the result of the structures passing over the spacecraft.

    
        The angular distribution of magnetic fluctuations reveals an isotropic distribution of the smallest deflections \citep{DudokdeWit2020, Fargette2022}. Large switchbacks in early Parker data (E1, E2), however, tend to deflect in the +T direction of the RTN frame \citep{Horbury2020, Meng2022}. 
        \citet{Fargette2022}, by splitting fluctuations into an isotropic core population and a non-isotropic tail, shows that this is statistically verified for several encounters (E1--E9), as the switchback magnetic field rotates preferentially in the clockwise direction in the ecliptic plane (i.e., $+$T when the magnetic field polarity is negative, and $-$T when the polarity is positive). 
        Such an asymmetry also shows up in Helios observations \citep{Macneil2020}. 
        This bias is typically what would be expected from an interchange reconnection process acting lower in the corona \citep{Fisk2020}, and has been shown to be consistent with Alfvén waves propagating along the Parker spiral \citep{Squire2022, Johnston2022}. In each of these cases, the key underlying symmetry-breaking is the sense of solar rotation.
    
        Over the time that a switchback is measured, a spacecraft travels at a constant velocity. 
        Historically, this has been small compared to the solar wind speed, but for Parker near perihelion with spacecraft speeds over 150~\kms\ and rather slow wind speeds, this is not the case.
        We must therefore consider both the spacecraft and plasma velocities when determining how a switchback is sampled. 
        With this in mind and motivated by the observation that switchbacks during E1 had on average a longer duration at the spacecraft when the field within them deflected in the $+$T direction, \cite{Horbury2020} calculated the velocity with which the spacecraft crosses through each structure (the vector sum of the spacecraft velocity with respect to the Sun and the plasma velocity within it). This analysis showed that the distribution of durations was consistent with approximately radially aligned structures, which were tens of thousands of km across, with high aspect ratios -- that is, they were long and thin. 
        This analysis is possible because of the Alfv\'enic nature of switchbacks, where, within a stream of a particular magnetic polarity, there is a consistent sense of correlation or anti-correlation between vector magnetic field and velocity deflections. That is, both the switchbacks and the local solar wind have well-defined, fixed common deHoffman-Teller frames \citep{Horbury2020, Agapitov2023ApJ}. 
        \cite{Laker2021} extended this work to multiple encounters, and with more detailed analysis, demonstrated that the apparent durations were consistent with structures with widths around 50\,000~km and aspect ratios of order 10, with the major axis deflected away from the radial direction but in the sense of the Parker spiral -- with consistent behavior within a stream, but quite different behavior from one stream to another. 
        They therefore concluded that the switchbacks were approximately aligned with the Parker spiral, and, hence, could be propagating along it. 
        Unlike in E1, there was not a preponderance of switchback deflections in the $+$T direction, with different streams having different average deflection directions. 

       Independent estimates of the transverse size scale of individual switchbacks were also determined by \citep{Krasnoselskikh2020}, \citep{Larosa2021} and \citep{Bandyopadhyay2021}, all yielding values in the range of 10$^4$ km. See \citet{Raouafi2023a} Table 1 for a numerical comparison. Lacking estimates which use more recent data (the latest data used for this purpose was up to E5 in \citet{Bandyopadhyay2021}), it remains an open question as to how these size scales evolve with distance from the Sun and with Alfv\'en Mach number.
    
        The importance of the deflection direction of switchbacks motivated \cite{Laker2022} to consider them in more detail. 
        The deflection direction of the magnetic field within a particular switchback is typically approximately constant: that is, they are arc-polarized structures, deflecting away from the background and returning along the same direction. 
        This makes it possible to define a single deflection direction of each switchback and study its distribution. 
        These directions are not isotropic, but within particular streams, they have a preponderance to particular directions \citep{Laker2022, Fargette2022}. 
        As was seen in the first encounter, switchbacks that are close together in time often have the same deflection direction. Surprisingly, over the first eight Parker encounters,  there seemed to be a preference for deflections in the tangential ($+$T and $-$T) directions. 
        \cite{Laker2022} interpreted the clustering of deflection directions as possible evidence for a reconnection origin of switchbacks. 
        Most recently, \cite{Laker2024} interpreted this clustering as an indication that individual switchbacks can be part of a larger quasi-periodic structure with the same deflection sense and noted that this is consistent with simulations of the outflows from magnetic reconnection events \citep{Wyper2022}. 
        Importantly, during the period of consistent switchback deflections, there was an overall increase in proton field-parallel temperature, as first noted by \cite{Woodham2021}, and this was above that expected from the well-known temperature-speed relation \citep{Woolley2020}. 
        Again, this was interpreted as evidence of a possible reconnection-related origin of these events.

    \subsubsection{Non-Alfv\'enic Boundary Features} \label{subsubsec: 4.2_SB_boundaries}
                
        \begin{figure}
            \centering
            \includegraphics[width=\textwidth]{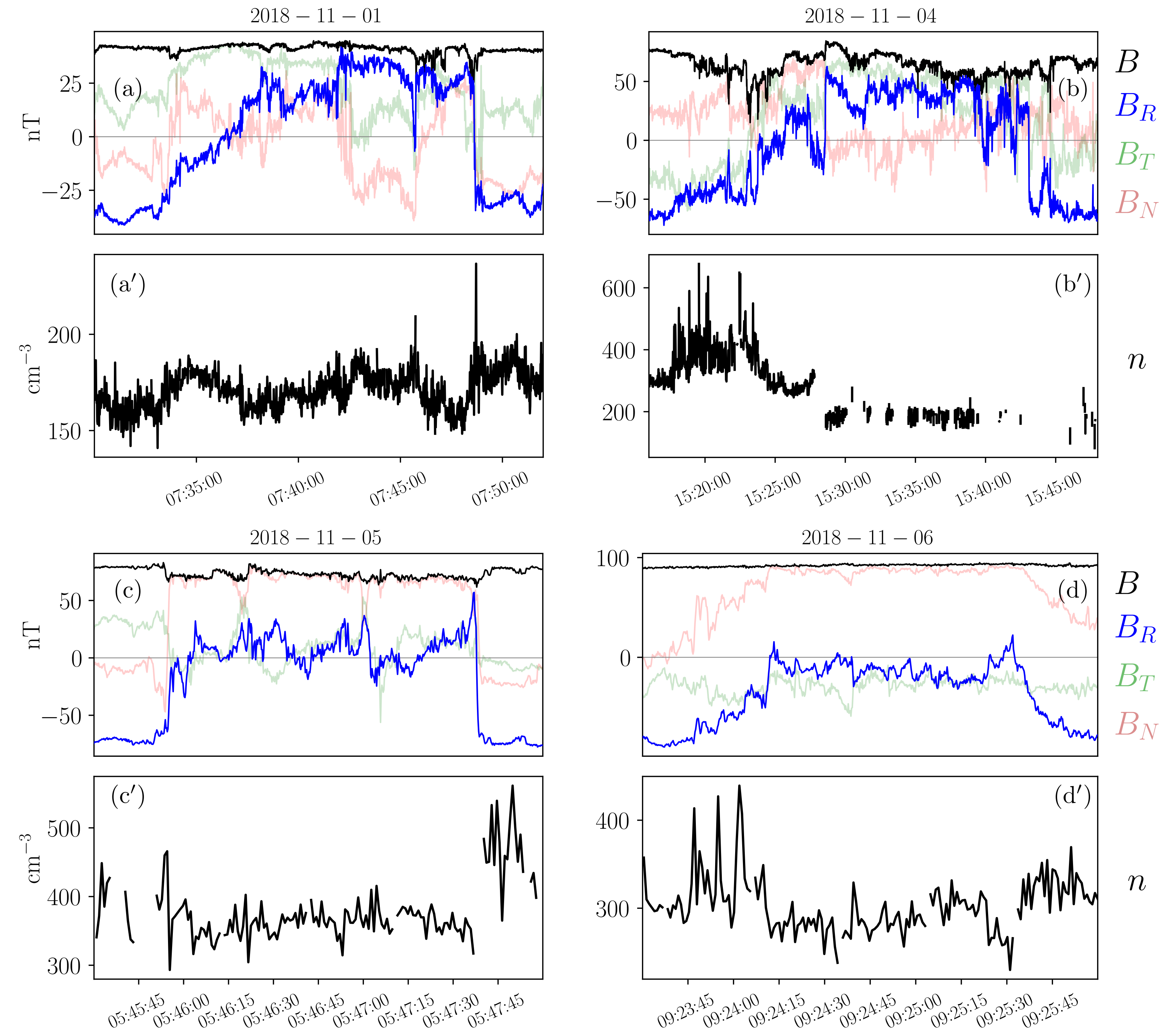}
            \caption{Example of switchbacks observed with Parker, with a range of sharpness and properties observed at their boundaries, including substructures, density enhancements, and dips in magnetic field strength. For each example, we display the magnetic field components in the RTN frame from the FIELDS instrument (panels $a$, $b$, $c$, $d$) and the proton density from SWEAP/SPC (panels $a'$, $b'$, $c'$, and $d'$). }
            \label{fig: 4.2_SB_boundaries}
        \end{figure}
        
        The boundaries of switchbacks, marking the separation between the switchback's internal structure and the ambient solar wind, are a key feature potentially holding clues for understanding the mass and energy exchange with the solar wind as well as the switchback formation mechanism and origin. 
        As magnetic switchbacks are ``deflections", their boundaries sometimes exhibit non-Alfv\'enic properties indicative of current sheets. 
        As detailed in Sect.~\ref{sec: 2_SB_definition}, the boundaries are typically sharp, i.e., have small apparent duration in the magnetic field time series compared to the total duration of the structure crossing. 
        However, more complicated boundaries can be encountered, as illustrated in Fig.~\ref{fig: 4.2_SB_boundaries}, where we show a range of examples.
        The magnetic field rotation can sometimes be gradual, resulting in ``blurry" edges in the magnetic field time series (see the leading edge of the structures in panels (a) and (b) of Fig.~\ref{fig: 4.2_SB_boundaries}, and both edges of the switchback in panel (d) of Fig.~\ref{fig: 4.2_SB_boundaries}).       
        Some switchbacks can exhibit an asymmetry between the leading and trailing edges, sometimes called a transition region \citep[e.g.,][]{AkhavanTafti2021, AkhavanTafti2022, HuangJ2023a, HuangJ2023b}.
        The thickness of these regions was found to have sizes ranging from \SIrange{10}{10000}{km}, with most cases featuring boundaries with widths on the order of the MHD scale \citep{Larosa2021, Bizien2023}. 
        These results are, to date, mostly based on the analysis of Parker's E1 observations. 
        This limitation may bias findings such as determining the normal to the switchback boundary, treated as an MHD discontinuity, which strongly depends on the method used, as explained later in the next paragraph (Sect.~\ref{subsubsec: 4.2_discontinuities}).
        
        Compressive features are frequently observed at switchback boundaries, where 
        the proton density often increases \citep{Agapitov2020, Farrell2020, Froment2021, Larosa2021}.         
        Examples of this are illustrated in Fig.~\ref{fig: 4.2_SB_boundaries} at the trailing edges of the top left (panel (a')) and bottom left (panel (c')) switchbacks, as well as the leading edge of the bottom right example (panel (d')). 
        Superposed Epoch Analysis of E1 switchback observations shows that the proton density tends to increase by about ~30\% at switchback edges \citep{Farrell2020}, which may be related to the ballistic propagation of switchbacks relative to the background solar wind speed \citep{Agapitov2023ApJ}. 
        This density increase could be explained by a compression region around a fast ejecta \citep{Liu2022} or by the possible merging of switchbacks \citep{Agapitov2022}
        However, these results on density variations must be interpreted with caution, as a more recent study using electron density measurements from the QTN method did not observe similar density variations \citep{Rasca2021}. 
              
        Frequent magnetic field magnitude depletions are also observed at switchback boundaries, sometimes named ``magnetic dropouts" \citep{Farrell2020, Rasca2022} or ``magnetic dips" \citep{Agapitov2020}.
        These dips are visible in panels $a$, $b$ and $c$ of Fig.~\ref{fig: 4.2_SB_boundaries}.
        These dropouts are generally close to 10\% \citep{Farrell2020}, but more dramatic dips can also be observed \citep{Froment2021, Larosa2021}. 
        The dips are usually not related to magnetic reconnection, as discussed in Sect.~\ref{subsubsec: 4.2_reconnection}. 
        They may be, however, the key ingredient for the generation of some of the waves observed at switchback boundaries (see Sect.~\ref{subsubsec: 4.2_waves_boundaries}). 
        
        Several explanations exist to explain the observation of these magnetic dropouts: 
        They might originate from diamagnetic currents forming at switchback boundaries \citep{Krasnoselskikh2020, Farrell2020, Farrell2021, Rasca2023}, which would be favored when the magnetic field vector rotates to an orientation quasi-perpendicular to the pressure gradient.
        Another model relates the dips to the development of switchback from the growth of Alfvén waves propagating in the expanding solar wind \citep{Squire2020, Mallet2021, Johnston2022, Squire2022}.
        By studying the evolution of large amplitude Alfvén waves with a gradient scale of the order of the ion \textbf{inertial} length in a low $\beta$ environment, \cite{Mallet2023PhPl} showed that the boundary dips occur in order to avoid singularities in the solutions of the wave equation when the magnetic field undergoes a sharp rotation. 
        Currently, the compressible nature of switchback boundaries is a property that many models do not address except for the Alfvén wave based ones \citep{Mallet2021, Mallet2023PhPl}, and warrants deeper investigation overall. 
        We further discuss the relationship between magnetic dropouts and the occurrence of electromagnetic waves in Sect.~\ref{subsubsec: 4.2_waves_boundaries}.            
        
    \subsubsection{Discontinuity Classification} \label{subsubsec: 4.2_discontinuities}
    
        One straightforward question when studying switchback boundaries is to wonder what type of discontinuity is involved. 
        Are switchbacks bound by rotational discontinuities (RDs), which are open boundaries propagating in the solar wind, or by tangential discontinuities (TDs), which are closed boundaries advected in the solar wind \citep{Hudson1970}?
        We know that switchbacks are Alfvénic structures and that their magnetic field vector is constrained to rotate on the surface of a sphere. 
        Their boundaries are, moreover, contained in planes and thus are usually referred to as arc-shaped structures \citep{Tsurutani2018, Horbury2020, McManus2022, Bizien2023}.
        We detail here the latest results regarding the nature of their boundaries as rotational or tangential discontinuities.           
        
        RDs and TDs can be distinguished through measurements of magnetic field fluctuations. Indeed, in ideal MHD, the magnetic field of RDs should present a component normal to the plane of the discontinuity, while TDs should not. 
        The nature of the switchback boundary therefore relies on the estimation of $B_n$, where $n$ designates the boundary normal.             
        The discontinuity plane and its associated normal direction are usually obtained through minimum variance analysis \citep[MVA,][]{Sonnerup_and_Cahill_1967}.      
        The classification of discontinuities can then be achieved by comparing the degree of collinearity of $B_n /B$ with the variation in field magnitude across the discontinuity $\Delta B/B$ \citep{Neugebauer1984}.
        TDs classically present  $B_n /B <0.4$ and $\Delta B/B \geq 0.2$, while RDs present the opposite, with $B_n /B \geq 0.4$ and $\Delta B/B <0.2$. 
        Discontinuities falling outside of these parameter regions are unclassified and named ``either'' ($B_n /B <0.4$ and $\Delta B/B < 0.2$) or ``neither'' discontinuities \citep{Neugebauer1984}.  
        This classification applied to switchback boundaries, however, led to contrasting results. 
        Switchbacks usually present low magnetic field magnitude variability on the timescale of the switchback itself, landing most of them as rotational discontinuities \citep{AkhavanTafti2021} or as ``either'' discontinuities \citep{Larosa2021}, consistent with Ulysses observations at larger distances \citep{Yamauchi2002}.

\begin{figure}
            \centering
            \includegraphics[width=\textwidth]{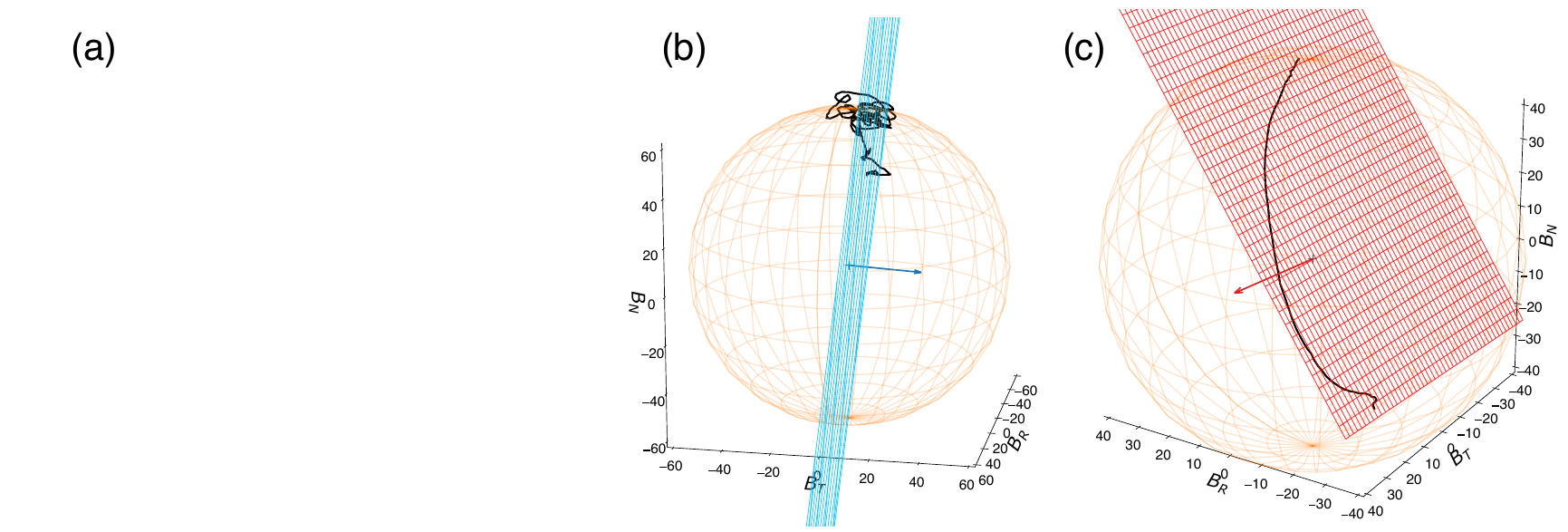}
            \caption{Evolution of the magnetic field vector on a sphere for TDs and RDs. 
            Panel (a) shows a schematic view of an arc-polarized switchback boundary. 
            A discontinuity plane is represented in blue with its normal $\mathbf{n}$ in green. 
            For a TD, the tip of the field vector rotates in that plane, which includes the origin in that particular case. 
            More generally, any arc-polarized boundary would be identified by the red plane, with the same normal but excluding the origin.
            The blue plane corresponds to a tangential-like boundary (small $B_n/B$) while the red plane corresponds to a rotational-like boundary (large $B_n/B$).  Panels $b$ and $c$ show two switchback boundaries, the first (panel (b)) was observed on November 7, 2018, at 20:53:15~UT and corresponds to a TD, while the second (panel (c) was observed on November 2, 2018, at 18:33:33~UT and corresponds to an RD. Figures reproduced with permission from \cite{Bizien2023}, copyright by AAS}
            \label{fig: 4.2_distinct_TD_RD}
        \end{figure}

        The MVA method, however, presents some serious limitations that lead to high uncertainties when computing the normal direction \citep{Hausman2004, Knetter2005, Wang2023}.
        The geometry of switchback boundaries, with their arc shape and the superimposed waves, prevents the use of MVA only for their classification.
        In a recent study, \cite{Bizien2023} suggest that the use of the singular value decomposition \citep[SVD,][]{Golub2013}, instead of MVA, sheds new light on the nature of switchback boundaries.
        While both methods rely on a diagonalization of the magnetic field covariance matrix, MVA additionally subtracts the mean value of each magnetic field component.
        MVA, therefore, finds a plane centered on the orbit of the magnetic field and usually tangent to the sphere, while the SVD plane will always include the origin.  
        In Fig.~\ref{fig: 4.2_distinct_TD_RD}, we illustrate how switchback boundaries evolve on a sphere and the difference between TDs and RDs. 
        Panel \ref{fig: 4.2_distinct_TD_RD}a shows the two types of boundaries in the ideal case, without any superimposed fluctuations. 
        The blue plane includes the origin and corresponds to a TD, while the red plane represents an RD.
        Panels \ref{fig: 4.2_distinct_TD_RD}b and \ref{fig: 4.2_distinct_TD_RD}c show two examples of switchback boundaries measured during the first encounter of Parker, both presenting arc-shaped structures \cite[see][]{Bizien2023}.
        Panel \ref{fig: 4.2_distinct_TD_RD}b shows the boundary of a switchback measured on November 7, 2018, at 20:53:15~UT, whose plane includes the origin and that is classified as a TD.
        Panel \ref{fig: 4.2_distinct_TD_RD}c shows a second example measured on November 2, 2018, at 18:33:33~UT, with a plane which clearly does not include the origin.
        This discontinuity thus presents a large normal component and corresponds to an RD. 
        The use of SVD over MVA leads \cite{Bizien2023} to conclude that the number of RDs has been previously overestimated compared to TDs, and that many switchback boundaries are TD-like discontinuities. 
        These results have important implications for the evolution of the switchbacks in the solar wind \citep[see][]{Mallet_this_issue}. TD-like discontinuities are closed boundaries, which seems to indicate that switchbacks are stable structures that could survive as they propagate at larger heliocentric distances. The results of \citet{Bizien2023} thus indicate that the nature of switchback boundaries is compatible with a solar origin in these structures. However, this is not enough evidence to rule out an \textit{in situ} formation.        
        Another important remark is that though not completely Alfv\'enic, switchback boundaries do preserve an outwardly propagating fluctuation correlation between velocity and magnetic field, a condition which is part of the definition of an RD, but a completely additional constraint for a TD. 
        In other words, classifying the boundary as a TD does not use the full properties of the switchback boundary, and calls into question this interpretation. Electron temperature behavior across the discontinuities are another potential discriminator with TDs able to support a temperature jump and RDs not. \citet{Rivera2024b} showed electron core temperature behavior across several switchback patches and did not observe variation correlated with the switchback deflections. Further statistical behavior of electron properties across switchbacks would be useful to further constrain the nature of their boundaries.

        Having these contradictory results in mind \citep{Larosa2021, AkhavanTafti2021, Bizien2023}, additional properties of switchback boundaries were retrieved. 
        The nature of switchback boundaries appears to be linked to properties of the solar wind, with apparent proton temperature sharply increasing by 29\% across RD-type leading boundaries, enhancing the thermal pressure gradient while the plasma density drops slightly \citep{AkhavanTafti2022}, although as noted in Sect.~\ref{subsubsec: 4.1_proton_alpha} temperature signatures during large flow deflections measured by Parker remain somewhat controversial. 
        Nevertheless, this evidence suggests switchback transition regions may not be in thermal equilibrium with their neighboring plasma, likely driven by plasma instabilities rather than magnetic reconnection. 
        Significant heating mechanisms are possible, and the topic requires further investigation \citep{AkhavanTafti2022}.             
        These conclusions are based, however, on the rotational nature of the boundary retrieved through the MVA method, and results may change if the SVD method was used.    
        In parallel, the deflection does not seem to influence the nature of the discontinuity \citep{Bizien2023}, suggesting that switchbacks are self-similar, which is consistent with earlier studies \citep{DudokdeWit2020}.

    \subsubsection{Switchbacks and Magnetic Reconnection} \label{subsubsec: 4.2_reconnection}
        
         \begin{figure}
            \centering
            \includegraphics[width=0.9\textwidth]{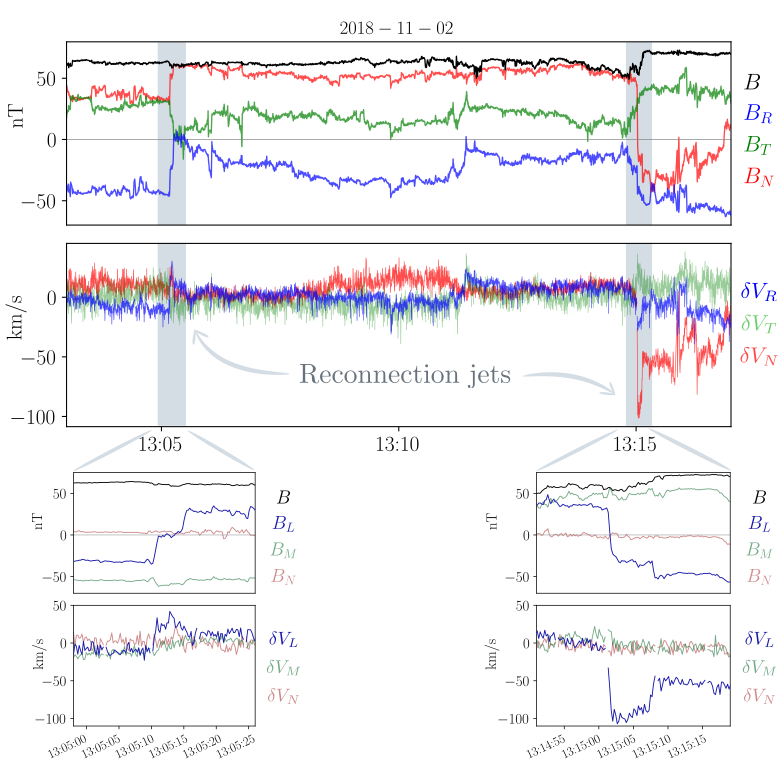}
            \caption{Magnetic reconnection observed at switchback boundaries by Parker, first reported by \cite{Froment2021}.
            The top panels shows the magnetic field for the full switchback encounter as well as the ion velocity variation measured by SPC $\delta \mathbf{V} = \mathbf{V} - \overline{\mathbf{V}}$, where the average background solar wind speed  $\overline{\mathbf{V}}$ is [322, 6, 26]$_{RTN}$~\kms\ over this time interval.
            The shaded regions show the locations of the switchback boundaries. 
            The insets zoom-in on the switchback boundaries and display $\mathbf{B}$ and  $\delta \mathbf{V}$ in the $LMN$ current sheet aligned frame, where $l$ points in the outflow direction, $m$ in the X-line direction, and $n$ is the current sheet normal direction.              
            }
            \label{fig: 4.2_Rx_at_boundaries}
        \end{figure}
        
        Magnetic reconnection is a fundamental plasma process that occurs at thin current sheets, where plasma is heated and accelerated by the release of energy stored in the magnetic field due to changes in the field topology \citep[e.g.,][]{Gosling2008, Gosling2012}. 
        Yet, the very nature of magnetic switchbacks as sudden deflection of the magnetic field goes hand in hand with an increased presence of current sheets in the solar wind.
        Switchback boundaries, therefore, seem like a potential favored place for magnetic reconnection to occur, and it is natural to wonder whether this process of magnetic reconnection, in addition to potentially playing a role in the formation of switchbacks, could also be responsible for their erosion and, ultimately, their dissipation. 
        Here, we review results relating to magnetic reconnection and switchbacks in the solar wind.
        
        \begin{table}
            \caption{Table summarizing the properties of the rare instances of reconnection at switchback boundaries observed by Parker and Solar Orbiter. For further details, see \citep{Froment2021} and \citep{Suen2023}.}%
            \begin{tabular}{@{}cccccc@{}}
            \toprule
            Spacecraft & \makecell{Heliocentric\\distance (AU)} & \makecell{Boundary\\} & \makecell{$\mathbf{B}$ shear \\($^\circ$)} & \makecell{$|\mathbf{B}|$ dip \\ (\%)} & \makecell{$V$ increase \\(\%$V_A$)}\\
            \midrule
            Parker & 0.22 & \makecell[c]{Leading\\Trailing} & \makecell[c]{58\\81} & \makecell[c]{6\\27} & 9  \\
            Parker & 0.22 & \makecell[c]{Leading\\Trailing} & \makecell[c]{67\\60} & \makecell[c]{16\\12} & 10\\
            Parker & 0.22 & \makecell[c]{Leading\\Trailing} & \makecell[c]{61\\168} & \makecell[c]{22\\90}& 39  \\
            \hline
            SolO & 0.72 & Trailing & 117  & 23 & 35  \\
            SolO & 0.61 & Trailing & 134  & 8 & 31  \\
            SolO & 0.61 & Trailing & 95  & 6 & 17  \\
            \botrule
            \end{tabular}
            \label{tab: 4.2_rx_prop}
        \end{table}           
        
        Only a handful of switchback boundary reconnection events have been reported, suggesting that they are rare compared to reconnection at the heliospheric current sheet and ICMEs \citep{Phan2020}.
        Some properties of the reported reconnection at switchback boundaries \citep{Froment2021, Suen2023} are listed in Table~\ref{tab: 4.2_rx_prop}. 
        The few observations suggest that reconnection neither favors the leading nor trailing edge boundary of the switchback and can sometimes occur simultaneously at both boundaries, as illustrated in Fig.~\ref{fig: 4.2_Rx_at_boundaries}.
        Most of the reconnection events observed by Parker occur across boundaries with moderate magnetic shear angle, whereas events observed by the Solar Orbiter spacecraft tend to have larger shear angles \citep[$>100^\circ$,][]{Phan2010}. 
        Interestingly, switchbacks with reconnection occurring at their boundaries exhibit modest velocity enhancements in their interior, especially when compared to those without boundary reconnection \citep{Froment2021, Larosa2021}. 
        Whether this is a consequence or a prerequisite of magnetic reconnection remains to be determined.             
        
        Figure \ref{fig: 4.2_Rx_at_boundaries} shows an example of switchback undergoing boundary reconnection observed by Parker on November 2, 2018, at $\sim0.2$~AU (first event of Table~\ref{tab: 4.2_rx_prop}).          
        Reconnection outflows in the solar wind are characterized by a bifurcation of the reconnecting current sheet (i.e., a plateau in the field rotation), wherein the outflow region is bound by a pair of standing Alfv\'enic rotational discontinuities in the magnetic field \citep{Gosling2005}.
        In Fig.~\ref{fig: 4.2_Rx_at_boundaries}, both the leading and trailing edge boundaries of the switchback exhibit an increase in $\delta \mathbf{V}$ confined within a bifurcated current sheet layer. 
        Fluctuations in $\mathbf{B}$ and $\delta \mathbf{V}$ are correlated on one side of each outflow and anti-correlated on the other. 
        These are characteristic signatures of magnetic reconnection observed \textit{in situ}, as described by \cite{Gosling2005}. 
        
        \citet{Froment2021} postulates that magnetic reconnection at switchback boundaries may destabilize the switchback structure, thus causing the velocity enhancements to decay.
        Using examples of reconnecting switchbacks observed at 0.6--0.7 AU by Solar Orbiter, \citet{Suen2023} finds that boundary reconnection could erode a switchback in minutes or hours, making it an efficient erosion process. 
        This short timescale could explain why switchback boundary reconnection is so rarely observed. A more detailed discussion of the erosion of switchback by magnetic reconnection can be found in {\cite{Mallet_this_issue}}
        
        The lack of observations of magnetic reconnection at switchback boundaries may also suggest that reconnection is suppressed.
        This is confirmed by recent statistical studies of magnetic reconnection occurrence in the solar wind, which suggest that reconnection jets are not observed in the highly Alfvénic wind dominated by switchbacks \citep{Fargette2023, Eriksson2024}.
        Again one must remember that switchbacks are Alfv\'enic structures, and therefore across the boundary the current sheet coexists with a vorticity sheet and a strong velocity-magnetic field correlation. 
        Shear flows across current sheets of magnitude comparable to the Alfv\'en speed suppress reconnection \citep{Einaudi86, Owen1987, Chen1997, Cassak2011, Shi2021}.
        The reason is that, in incompressible MHD, the purely unidirectional Alfv\'en wave is an attracting nonlinear state, and should therefore be completely stable. However, including compressibility and only partial Alfv\'enicity, the tearing instability is quenched but not completely suppressed \citep{Shi2021}.
        This 
        would explain the moderate velocity enhancements inside switchbacks undergoing reconnection at their boundaries (see Table~\ref{tab: 4.2_rx_prop}).
        Flow shears may also induce diamagnetic drifts that suppress reconnection if the shear angle across the current sheet $\theta$ is sufficiently large through the $\Delta\beta-\theta$ condition, where $\Delta\beta$ is the difference in plasma $\beta$ across the sheet \citep{Swisdak2003, Phan2010}.
        \citet{AkhavanTafti2021} show that the three Parker events of Table \ref{tab: 4.2_rx_prop} fall within the regime set by the $\Delta\beta-\theta$ condition where reconnection is favored. The authors further argue that the negligible magnetic tension and the presence of strong $|\mathbf{B}|$  prevent instability formation and growth, therefore suppressing the triggers of magnetic reconnection. 
        However, a recent study by \citet{Vasko2021} shows that all current sheets, both reconnecting and non-reconnecting, fall within the $\Delta\beta-\theta$ regime where reconnection is favored. 
        This suggests that the diamagnetic drift mechanism does not play a significant role in suppressing switchback boundary reconnection.

    \subsubsection{Waves at Switchback Boundaries} \label{subsubsec: 4.2_waves_boundaries}
        
        Coherent electromagnetic fluctuations from about the Hz to the MHz ranges have been reported inside and at the boundaries of switchbacks \citep[e.g.,][]{Krasnoselskikh2020, Larosa2021, Malaspina2022}. 
        Most of the wave activity detailed in this section is linked to wave-particle interactions involving, in particular, electron beams (whistler and Langmuir waves) or particle transfer between the ambient solar wind and the switchbacks (surface waves).
        
\begin{figure}
            \centering
            \includegraphics[width=\textwidth]{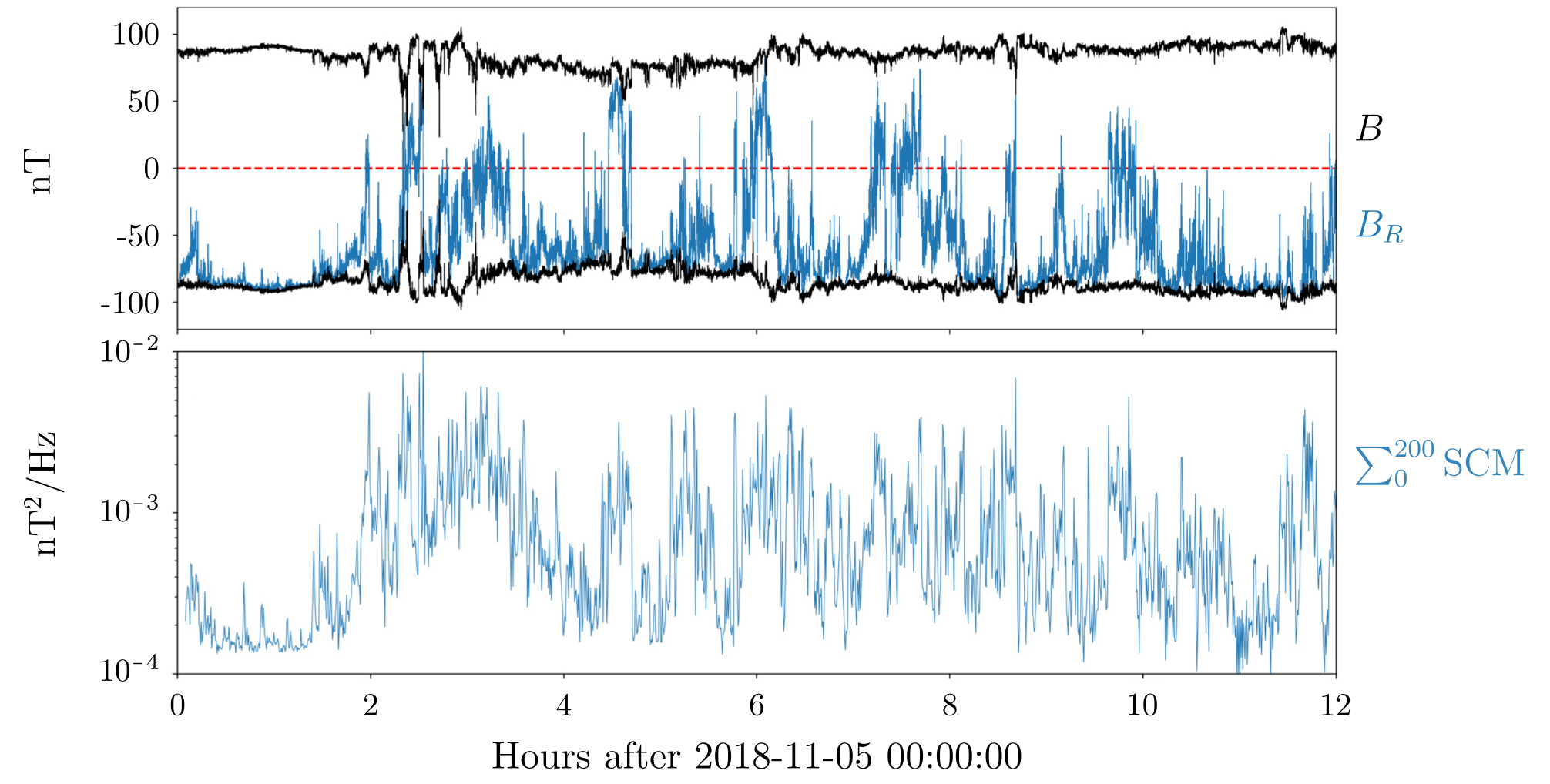}
            \caption{Top panel: Radial component of the magnetic field with FIELDS/MAG Parker observations, showing a switchback interval. Bottom panel: Time series of the magnetic power spectral density, summed from 20 to 200 Hz from FIELDS/SCM data. Figure reproduced with permission from \citet{Malaspina2022}, copyright by AAS} 
            \label{fig: 4.2_waves_Malaspina}
        \end{figure}

        \citet{Larosa2021} used FIELDS Search Coil Magnetometer \citep[SCM;][]{Bale2016,Bowen2020scam} data to show that, during Parker's first encounter, the amplitude of the broadband magnetic fluctuations above 3~Hz increased by 65\% on average inside the structures.
        This increase of electromagnetic activity is illustrated in Fig.~\ref{fig: 4.2_waves_Malaspina}, which shows the enhancement of the magnetic power spectral density across a switchback interval.            
        The wave activity then seems to decay with the heliocentric distance \citep{Mozer2020_SB, Farrell2021, Mozer2020_Efield}, a topic covered in more depth in \citep{Mallet_this_issue}. 

        Different types of waves have been found at switchback boundaries and inside switchbacks, such as ultra-low frequency (ULF) waves \citep{Farrell2021}, surface waves \citep{Krasnoselskikh2020}, as well as whistler waves discussed in the next paragraph.
        ULF waves were observed while considering the range \SIrange{0.7}{1}{\hertz} \citep{Farrell2021}. A large portion of the detected ULF waves presents a low amplitude compared to the magnitude of the background magnetic field (less than 3\%). These waves might play a role in switchback degradation \citep{Farrell2021}. 
        In parallel, waves with frequency close to the ion cyclotron frequency $f_{ci}$ (i.e., a few Hz) were reported at switchback boundaries
        \citep{Krasnoselskikh2020}. 
        The angle of about 60$^{\circ}$ between the wave normal angle and the normal to the boundary seems to indicate these are surface waves, which could increase the particle transfer across switchback boundaries.
        Both ULF waves and surface waves have been largely underexplored to date.
        More work is needed to understand their occurrence, generation, and importance for the evolution of switchbacks in the solar wind.

        Parker observations revealed the occurrence of localized bursts of low-frequency whistler wave packets, collocated with magnetic field dips associated with switchback boundaries (Sect.~\ref{subsubsec: 4.2_SB_boundaries}) with a typical frequency range of $100$~Hz to $300$~Hz ($0.05$--$0.2~f_\mathrm{ce}$) in the spacecraft frame \citep{Agapitov2020, Karbashewski2023, Froment2023, colomban_reconstruction_2023, Colomban2024}.
        However, whistler waves seem to only be observed in 15\% of switchbacks, where they tend to have a shorter duration with respect to non-switchback intervals \citep{jagarlamudi_whistler_2021}.            
        
        Whistler waves observed in the solar wind (not only at switchback boundaries) present wave amplitudes that peak at $2$~nT, representing up to $5\%$ of the background magnetic field magnitude. 
        The polarization of these waves was found to be predominantly quasi-parallel to the background magnetic field \citep{Froment2023, Karbashewski2023}.
        Quasi-parallel propagation appears to be the predominant mode of whistler waves in the solar wind. 
        Indeed, over the heliocentric distances covered by Parker and Solar Orbiter, the occurrence of quasi-perpendicular whistler waves (normal wave angle greater than $50^\circ$) was less than 1\% \citep{kretzschmar_whistler_2021}. 
        The direction of propagation is therefore the key factor in the efficiency of the wave-particle interaction: scattering of strahl electrons by counter-propagating (sunward) whistler waves is an order of magnitude more efficient than scattering by anti-sunward quasi-parallel waves \citep{Colomban2024}. 
        Also, all whistler waves observed by Solar Orbiter at $0.5-1.0$~AU were found to propagate from the Sun \citep{kretzschmar_whistler_2021}. Thus, sunward propagating whistler populations in the young solar wind (observed by Parker at 25--75~R$_\odot$) are potentially a key factor for the diffusion of the strahl electron population, which raises the question of the mechanism behind their local generation \citep[to explain the range of generation of these waves reported by][]{cattell_parker_2022}.    
        \begin{figure}
            \centering    
            \includegraphics[width=\textwidth]{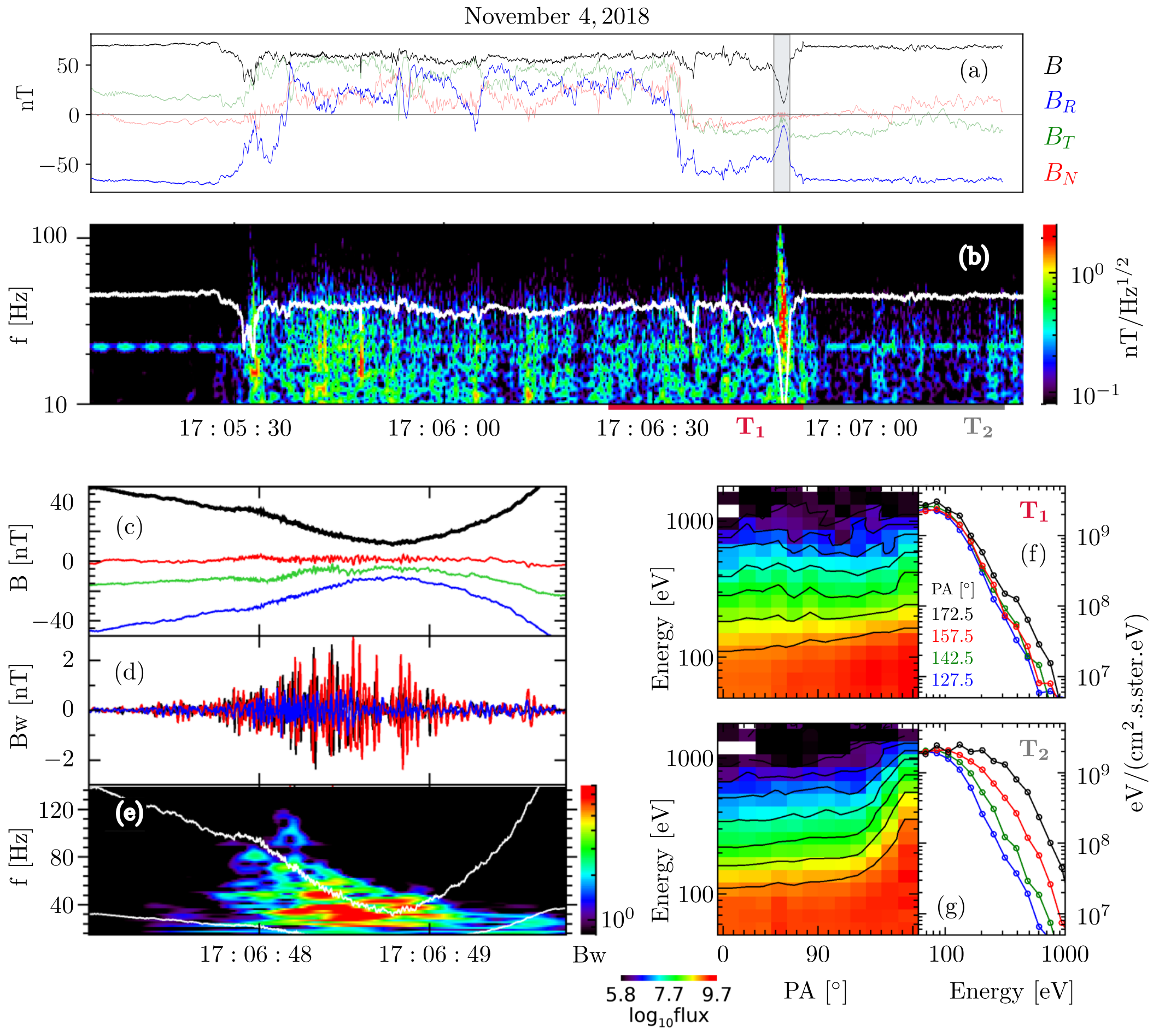}
            \caption{Magnetic field dynamics associated with a switchback observed on November 4, 2018, from 17:05 to 17:07~UT. The magnetic field (a) exhibits an almost complete reversal of the radial component. Panel (b) represents the dynamic spectrum of magnetic field waveforms estimated from the combined MAG and SCM data \citep{Bale2016,Bowen2020scam}, in which the solid white curve indicates the local lower hybrid frequency. Panels (c-e) present a zoomed interval covering the magnetic dip observed at the trailing edge of the switchback (shaded area in panel a) around 17:06:49~UT with (c) -- the magnetic field components and magnitude from MAG; (d) the magnetic field high-frequency ($>$\SI{10}{\hertz}) waveforms; (e) -- the spectral power of the magnetic field perturbations. Panels (f) and (g) present the pitch-angle distribution electron functions in the vicinity of the magnetic dip (T1 in panel (b)) and the unperturbed solar wind (T2 in panel (b)). Figures reproduced with permission from \cite{vo_enhanced_2024}, copyright by AAS}
            \label{fig: 4.2_SB_Waves}
        \end{figure}
        
        The generation mechanism for these sunward and anti-sunward propagating quasi-parallel whistler waves is presumably the cyclotron instability (suggested by the simultaneous generation of counter-propagating wave bursts). The latter could be driven by a population of suprathermal electrons ($\sim 100$~eV) trapped in magnetic field dips and/or by the strahl population, whose geometry is disturbed by local inhomogeneities in the magnetic field (such as abrupt rotations) at switchback boundaries.             
        The derived difference in frequency of sunward and anti-sunward counter-propagating waves confirmed their simultaneous generation in the source region moving in the solar wind frame \citep{Karbashewski2023}.            
        
        A statistical study of whistler wave occurrence rates and their parameters in the young solar wind showed that the high-amplitude waves observed between 25--45~R$_\odot$ were mostly tied to the magnetic dips or sharp rotations often associated with switchbacks and were propagating toward the Sun (antisunward waves were observed mostly above 40 R$_\odot$ and without any relationship to magnetic field inhomogeneities) \citep{choi_whistler_2024}. This co-existence of sunward whistlers and magnetic field gradients (Fig.~\ref{fig: 4.2_SB_Waves}c, d) provides favorable conditions for the fast local scattering of the strahl electrons by the waves and can lead to broadening of the superthermal pitch-angle electron distribution seen in Fig.~\ref{fig: 4.2_SB_Waves}f \citep{vo_enhanced_2024}.
        
        \citet{Rasca2023} found that magnetic dips, or dropouts, are both less frequent and of lower amplitude in the sub-Alfvénic solar wind. These dropouts may originate from diamagnetic effects stemming from a large magnetic field rotation at switchback boundaries (see Sect.~\ref{subsubsec: 4.2_SB_boundaries}). They hypothesize that the rarer observations of magnetic dips at switchback boundaries below the Alfvén critical surface may be linked to smaller magnetic field deflections (see also Sect.~\ref{subsec: 5_low_MA}). This may be an avenue to explore to explain the apparent lack of whistler waves close to the Sun \citep{cattell_parker_2022}.
   
        Finally, we note that it remains unclear whether whistler wave activity at switchback boundaries is related to the emission near the electron plasma frequency $f_{pe}$ reported by \cite{Rasca2022}. 
        \citet{Rasca2022} find an enhanced emission in LFR radio spectrograms near $f_{pe}$ at switchback boundaries with a significant magnetic field dropout. They argue that these dropouts are creating electron currents that stimulate Langmuir waves, showing as $f_{pe}$ emissions. 
        Indeed, \citet{jagarlamudi_whistler_2021} reported that whistlers were detected together with Langmuir waves about 85\% of the time in Parker data, which is consistent with previous solar wind observation \citep{Kennel1980}.
        The relationship between these different types of waves at switchback boundaries remains to be investigated and will certainly shed light on the properties of switchbacks, their generation mechanisms, and evolution in the solar wind. 
        
    \subsubsection{Geometry and Boundary Takeaways} \label{subsubsec: 4.2_takeaways}
        
        The phenomenon that we call magnetic switchback is characterized by a self-similar distribution of deflection angle, where small and large deflections share the same properties (Sect.~\ref{subsubsec: 4.2_occurrence}). 
        It is thus important to consider all deflections as relevant, and not focus only on ``polarity reversal'' switchbacks of more than \ang{90}. 
        Switchbacks also seem to have a tendency to deflect in a preferential direction, and their orientation presents a clustering behavior (Sect.~\ref{subsubsec: 4.2_SB_geometry}).
        Finding a physical mechanism able to reproduce the full distribution of deflection angles through simulations will be key to understanding the origins of switchbacks. 
        This distribution may also vary depending on solar wind conditions or as the solar wind evolves, and clearly, establishing what causes the distribution of $z$ to vary will be crucial for constraining the generation mechanisms of switchbacks.

        Much remains to be studied about switchback boundaries, and understanding their properties and evolution will shed light on their interaction with the ambient solar wind.   
        The compressible nature of switchback boundaries is, for now, only addressed by  wave-based models and requires further investigation (Sect.~\ref{subsubsec: 4.2_SB_boundaries}). 
        Arriving at a consensus on whether switchbacks are bound by rotational or tangential discontinuities, in combination with establishing if switchbacks contain distinct plasma populations, would provide evidence for determining a solar or \textit{in situ} origin (Sect.~\ref{subsubsec: 4.2_discontinuities}).
        The lack of magnetic reconnection at switchback boundaries deserves more attention and could be explained by either it being an efficient erosion mechanism for switchbacks, or it being suppressed by \textit{in situ} local conditions (Sect.~\ref{subsubsec: 4.2_reconnection}). 
        Finally, enhanced wave activity has often been observed at switchback boundaries, often associated with whistler waves. ULF and surface waves have been underexplored to date. More work is needed to understand their occurrence, generation, and importance for the evolution of switchbacks in the solar wind (Sect.~\ref{subsubsec: 4.2_waves_boundaries})
    
        Overall, most studies on switchback geometry and boundaries focus on the early encounters of Parker with the Sun. 
        Seeing how their properties (deflection direction, occurrence, compressibility, wave activity, magnetic reconnection occurrence, etc.) evolve in the latest encounters, i.e., closer to the Sun, could be an important step in understanding switchbacks' origins. 
    
\subsection{Switchbacks, Turbulence, and Ion-scale waves} \label{subsec: 4.3_turbulence}  
In the regions explored by Parker and other spacecraft, switchbacks are typically embedded in the young solar wind's highly Alfv\'enic turbulent fluctuations \citep{BrunoCarbone2013,chen2020evolution,Sioulas_2023, Wu2023,Brodiano2023,Sorriso-Valvo2023}. 
Irrespective of their origin, the large magnetic and velocity gradients associated with switchbacks, as well as their modulation in patches, provide a source of energy that is expected to feed the turbulent cascade. 
They might, in fact, have a relevant role in setting or modulating the properties of turbulence in the expanding solar wind. 

One major basic question arises: to what degree can switchbacks be considered as entities that are distinct from the turbulent field in which they are embedded, since when propagating away from the Sun, they eventually merge with it and become indistinguishable?
Determining how turbulence and switchbacks interact and develop as the solar wind expands may provide answers to such questions.
    
\subsubsection{Background on Alfvénic Turbulence} \label{subsubsec: 4.3_alfv_turb} 
    Various mechanisms, possibly involving photospheric motions or magnetic reconnection, lead to a predominance in the solar wind of anti-sunward propagating Alfv\'enic perturbations, a condition known as ``imbalance''. 
    Such imbalance is usually quantified through the normalized cross-helicity spectrum $\sigma_c = (E^{+} - E^{-}) / ( E^{+} + E^{-})$, where $E^\pm$ are the power spectral densities of the inward and outward Alfv\'enic modes which are defined in terms of the Elsasser fields \citep[][ $\boldsymbol{z}^\pm=\boldsymbol{v}\pm\boldsymbol{b}$]{elsasser_1950} as $E^\pm(\tau) = \frac{1}{2}\langle (\boldsymbol{z}^\pm)^2 \rangle_\tau$ where $\langle \rangle_\tau$ indicates an ensemble average over a time window $\tau$, $\boldsymbol{v}$ is the bulk velocity fluctuation vector (also referred to as $\delta v$ earlier) and $\boldsymbol{b}=\delta\boldsymbol{B}/ \sqrt{\mu_0 \rho}$ is the magnetic field fluctuation vector in velocity (Alfv\'en) units. Note that in addition to $\tau$, the background vectors with which fluctuations are computed with respect to can vary between studies.
    Another closely related quantity is the residual energy spectrum $\sigma_r = (E^{k} - E^{b}) / ( E^{k} + E^{b})$, representing the level of balance between magnetic ($E^{b}$) and kinetic ($E^{k}$) energy in the fluctuations, such that $\sigma_r=0$ if they are pure outwards traveling Alfvénic fluctuations and $\sigma_r<0$ if there is an excess of magnetic with respect to kinetic energy. Here, $E^k = \frac{1}{2}\langle \boldsymbol{v}^2\rangle_\tau$ and $E^k = \frac{1}{2}\langle \boldsymbol{b}^2\rangle_\tau$. A related quantity is the Alfv\'en ratio, $r_A = E^k/E^b$, which can be used to describe the `Alfv\'enicity' of the fluctuations, where the limit of $r_A = 1$ corresponds to perfect Alfv\'enic correlation between magnetic and kinetic fluctuations.
    Linear couplings of the outgoing modes to the large-scale inhomogeneity result in non-WKB (Wentzel-Kramers-Brillouin) reflections \citep{velli_turbulent_1989}, providing the sunward propagating component that is necessary for the MHD non-linear interactions to be activated and for turbulent characteristics to emerge \citep{iroshnikov_turbulence_1963, kraichnan_inertial-range_1965}. 
    
    When the turbulence is fully developed, nonlinear interactions among counter-propagating modes result in an energy cascade from large to small scales \citep{Montgomery_turner_1981, shebalin_matthaeus_montgomery_1983}, highlighted by a broad range of power-law scaling forms for different statistical quantities as a function of fluctuation frequency, such as the power spectral density (PSD) and the high-order moments of the fields' fluctuations. 
    The PSD (e.g., of magnetic fluctuations) is one of the most commonly studied turbulent quantities of space plasmas and exhibits key features such as a broken power law slope and a break frequency separating the two behaviors. The PSD amplitudes and power law exponents determines an apparent energy cascade rate (a rate of transfer of energy from one scale to another), and the location of the break frequency indicates a change in physics e.g. a transition from dissipation free (inertial-range) to dissipative scales where fluctuation energy can be efficiency transferred to random particle thermal motion (heating). The amplitude of the PSD also encodes the overall turbulent fluctuation strength at all scales. 
    Various degrees of intermittency (most simply put, the departure of fluctuation statistics from a Gaussian distribution) also emerge~\citep[see, e.g.,][]{Frisch1995}, associated with the inhomogeneous generation of highly energetic small-scale structures, and typically estimated using the anomalous scaling of the magnetic field and velocity fluctuations' statistics \citep{Sorriso-Valvo1999}.
    The cascade culminates at ion scales with the collisionless dissipation of fluctuating energy and ensuing heating of the plasma \citep[see e.g.,][]{Verscharen2019}.

    \begin{figure}
        \centering         \includegraphics[width=0.99\textwidth]{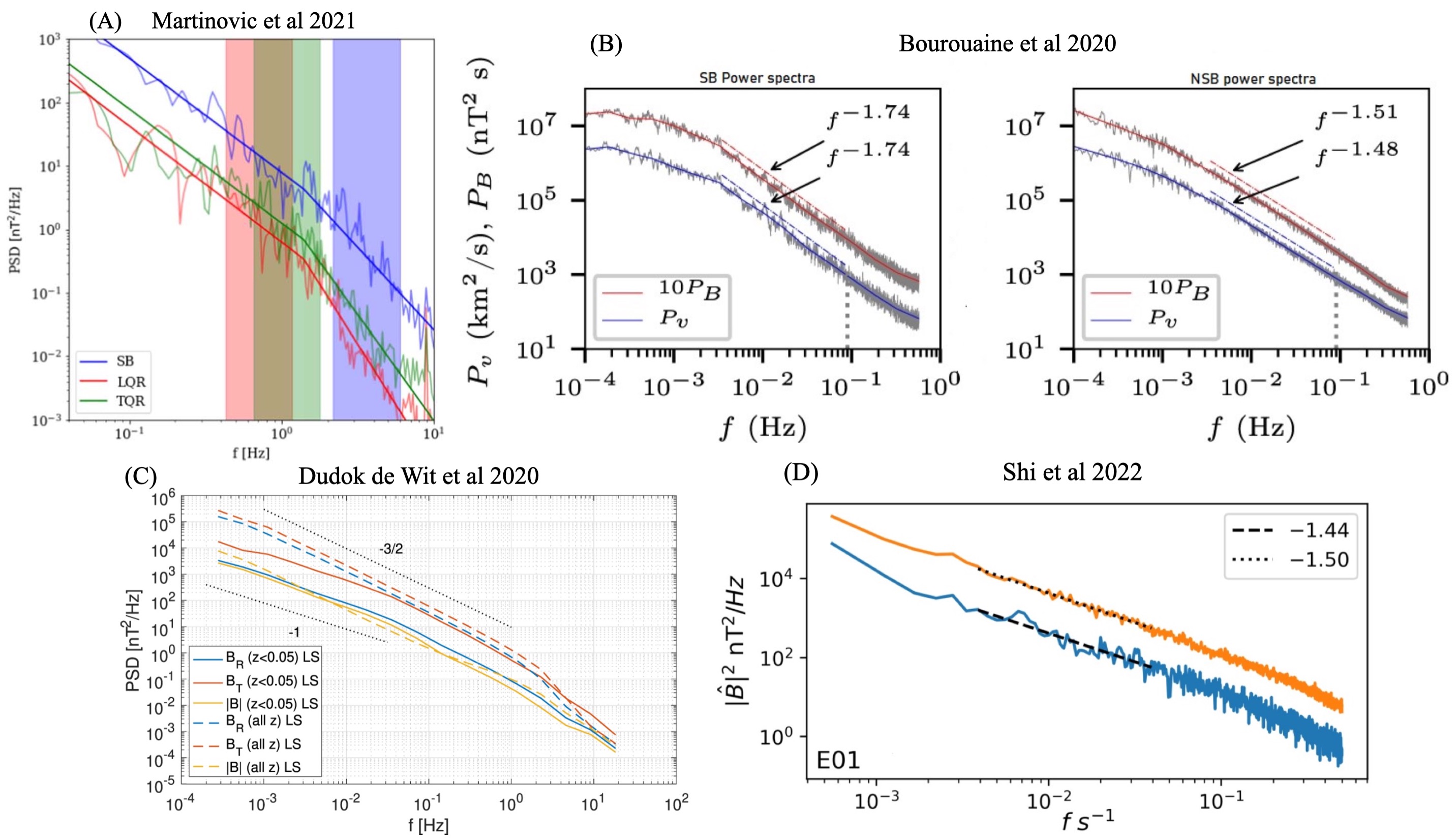}
        \caption{ 
        Power spectral density (PSD) of magnetic field fluctuations during switchback and non-switchback intervals. 
       In (A), \citet{Martinovic2021} compare the PSD of a switchback interval (blue) with the PSD of the surrounding leading quiet region (LQR, in red) and trailing quiet region (TQR, in green).
        In (B), \citet{Bourouaine2020} show the PSD of both magnetic field (red) and velocity (blue) fluctuations, for switchback intervals (left panel) and non switchback intervals (right panel).
        In (C), \citet{DudokdeWit2020} compare the spectral properties of the quiescent periods where switchbacks have been removed (full lines, computed through the Lomb–Scargle (LS) method), and the spectral properties of the full time series, including switchbacks (dotted lines).
        In (D), \citet{Shi2022} show the PSD a quiescent interval (blue) and a switchback interval (orange) observed during E01. 
        Linear fits  of the PSD and their associated slopes are indicated in each figure.
        See Sect.~\ref{subsubsec: 4.3_spectra} for discussion. Figures reproduced with permission  from 
        \cite{Martinovic2021, Bourouaine2020, DudokdeWit2020, Shi_2022_patches}, copyright by AAS}
        \label{fig: 4.3_spectra}
    \end{figure}
    
    \subsubsection{Approach to Turbulence Studies}  \label{subsubsec: 4.3_approaches} 
        
        In order to probe the relationship between switchbacks and turbulence, current research focuses on comparatively characterizing and modeling the major properties of the turbulent fluctuations (including those described in the previous section) in solar wind switchbacks and non-switchback intervals (see Sect.~\ref{sec: 3_methodology} for a discussion on different definitions of switchbacks). 
        This can be done using different methodologies, so that the broadly used expressions ``switchbacks interval'' and ``non-switchbacks interval'' may correspond to at least four different selection criteria, as briefly summarized in the following.
        (i) \citet{Martinovic2021} examined the properties of turbulence in stationary intervals completely contained within each switchback structure, and in regions completely outside of them, then performing ensemble statistics over the different categories. This procedure ensures the study of homogeneous plasma, avoiding the sharp switchback boundaries. However, it severely reduces the low-frequency reach of the spectral analysis, preventing the study of large-scale properties.
        (ii) In a similar approach, \citet{Bourouaine2020} performed a threshold-based conditioned correlation function analysis to study the turbulence inside and outside switchbacks in longer time series, retaining the homogeneity of the sample but also gaining access to lower frequencies. 
        The two methodologies described above are appropriate for studying in a well-separated way the properties of turbulence in parcels of plasma entirely embedded in a switchback and in the regions with no switchbacks. 
        (iii) On the other hand, \citet{DudokdeWit2020} separated quiescent regions (i.e., with very small magnetic field rotations), and then employed a technique for spectral analysis of unevenly sampled data (the Lomb–Scargle method) to compare the spectral properties of the quiescent ensemble with those of the whole time series, which included switchbacks. 
        In this case, the switchback sharp boundaries were retained in the ``switchback intervals''.
        (iv) Alternatively, \citet{Shi_2022_patches} and \citet{Hernandez2021} used hand-picked intervals or fixed-size running windows to obtain turbulence parameters in samples where regions inside, outside, and at the boundaries of a variable number of switchbacks could be mixed together. 
        The latter two approaches are more appropriate for the study of global or regional interplay between turbulence and switchbacks \cite[][]{Mallet_this_issue}. 
        The different methodologies described above may have profound differences, as for example (and primarily) the inclusion or exclusion of the sharp switchback boundaries, which can affect the statistical properties of the turbulent fields.
        It is therefore of utmost importance to give a clear and explicit description of the chosen methodology in order to avoid misinterpretations of the observations and inappropriate comparisons between different studies. 
        
    \subsubsection{Turbulent Spectra}  \label{subsubsec: 4.3_spectra} 
        Several studies based on the above methodologies were performed to compare the turbulence properties inside and outside, or with and without switchbacks. Such studies
        revealed substantial differences and similarities in spectral properties, intermittency, Alfv\'enicity, and turbulent cascade rate.  
        Fig.~\ref{fig: 4.3_spectra} shows four examples of power spectral densities obtained from various Parker encounters in the works described above.
        The top-left panel (A) shows that, using several short intervals, \citet{Martinovic2021} found no major differences between the plasma inside the switchbacks and outside of them, except for a moderate excess of magnetic power. 
        Similar larger magnetic power was found by \citet{Bourouaine2020} (top-right panels, (B)), but in this case, the spectral index inside the switchbacks was $\sim 1.7$ (close to the Kolmogorov value $5/3$), while in the quieter intervals it was $\sim1.5$. 
        \citet{DudokdeWit2020} (bottom-left panel, (C)) showed that a broader and steeper inertial range is present in the whole time series with respect to the extracted quiescent intervals between switchbacks. The magnetic power was also reduced in the quiescent regions, as expected when removing the larger fluctuations.
        Finally, \citet{Shi_2022_patches} (bottom-right panel, (D)) obtained slightly steeper spectral exponents and one order of magnitude enhanced power in the intervals containing switchbacks.
        While the above differences highlight the strong sensitivity of the selection methodology, the general picture that emerges is that solar wind intervals that include switchbacks have generally more developed turbulence, with enhanced power and steeper spectra.  
        The magnetic spectral power also appears more isotropically distributed between the parallel and perpendicular direction to the magnetic field than in the ambient solar wind \citep{saksheeSBs2022}. 
        However, it is important to note that due to the short duration of switchback intervals, large uncertainties may arise \citep{Dudok_de_wit_Samples_Rule}. Consequently, a larger statistical dataset is required to further corroborate this observation. 
        Additionally, \cite{Tatum2024} have recently cautioned that the power spectral enhancement may be an effect related to the modulation of the sampling direction, following relative changes between the local magnetic and flow velocity. During large deflections (switchbacks), the stronger turbulent power typically associated to the perpendicular direction \citep{Horbury2008,Duan_2021}, would contribute more to the instantaneous single-spacecraft measurement of the power spectrum. This effect also likely complicates determinations of the turbulent anisotropy during switchbacks. Disambiguation of single spacecraft effects from such measurements will likely require future multi-spacecraft missions such as Helioswarm.   
        
    \subsubsection{Intermittency and Dissipation}  \label{subsubsec: 4.3_intermittency} 
        The above results are supported by  observations suggesting that both intervals containing switchbacks and intervals within switchbacks exhibit elevated levels of intermittency, as compared to quiet intervals \citep{Hernandez2021}. This picture seems to be also confirmed by more recent Solar Orbiter observations \citep{Perrone2025}. 
        The increased intermittency is evidenced by the enhanced occurrence of small-scale current sheets \citep{Martinovic2021,HuangJ2023a}, also resulting in the fast deviation from Gaussian statistics of the scale-dependent field increments \citep{Hernandez2021}. 
        An additional signature of more developed turbulence is evident in the positive correlation between the cross-scale energy transfer, obtained using third-order moment scaling laws \citep{Marino2023}, and the occurrence rate of switchbacks \citep{Hernandez2021}. 
        This observation suggests that the enhanced energy injected by the strong magnetic and velocity shears is rapidly redistributed to ion-scale structures, where it is eventually dissipated. Evidence suggests that the physical mechanisms behind turbulent dissipation may depend significantly on the level of imbalance, with studies indicating that strongly Alfvénic, high cross-helicity intervals lead to the generation of ion-scale waves that impart their energy into the plasma. Conversely, more balanced, low cross-helicity intervals preferentially cascade to form sub-ion scale structures \citep{Bowen2024}. These results connect MHD scales to kinetic dissipation and suggest that, as examples of highly Alfvénic, outwards propagating structures, switchbacks may be dissipated preferentially via cyclotron resonant processes.

    \subsubsection{Cross-helicity, Residual energy and Alfv\'enicity of the fluctuations}  \label{subsubsec: 4.3_cross_hel} 
        
\begin{figure}
            \centering    
            \includegraphics[width=0.8\textwidth]{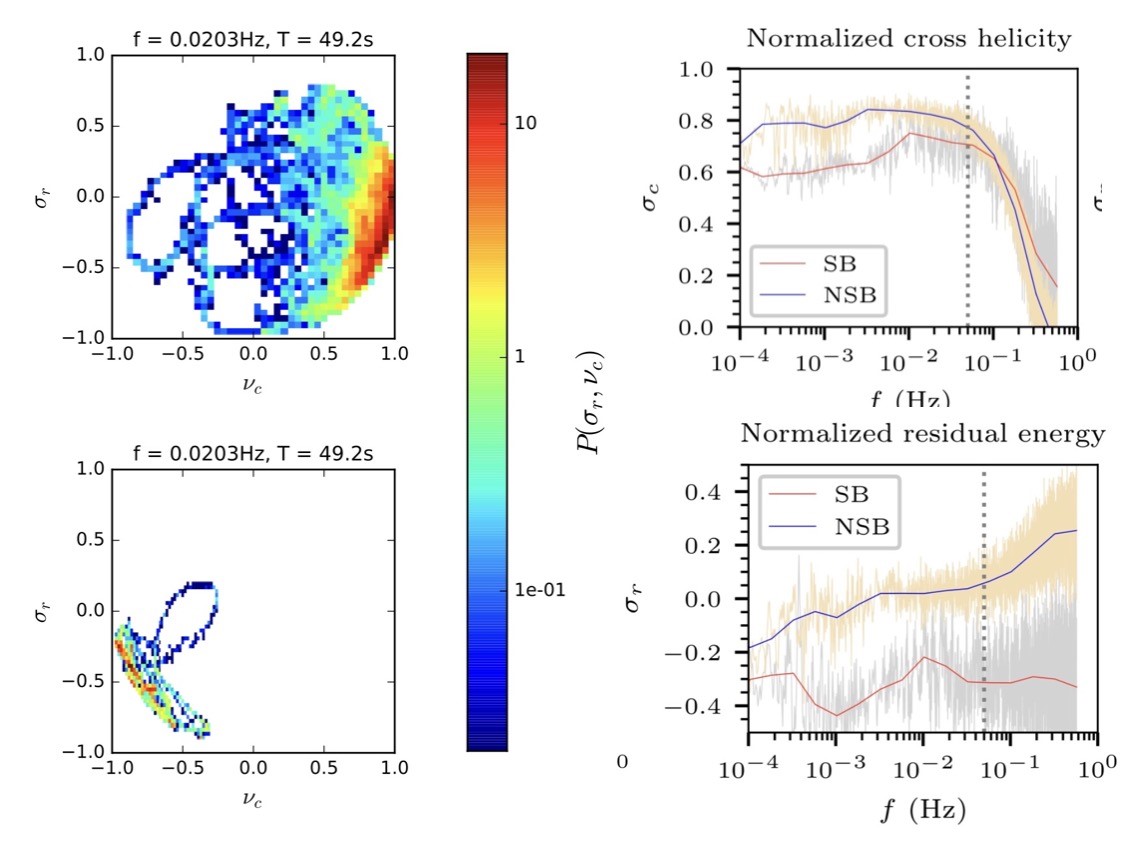}
            \caption{Left: 
            Distribution in the $\nu_c-\sigma_r$ plane for background turbulence (top) and switchbacks (bottom) where $\nu_c$ is the rectified cross-helicity. Figure reproduced with permission from \citep{McManus2020}, copyright by AAS. 
            Right: Normalized Cross-Helicity ($\sigma_c$, top) and Normalized Residual Energy ($\sigma_r$, bottom) for different solar wind conditions: Switchback (SB, in red) and Non-Switchback (NSB, in blue) intervals, plotted against frequency. For frequencies $f \geq 5 \times 10^{-2}$ Hz (marked by a vertical dotted line), results are potentially subject to contamination from noise in velocity measurements. Figure reproduced with permission from \citet{Bourouaine2020}, copyright by AAS
            }
            \label{fig: 4.3_cross_helicity}
        \end{figure}

        \cite{Balogh1999} have discussed how the direction of propagation of high-frequency Alfvén waves during switchbacks (where high-frequency means much higher frequency than the frequency of the switchback itself) can be used as a diagnostics for their magnetic connectivity. In fact, they have shown that during switchbacks observed by Ulysses in the polar wind, the otherwise anti-Sunward flux of Alfvén waves reverses, becoming Sunward, following the bent field line. Such dynamics was confirmed in Parker observations \citep{McManus2020}, shown in  Fig.~\ref{fig: 4.3_cross_helicity}, in terms of $\nu_c$ and $\sigma_r$ where $\nu_c$ is the rectified cross-helicity such that the sign indicates fluctuations in the sunward or antisunward direction (as compared to $\sigma_c$ whose sign reflects the direction with respect to the magnetic field orientation), with $\nu_c>0$ indicating anti-sunward direction. As typically observed in the Alfvénic solar wind, most of the background fluctuations (top panel) lie near the edge of a circle in the $\sigma_c-\sigma_r$ plane. The same pattern is observed in switchbacks (bottom), once the reversal in the propagation direction, i.e., change of sign in $\nu_c$, is accounted for.

\begin{figure}
        \includegraphics[width=\textwidth]{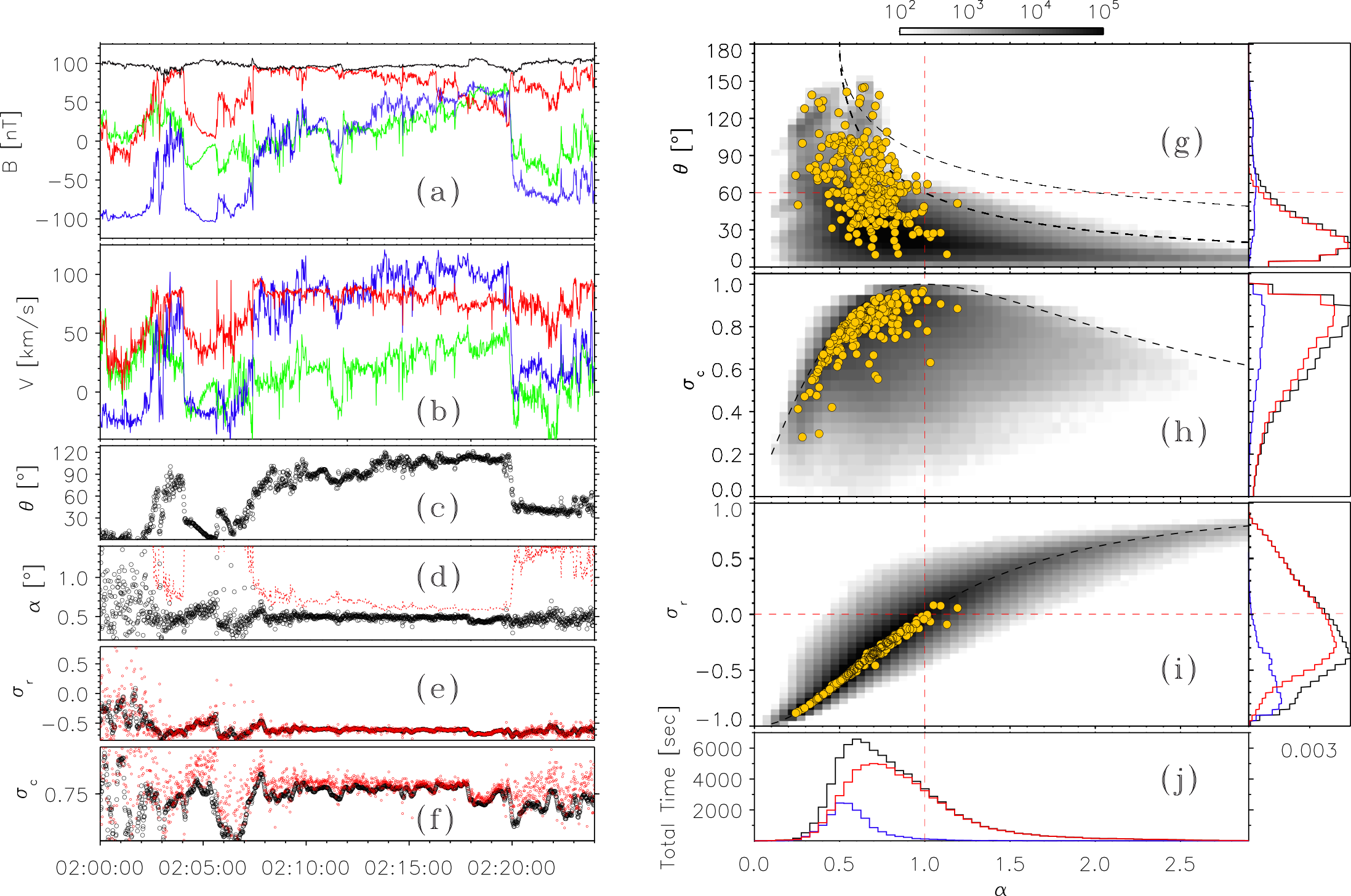}
        \caption{Example and statistics of Alfvénicity, cross , and residual energy of switchback events from Parker E1. 
        In the left hand column, panels (a)--(f) show timeseries of a switchback measured on Nov. 6, 2018 showing respectively the magnetic field in RTN (R blue, T green, R red and black shows the magnitude), proton bulk velocity with the DeHoffman-Teller velocity subtracted \citep[see][]{Agapitov2023ApJ}, the angular deflection of field relative to the mean magnetic vector prior to the switchback, the instantaneous measured Alfv\'enicity parameter ($\alpha$, black) and the geometric limit which would correspond to a velocity fluctuation of $V_A$ (red), the instantaneous measured residual energy and cross helicity. 
        In the right-hand column, statistics of individual events (yellow points) and instantaneous measured values (grey/black shading) are shown as a function of $\alpha$ for panels (g) deflection angle, (h) cross helicity, and (i) residual energy. 
        The inset panels show 1D histograms of the different parameters subdivided into switchbacks (blue), non-switchbacks (red), and all measurements (black). Panel (j) shows the total time different values of $\alpha$ are sampled. The thicker dashed curve in panel (g) shows the geometric limit mentioned above. 
        Figures reproduced with permission from \citet{Agapitov2023ApJ}, copyright by AAS
        }
        \label{fig:agapitov2023}
        \end{figure}

        More detailed Parker studies suggest that intervals with switchbacks present overall a slightly lower cross helicity \citep{Bourouaine2020,Wu2021}, as well as more negative residual energy \citep{Bourouaine2020, Wu2021, Agapitov2023ApJ}, in comparison to quiet intervals, as illustrated in the right-hand panels of Fig.~\ref{fig: 4.3_cross_helicity}. 
        \cite{Agapitov2023ApJ} further investigates cross-helicity, residual energy and `Alfv\'enicity' at the scale of switchbacks themselves. They introduce an `Alfv\'enicity' parameter ($\alpha$) which is defined as $\alpha=\boldsymbol{v}\cdot\boldsymbol{b}/b^2$ where the fluctuation vectors $v$ and $b$ (in Alfv\'en units) are computed by a frame transformation to the deHoffman-Teller frame. Figure~\ref{fig:agapitov2023} illustrates both the detailed evolution of this quantity (in addition to the instantaneous\footnote{Note: Here instantaneous means that the fluctuation vectors $\boldsymbol{b}$ and $\boldsymbol{v}$ are computed relative to a single background vector for each switchback event and $\sigma_c$ and $\sigma_r$ are no longer spectral/ensemble-averaged quantities but computed from the instantaneous variations of $\boldsymbol{b}$ and $\boldsymbol{v}$.} values of $\sigma_c$ and $\sigma_r$) for a single event (left hand column) and statistics (right hand column) over many switchback events (yellow circles) in comparison to the general measurement population (black shading) from Parker E1. For a purely Alfvénic fluctuation, $\alpha=1$ which  corresponds to a fluctuation where  $\boldsymbol{v} = \boldsymbol{b}$.

        Figure~\ref{fig:agapitov2023}(f, i) shows that $\sigma_c$ clusters close to the geometric expectation for pure Alfv\'enic  fluctuations (dashed black curves) compared to the solar wind, meaning $\boldsymbol{b}$ and $\boldsymbol{v}$ remain closely aligned inside switchbacks. However, the mean value of $\sigma_{c}$ inside switchbacks is $0.66\pm 0.03$, which is close but slightly lower than the mean value of $\sigma_{c}$ in the solar wind, $0.69\pm 0.01$ (consistent with time scale decomposed values reported by \citet{Bourouaine2020} and \citet{Wu2021}). Other studies \citep{Shi_2022_patches} did not observe significant differences inside and outside of the switchbacks, but rather reduced cross-helicity at the switchback boundaries. The reason for the difference between those studies is not clear and should be addressed in future works. 
        
        The statistical behavior of $\sigma_r$ is shown in Fig.~\ref{fig:agapitov2023}(e,j). Switchback-related values are distributed from -1 to 0 with the mean value of $-0.58\pm0.07$, the solar wind-related intervals demonstrate values distributed around 0 with the mean value of $-0.18\pm0.04$. 
                
        More generally, a reduced cross-helicity and larger excess of negative residual energy could be attributed to an enhanced level of non-linear interactions between counter-propagating Alfv\'en waves within the switchbacks structures \citep{muller2005spectral, boldyrev2012residual, Gogoberidze2012PhPl, chandran2015intermittency}, consistent with observations of the enhanced Alfv\'enic turbulence levels attributed to switchbacks close to the Sun \citep[e.g.,][]{DudokdeWit2020, Mozer2020_SB}. This scenario would also be consistent with  switchback periods presenting a more developed inertial range. 

        Returning to the Alfv\'enicity parameter itself, Fig.~\ref{fig:agapitov2023}(g) shows evidence that switchbacks have a systematic reduction in $\alpha$ with increasing deflection angle. Moreover, these results suggest that the Alfv\'en speed may play a role in modulating this behavior. The black dashed curve in panel (g) and red dashed data in panel (c) show a geometric construction corresponding to the value $\alpha$ that would take as a function of deflection angle such that $|\boldsymbol{v}| = V_A$. The example event shown in panels (a--f) is seen to approach but remain below this limit. Statistically, the majority of the events (yellow markers) remain below this limit, but a significant number of them are observed above the limit. The distribution of $\alpha$ values increases towards 1 with decreasing deflection angle, and overall, $\alpha$ is below 1 in switchbacks \citep[consistent with their generally exhibiting negative residual energy, and correspondingly having an Alfv\'en ratio $<1$][]{Wu2021}, with a maximum close to 0.8 for the largest deflections and approaching 1 for smaller deflections.  
        
        On the other hand, it is worth noting that high Alfv\'enicity ($\alpha\sim1$) in large deflections has been reported elsewhere \citep[e.g.,][]{Woolley2020,  McManus2022, Bowen2025}, including cases of switchbacks with speed enhancements above $V_A$. While then the Alfv\'en speed may be a reasonable term of comparison for the statistical behavior observed here, it is unlikely to be a true physical limitation, as can also be appreciated e.g. in Fig.~\ref{fig: 5_subAlfvénic_SB}. Nevertheless, this remains an interesting direction for further study, with particular attention to any changes in this behavior with distance from the sun, which may shed light on the interaction of Kelvin-Helmholtz or other instabilities on switchback evolution \citep{Larosa2021, Mozer2020_SB}. 

    \subsubsection{Interaction with Ion Scale Waves}  \label{subsubsec: 4.3_ion_scale_waves} 

        The suggestion of a modulation of solar wind turbulence properties with the presence of switchbacks hints at a connection to how switchback and turbulent energy is dissipated at small scales in the solar wind.

        At ion-kinetic scales, the equations of MHD break down in favor of kinetic-plasma descriptions. Observations of the onset of kinetic effects include various signatures such as kinetic plasma waves and coherent structures that are thought to contribute to solar wind heating and turbulent dissipation. Signatures of ion-scale waves in the presence of switchbacks may help constrain processes behind their kinetic evolution and decay, and consequently their impacts on solar wind heating. 
        
        One primary signature associated with plasma waves is the magnetic helicity of the electromagnetic fields. Helicity signatures of turbulence at ion-kinetic scales have been studied as a constraint on turbulent heating mechanisms \citep{Goldstein1994,Leamon1998a,Leamon1998b}. Studies at 1 AU show that ion-kinetic scale helicity measurements vary depending on the angle between the solar wind and mean magnetic field \citep{He2011,Podesta2011}. A population of left-hand polarized waves is often observed when the mean magnetic field is roughly parallel to the flow direction, i.e., $\theta_{VB}\approx 0$, while a right-handed helicity emerges at perpendicular sampling angles, i.e.,  $\theta_{VB}\approx 90^\circ$.
        
        It has been proposed that the right-handed helical signatures that appear at  $\theta_{VB}\approx 90^\circ$ arise from a spectrum of oblique kinetic Alfvén waves (KAW) \citep{HowesQuataert2010,Podesta2011,He2011}. \cite{Huang2020_turbulence} reproduced results similar results during Parker's first encounter, demonstrating that right-handed polarization, consistent with KAWs, is evident when the local magnetic field is perpendicular to the mean flow direction, which may be preferentially during large-scale inversions associated with switchbacks, such that $\theta_{VB}\approx 90^\circ$.
        
        In addition to right-handed ion-kinetic scale signatures that have been associated with anisotropic turbulence, observations at 1 AU show the presence of circularly polarized waves occurring at ion scales when $\theta_{VB}\approx 0^\circ$. These waves have been interpreted as a population of left-handed waves located near $k_\parallel d_i\sim1$ \citep{He2011,Podesta2011,Klein2014} where $k_\parallel$ is the component of the wavevector parallel to the magnetic field and $d_i$ is the ion inertial length. Statistical studies of parallel-propagating ion-scale waves consistently suggest a preference for the occurrence of these waves during intervals where the mean field is parallel to the background solar wind flow \citep{Jian2010, Wicks2016,Boardsen2015}. This statistical preference for the occurrence of circularly polarized waves during radial field intervals, is consistent with observations from Parker \citep{Bale2019,Bowen2020a,Verniero2020,Liu2023_waves}. However, there are significant measurement effects associated with single-spacecraft sampling \citep{FredricksCoroniti,HowesQuataert2010} that complicate observation of parallel-propagating waves when the magnetic field and solar wind flow are not parallel (or anti-parallel). First, in the presence of anisotropic turbulence, increased amplitudes at perpendicular angles, i.e., in switchbacks, may obscure circularly polarized ion-scale waves \citep{Bowen2020a}. Second, the observed circularly polarized fluctuations in the solar wind are Doppler-shifted in the spacecraft measurement frame such that circularly polarized parallel propagating waves may not be evident at perpendicular sampling angles.
        
        \cite{Bowen2020a} constructed a simple 2D and slab model spectrum to test the observability of parallel-propagating waves in the presence of an anisotropic turbulent background. 
        Their synthetic spectra, as well as their properties, strongly resembled the observations. The similarity between the observed spectra and synthetic spectra suggests that parallel propagating waves may be present during switchbacks, but that the Doppler shift and strong anisotropy of the solar wind turbulence may inhibit their observation.
        
        Parker observations of radial (i.e., non-switchback) intervals, and especially near the heliospheric current sheet, demonstrate that pronounced proton beams exist simultaneously with ion-scale waves \citep{Verniero2020,Verniero2022,Bowen2022} indicating such waves are either generated by proton beam instabilities, or pre-existing wave energies are being damped to create such beams \textit{in situ} (a potential heating mechanism). 
        \cite{Krasnoselskikh2023}, hypothesize that the proton beam itself may be coming from interchange reconnection which would support the former hypothesis. 
        Given the interchange reconnection theory of switchback formation \citep[see][]{Wyper_this_issue}, it is therefore of interest to establish if such beam-wave correspondence is also present during switchback intervals. However, such an association is currently hampered by the difficulty of observing coincident wave signatures during switchback intervals. One potential path forward is to use the fact that switchbacks exist with a spectrum of deflection angles and examine less extreme events to identify if non-instrumental wave signatures can be teased out.
        
    \subsubsection{Switchbacks \& Turbulence Takeaways}  \label{subsubsec: 4.3_takeaways} 
        The joint observation of enhanced energy spectra, intermittency, third-order moment scaling, and cross-helicity in intervals containing switchbacks suggests that these are a crucial component of the turbulence at large scales and that the turbulence is energized by the switchbacks. 
        On the other hand, the recent results are not fully conclusive concerning possible intrinsic differences between the turbulent fluctuations inside and outside the switchbacks. In particular, evidence of enhanced wave-mediated turbulent dissipation during switchbacks is in question \citep{Tatum2024}. Evidence is also not yet conclusive on whether switchbacks of a certain size are meaningfully distinct from smaller amplitude fluctuations. 
        Further studies of the interplay between turbulence and switchbacks are necessary to determine the origin of switchbacks and their evolution, but also to understand if they are an effective source of turbulence and its modulations. 
        In carrying out these studies one should be always aware that the observations could be affected by the level of Alfv\'enicity, the presence of shears, and the geometry of the spacecraft crossing into the structures ~\citep[see, e.g.,][]{Sioulas2022_interm, Cuesta2022, Zank2022} as well as the limitations of single spacecraft measurements in an anisotropically fluctuating field \citep{Bowen2020scam, Tatum2024}.

\section{The Underlying Structure of the Solar Wind}\label{sec: 5_sw_structure}

Up to this point, with the exception of the turbulence paragraphs above, the properties of switchbacks have mostly been discussed in terms of the properties of individual events. However, their aggregate behavior is just as remarkable and important to discuss, and may contain further clues to the origin of these structures. In this section, we discuss first the general observation that switchbacks occur in so-called ``patches'' close to the Sun. We then review current results on the relationship between switchback occurrence in different types of solar wind, including sub-Alfv\'enic intervals from the most recent Parker perihelia.

\subsection{Switchback patches} \label{subsec: 5_patches}

    Visual inspection of Parker's first solar encounter, in November 2018, revealed almost two weeks of uninterrupted switchbacks in a slow Alfvénic solar wind (see Fig.~\ref{fig: 1_ubiquitous_SB}). A striking feature was their occurrence in ``patches'', with a series of large deflections that could last several hours \citep{Bale2019}. By examining the waiting times between successive switchbacks, \citet{DudokdeWit2020} found clear evidence of switchback aggregation. \citet{Chhiber_2020} and \citet{Sioulas_2022} found similar evidence by examining intermittent events using the  Partial Variance of Increments (PVI) method \citep{Greco_2018}.           
    The probability distribution of these waiting times has a power-law distribution, which is typical for sequences of correlated events. The distribution of switchback durations followed a very similar power-law distribution, suggesting that the physical mechanisms governing the rise and fall phases of each switchback were the same \citep{DudokdeWit2020}. Switchbacks were thus unlike instabilities that are triggered by one mechanism and then relax when another mechanism comes into play. It is also interesting to note that the patches' modulation in amplitude is more clear than their modulation in occurrence rate: that is small deflection events between patches occur almost as frequently as large deflection events occur in the middle of patches -- see Fig.~\ref{fig: 4.2_deflection_duration}).  Therefore, a subtle but important point is that when one identifies a switchback patch in Parker data, the key visual indicator is a rise and fall in switchback amplitude rather than a start or stop in switchback occurrence.
    
    However, these results should be treated with caution, since power-law distributions of waiting times are known to occur even in non-stationary systems with uncorrelated events \citep{Wheatland_2002}.    
    In addition, most analyses mixed different regimes of the solar wind and also different samplings of the switchbacks.
    During co-rotation periods, the spacecraft provided a temporal slice of their dynamics, while it provided a spatial slice elsewhere. 
    In addition, Parker sampled the solar wind transversely near perihelion and radially farther out. 
    Disentangling these different regimes is important to further understand switchbacks' structures and properties.
    
    \begin{figure}%
        \centering            
        \includegraphics[width=1.0\linewidth]{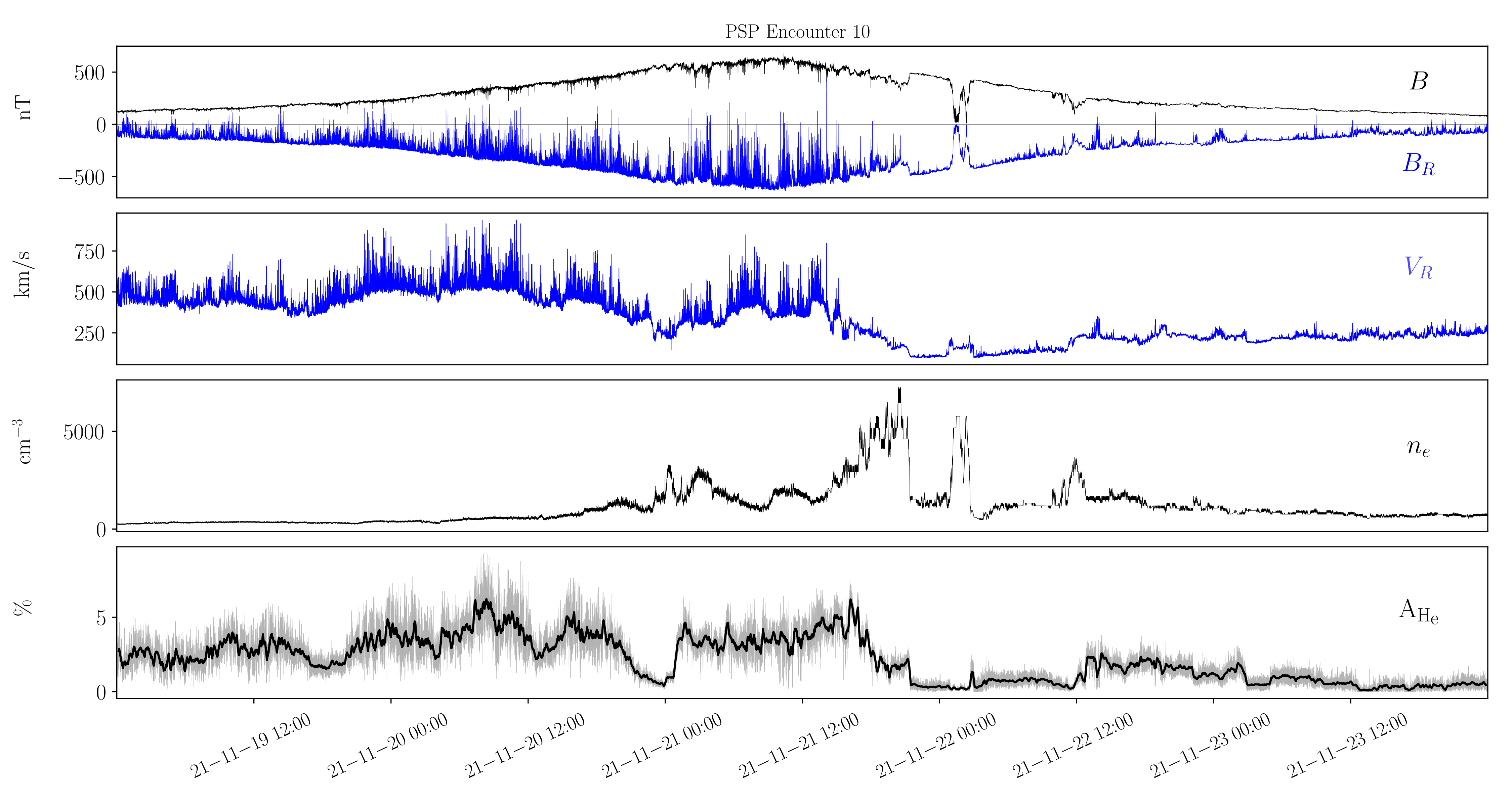}%
        \caption{Example \textit{in situ} data from Parker E10. From top to bottom: radial component and amplitude of the magnetic field,  solar wind radial velocity, electron density, and alpha particle abundance.}
        \label{fig: 5_patches_overview}
    \end{figure}
    
    \citet{Fargette2021} and \citet{Bale2021} studied the patch structure and its connection to solar source regions. In particular, both studies examined intervals where Parker moved rapidly prograde in Carrington longitude and therefore comprised a spatial cut through solar wind structure. This revealed a typical patch angular size of $5^o$, which is similar to the angular size of solar supergranular structure. 
    As seen in Fig.~\ref{fig: 5_patches_overview} and in \cite{Bale2021}, typically the velocity spikes and field deflections show obvious patch structure, however the alpha abundance is also seen to vary similarly in some cases \citep[although it is not always correlated, see][]{McManus2022}). The alpha abundance in particular was taken as further evidence of a connection to underlying coronal structure being the cause of the patches. 
    \citet{Bale2023} combined similar patch modulation from Parker's 10th solar encounter with simultaneous observations of suprathermal ion tails with particle-in-cell simulation to further interpret the switchback patches as the remnants of bursty interchange reconnection organized on the solar supergranular network. 
    
    More recently, \citet{Rivera2024b} used a well-aligned conjunction between Parker and Solar Orbiter \citep[the same as studied in][]{Rivera2024a} to use Solar Orbiter Heavy Ion Sensor \citep{Livi2023} measurements as a proxy of the compositional characteristics of switchback patches, as well as direct measurements of the electron core temperature by Parker SPAN-e. They found similar characteristics inside and immediately adjacent to patches,  with characteristics varying between patches, and also elemental abundance indicating source regions with open and closed field in close proximity to each other. 

    \citet{Horbury2023} and \citet{Soni2024} have similarly exploited Parker-Solar Orbiter alignments and observed patch-scale structure which appears to be preserved at consistent angular scales (at least out to distances probed by Solar Orbiter). They observe that these structures come to resemble ``microstreams'', a feature identified historically in Ulysses polar fast wind data \citep{Neugebauer1995}. 
   
    While these studies provide evidence on the longitudinal structure of switchback patches, there is evidence that modulation of switchbacks may exist in the radial direction too. Switchback patches typically exhibit slow modulation in $|B|$ and $V_R$ (see e.g., Fig.~\ref{fig: 5_patches_overview}). 
    Domains of nearly constant $|B|$ (i.e., magnetic pressure balance) often represent the space between switchbacks. 
    For example, figure~1 of \citet{Ruffolo2021} shows an example of the identification of such domains in Parker data near the 2nd perihelion. 
    In some cases, either one or both boundaries of a switchback patch disrupt the domain structure, while there is a domain of nearly constant $|B|$ within the patch. 
    \citet{Ruffolo2021} characterized the aspect ratio of such domains by considering their duration as a function of either the total solar wind speed or a perpendicular speed relative to Parker, where ``perpendicular'' was relative to a possible axis of systematic domain elongation. 
    They concluded that the domain shape could be isotropic (or spherical) and ruled out long aspect ratios either along the radial or Parker spiral direction. 
    This is consistent with prior observations of isotropic density fluctuations or flocculae in the super-Alfv\'enic solar wind \citep{DeForest16}. 
    Given that domains can be interrupted by switchback patch boundaries, this suggests that the spatial distribution of switchback patches (and the space between patches) is consistent with being isotropic, and the patches are not much more widely spaced along the radial or Parker spiral direction. This is in contrast to the switchbacks themselves, which are observed to have an elongated aspect ratio \citep{Laker2021}, which may play a role in the structures' stability \citep{Shi2024}, see Sect.~\ref{subsec: 4.2_geometry_boundaries}.

\begin{figure}%
        \centering
        \includegraphics[width=1.0\linewidth]{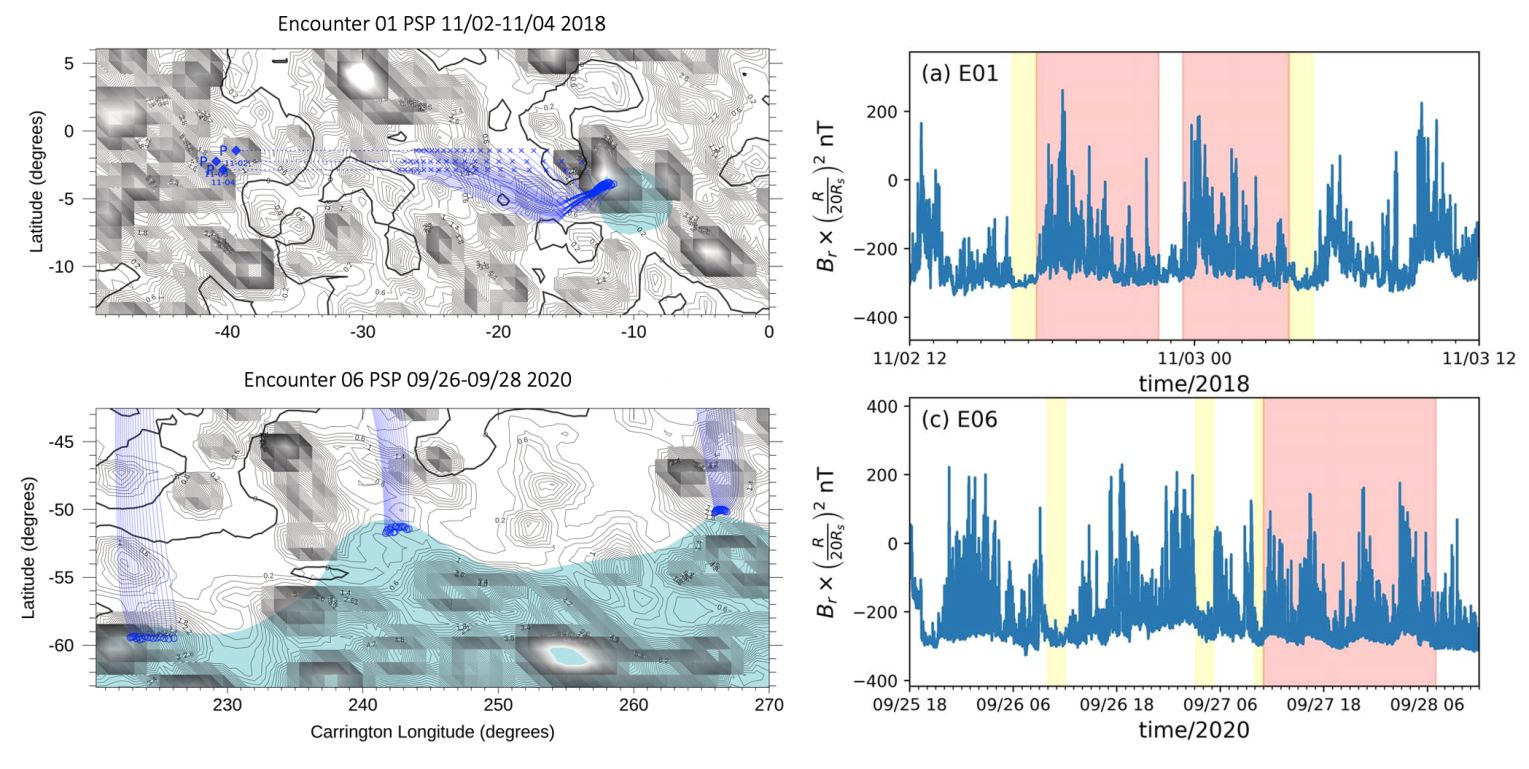}%
        \caption{Left hand column: Potential Field Source Surface mapping \citep{Panasenco2020} of a time interval where Parker moved mostly radially during E01 (top) and mostly longitudinally during E06 (bottom). Right hand column: Corresponding time profiles of the radial magnetic field measured by Parker and multiplied by the square of its distance from the Sun. Yellow and pink shading indicates ``quiescent'' and switchback patch structure, respectively. Switchback patches are observed in both cases. Figure adapted with permission from \cite{Shi_2022_patches}, copyright by AAS.
        }
        \label{fig: 5_patches_spatial_temporal}
    \end{figure}

    Similarly, \citet{Shi_2022_patches} examined data from Parker observations from E1--E10 during intervals both when the spacecraft is near co-rotation with the Sun and at intervals \citep[as in][]{Fargette2021,Bale2021,Bale2023} with rapid longitudinal motion. 
    In both cases, switchback patches were visually identified (see Fig.~\ref{fig: 5_patches_spatial_temporal}). For cases where the motion was primarily longitudinal (bottom row of Fig.~\ref{fig: 5_patches_spatial_temporal}), \citet{Shi_2022_patches} arranged spatially just as in \citet{Bale2021}.
    However, for intervals near the co-rotating interval of Parker’s orbit (top row in Fig.~\ref{fig: 5_patches_spatial_temporal}) the spacecraft's magnetic footpoint on the source surface moved extremely slowly in Carrington longitude (top left panel), suggesting spatial modulation alone does not explain patches, but rather there must also be a time-dependent effect that the authors associated with flux emergence at the solar source. 

    These additional results suggest some switchback patches may have finite radial \textit{and} longitudinal extent. While there is good physical motivation for coronal origin of the longitudinal extent, their radial behavior is a key area of further study that is necessary to understand their evolution with heliocentric distance and to explain why they are much less visually apparent farther from the Sun.
       
\subsection{Switchback Occurrence in Different Solar Wind Streams} \label{subsec: 5_streams}
        
    As discussed above, switchbacks do not occur randomly (Fig.~\ref{fig: 3_z_angle_annotated}) and are not omnipresent but rather are modulated in amplitude both on consistent spatial scales (see Sect.~\ref{subsec: 5_patches}) and on a less well understood basis with different solar wind streams: The solar wind, as determined from statistical observations from 1 AU, is composed of a ``fast'' and ``slow'' component with a wind speed of 500\kms\ at 1~AU used as a typical delineation \citep[e.g.,][]{Kepko2016}. Slow wind streams are observed to be further distinguishable according to their cross-helicity (or ``Alfv\'enicity'') into slow Alfv\'enic solar wind and slow non-Alfv\'enic solar wind \citep[e.g.,][]{D'Amicis2021}. These different types of streams are thought to originate from different source regions and processes back at the Sun. It is therefore of interest to study the occurrence of switchbacks in different streams since a consistent difference according to solar wind type could indicate that the source region and associated energization method play a role in the creation of switchbacks. 
        
    At the time of writing, there are limited studies that have attempted to relate switchback occurrence to the type of solar wind in which they are embedded. We review these here but remark that they are mostly qualified as case studies, while more systematic and statistical investigations should be carried out to corroborate these findings. Further, as discussed in Sect.~\ref{sec: 3_methodology}, there are numerous ways to quantify the presence of switchbacks, and in fact, the term ``occurrence'' depends on such a definition. 
       
    \citet{Rouillard2020} studied the properties of switchbacks inside and outside the solar minimum streamer belt using observations from the second perihelion pass of Parker. 
    They exploited coronagraphic imagery recorded by the Solar and Heliospheric Observatory  \citep[SoHO;][]{Domingo1995} to infer the types of solar winds measured by Parker during bursts of magnetic switchbacks. In this study, the proton speed measured by SWEAP was used to trace back (ballistically) the origin of the measured solar wind in the solar imaging taken by the Large Angle and Spectrometric COronagraph (LASCO; Brueckner et al. 1995) on SOHO.  This back-mapping showed that during most of E2, Parker remained on the edges of the streamer belt. This was confirmed by the measurements \textit{in situ} of high-density and highly variable wind flows, but no polarity inversion. As long as Parker remained inside the bright streamers, the density of the solar wind remained elevated, but as soon as it exited the streamers to enter solar wind flows produced by an isolated coronal hole, the wind density dropped by a factor of two to four. \citet{Rouillard2020} showed that while switchbacks were observed in both streamer and coronal hole flows, the patches of switchbacks had different properties between the two types of slow winds with more pronounced reversals of the magnetic fields \citep{Rouillard2020} and enhanced spectral power in the switchbacks transported in the streamer flows than in the coronal hole \citep{Fargette2021}. In addition, the patches of switchbacks in the streamer flows were associated with more significant compressive (density and magnetic field magnitude) variations than in the coronal hole flows. It is yet unclear if these density changes resulted from the formation of these switchbacks or were produced through distinct processes that are known to occur near the center of the streamer belt \citep{Wang1998,SanchezDiaz2017a, SanchezDiaz2017b, Tripathi_this_issue}. This type of study should be repeated for more encounters.

\begin{figure}
        \centering
        \includegraphics[width=0.7\linewidth]{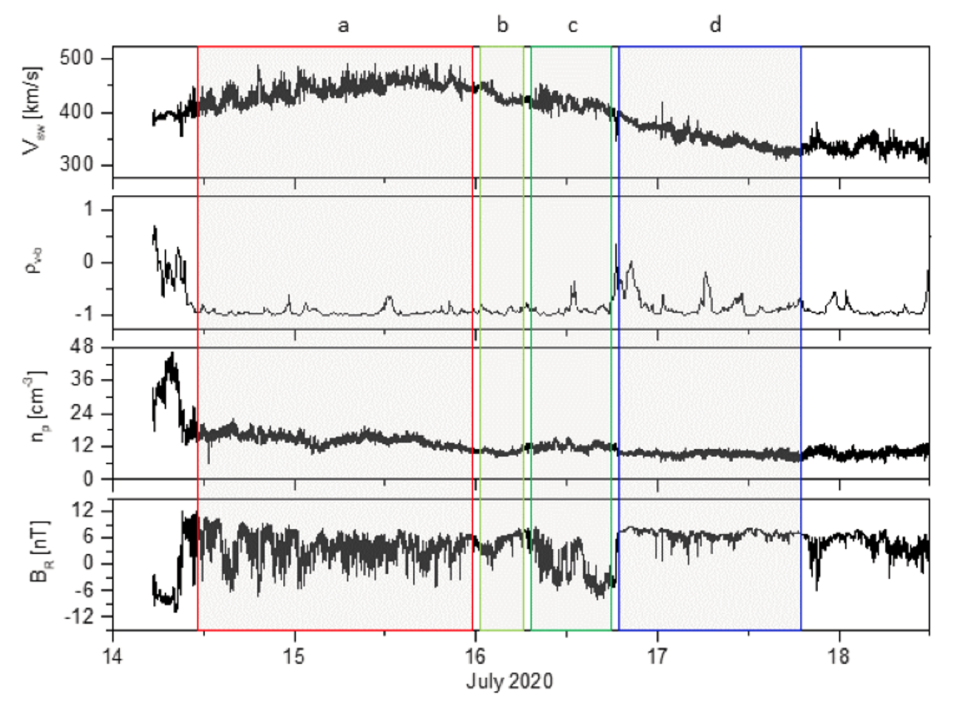}
        \caption{Four solar wind intervals observed by Solar Orbiter labeled by solar source region. (a) faster wind from a northern polar coronal hole extension (b) wind from its eastern boundary, (c) negative open field area at the equator and (d) a slow transition across a pseudostreamer. Figure reproduced with permission from \cite{D'Amicis2021}, copyright by ESO}        
        \label{fig: 5_SB_Orbiter}
    \end{figure}

    \citet{Panasenco2020} and \citet{D'Amicis2021} identified intervals where Parker and Solar Orbiter traversed different types of solar wind, including pseudostreamers. Fig.~\ref{fig: 5_SB_Orbiter} shows an example of \textit{in situ} solar wind data taken by Solar Orbiter around 0.64 AU with segment (d) corresponding to pseudostreamer-type wind. They observe lower amplitude Alfv\'enic fluctuations with almost no deflections past 90$^{\circ}$ during a slow Alfv\'enic stream as compared to a prior fast stream (segment a in \ref{fig: 5_SB_Orbiter}). Given this is based on data farther from the Sun, it is important that such studies be repeated with close approach Parker data when switchbacks are  more unambiguously identified. In one such recent study, \citet{Rivera2025} studied Parker E11 and showed the energy flux of large amplitude Alfv\'enic fluctuations was a larger fraction of the total solar wind energy budget in a slow Alfv\'enic wind stream as compared to a neighboring slow non-alf\'enic stream, but in both cases was lower than the fractional energy of such fluctuations in a neighboring fast stream. 
                
\begin{figure}%
        \centering
        \includegraphics[width=1\linewidth]{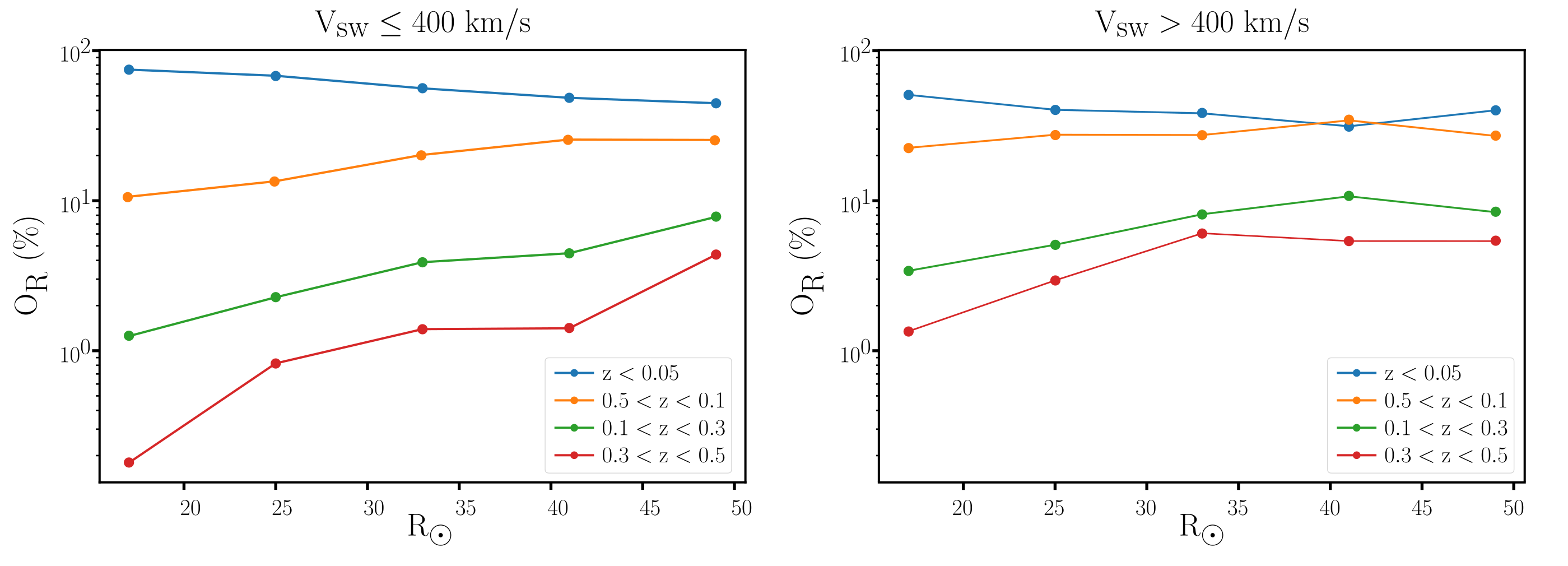}
        \caption{Occurrence percentage of deflections ($O_R$) as a function of radial distance (R$_{\odot}$). For each radial distance bin, we show the fraction of deflections lying with a certain range of deflectiona angles, $z$. In panel $a$ we showthis variation for the intervals with $V_{SW}\leq400$~\kms\ and in panel $b$ we show the variation of D for the intervals with $V_{SW}>400$~\kms\. Figure reproduced with permission from \citet{Jagarlamudi2023}, copyright by AAS}
        \label{fig: 5_Deflection_distribution}
    \end{figure}

    \citet{Jagarlamudi2023} performed a statistical study of switchback occurrence rates. While theirs and others' results examining the occurrence as a function of radial distance from the Sun are discussed at length in \cite{Mallet_this_issue}, we briefly examine their results pertaining to separating switchbacks according to solar wind speed. In Fig.~\ref{fig: 5_Deflection_distribution}, switchback occurrence measured using a deflection threshold of 37$^\circ$ (z $>$ 0.1) binned by deflection scale and radial distance, is shown for fast and slow wind speeds. The main difference observed is that the deflection angles (D) are overall skewed to larger values in the faster wind streams.
    
    The commonality in these studies is an observation that in slow Alfv\'enic solar wind, switchback amplitudes tend to be reduced, consistent with the inference that switchbacks are more energetically relevant to faster wind types \citep{Halekas2023,Rivera2024a}.
    These three studies show this in the context of case studies of the streamer belt wind, pseudostreamer wind, or simply binning by local wind speed. A systematic separation of slow wind into different sub-types may reveal different variations of switchbacks among solar wind from different sources.
        
\subsection{Switchbacks in Low Alfvén Mach Number Wind} \label{subsec: 5_low_MA}

    Parker is the first heliospheric mission to routinely sample low Alfvén Mach number and sub-Alfv\`enic solar wind close to the Sun \citep{Kasper2021}. Since the Alfv\`enic fluctuations comprising switchbacks typically move on a sphere in velocity space of radius near the Alfvén speed \citep{McManus2022}, their behavior as this parameter changes, and in particular becomes comparable to the background flow speed, is of keen interest in understanding their nature and origin.

    Early Parker observations and numerical simulations had predicted that switchbacks may be generated low in the corona \citep[see e.g.,][]{Fisk2020, Bale2023, Tripathi_this_issue, Wyper_this_issue}. For example, \citet{AkhavanTafti2022} used discontinuity classification of switchbacks from the first eight Parker encounters to argue that switchbacks were most likely generated in the solar corona via interchange reconnection. They further went on to hypothesize that switchback evolution in the solar corona could contribute significantly to coronal heating. However, a clear deficit of switchbacks with full magnetic field reversals (deflection angle $>90$ deg, see Sect.~\ref{subsec: 3_SB_methods}) is observed in the sub-Alfv\'enic solar corona both during observations in the earliest sub-Alfv\'enic streams \citep{Bandyopadhyay2022,Pecora2022,Jagarlamudi2023} and more recently over the first 14 full orbits \citep{AkhavanTafti2024}. 
    These observations indicate that switchbacks (as defined as reversals) are mostly either generated locally in the solar wind, or that magnetic switchbacks with deflection angle $<90$ deg are generated in the solar corona and evolve to become full magnetic reversals in the super-Alfv\`enic solar wind as predicted by several models \citep[see e.g.][]{Ruffolo2020, Squire2020,Schwadron2021, Shoda2021, Touresse2024}. See also \citet{Wyper_this_issue} for a complete review comparing and contrasting such model predictions.
        
    \begin{figure}
        \centering
        \includegraphics[width=1.\linewidth]{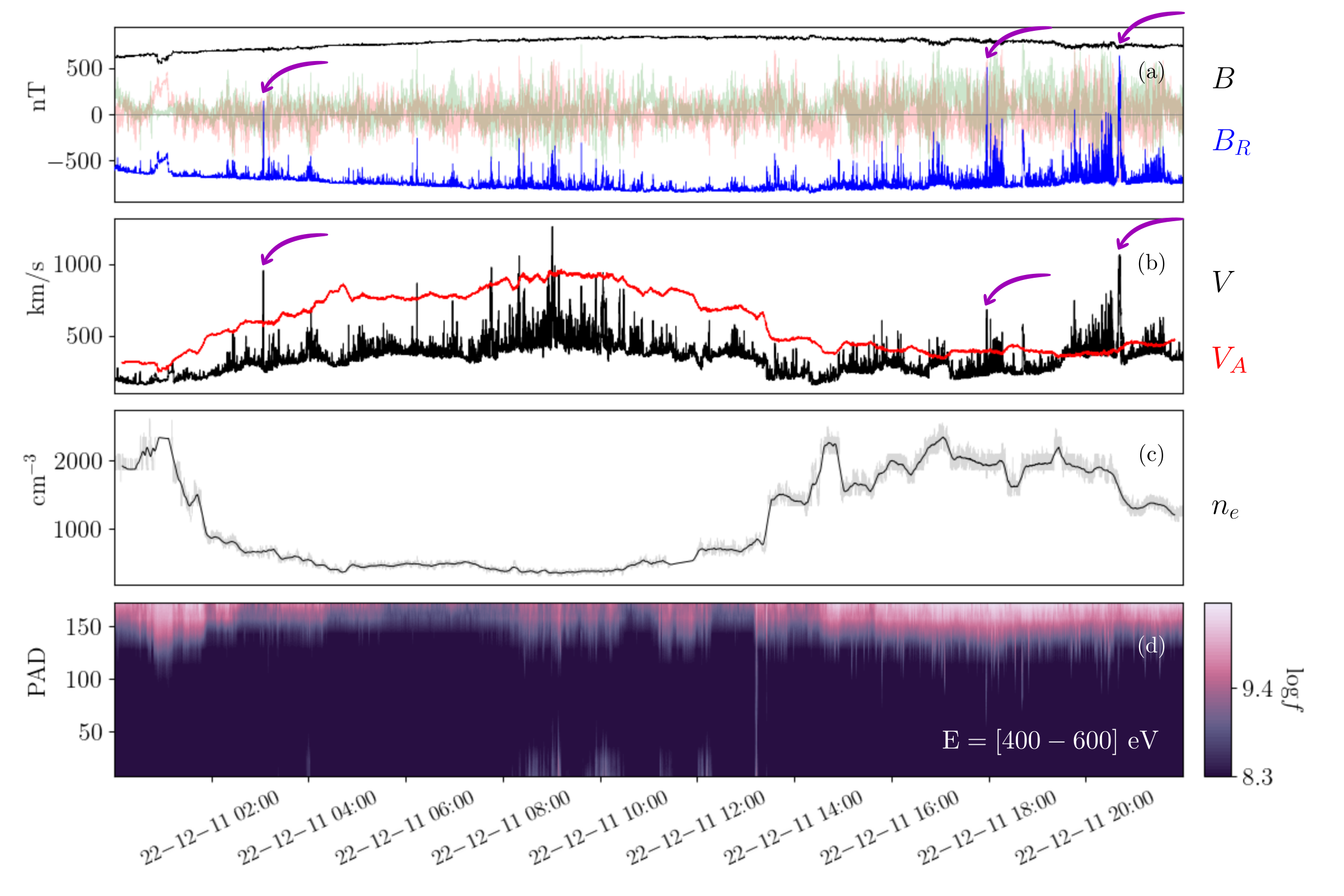}
        \caption{Examples of magnetic field reversals embedded in otherwise sub-Alfvénic solar wind, indicated by purple arrows. Panels show from top to bottom: the magnetic field, the solar wind speed $V$ and local Alfvén speed $V_A$, the electron density $n_e$ and PAD of electrons within the 400--600~eV energy range.}            
        \label{fig: 5_subAlfvénic_SB}
    \end{figure}

    Two subtleties of this inference should be noted, however. Firstly, switchbacks comprise local spikes in the radial velocity which become a sizable portion of the Alfvén speed for large deflections \citep[see][]{Matteini2014}. When the magnetic field becomes more transverse or is even reversed (due to $\delta B_R$), $\delta v_R$ is invariably positive, resulting in a localized increase in the wind bulk velocity. Therefore, even though switchbacks may be situated within a nominally sub-Alfv\'enic wind, the presence of these radial velocity jets can locally amplify the solar wind speed, temporarily pushing the local dynamics into a super-Alfv\'enic regime (Sioulas et al, AGU Fall Meeting 2024\footnote{See \url{https://ui.adsabs.harvard.edu/abs/2024AGUFMSH23F..03S}}). Such an example is presented in Fig.~\ref{fig: 5_subAlfvénic_SB}. Therefore, if simply filtering data timeseries according to the instantaneous Alfvén Mach number, large deflections in otherwise sub-Alfv\'enic streams may be excluded by construction. To avoid this pitfall, an average background Alfvén speed should be used when studying switchbacks in a sub-Alfvénic solar wind.

    Secondly, there are differences in the origin of sub-Alfv\'enic intervals in the Parker dataset to date. While several appear attributable to steady solar wind conditions and coronal source region properties \citep[e.g.,][]{Kasper2021, Bandyopadhyay2022, Liu2023_SB, Badman2023}, there is evidence that transient structures such as interplanetary coronal mass ejections (ICMEs) can also create long sub-Alfv\'enic wind intervals. Evidence is provided by \citet{Chane2021} and \citet{Romeo2023}. In the study of \citet{Romeo2023}, it was shown that the passage of the CME created a huge density depletion region in its wake, which resulted in the creation of sub-Alfv\'enic wind streams with Alfv\'en points signficantly farther out from the Sun than normal. So far, both types of intervals appear to be similar microphysically (i.e., the absence of magnetic reversals and high transverse fluctuations) but further work is needed to investigate whether this is true more generally, especially with the most recent, closest perihelia of Parker. It is important, therefore, to differentiate the source for the observation of sub-Alfv\'enic wind intervals before studying the properties of switchbacks in these intervals and connecting them to the global solar wind evolution.

\subsection{Switchbacks Collective Behavior Takeaways} \label{subsec: 5_takeaways}

    The organization of switchbacks into larger-scale patch structures is a key new observational feature from Parker Solar Probe that defines their appearance in the young solar wind and marks a difference from prior observations of individual field reversals farther from the Sun (although such reversals did appear within structures, called microstreams, in the Ulysses polar observations). The main takeaways are:
    \begin{itemize}
        \item Switchback patch longitudinal sizes match a key photospheric angular scale (supergranules) as judged by modulation of plasma parameters on patch scales not involved in the switchback fluctuations (i.e., density, potentially proton parallel temperature and sometimes alpha abundance). Temporal scales also match characteristics of supergranulation. This and compositional data indicative of open and closed field structures provide evidence that coronal magnetic field structure plays a role in their formation.
        \item Switchback patch properties (most clearly their amplitude) vary between different solar wind stream types (lower amplitudes typically observed in slower streams) suggesting their formation and evolution are affected by the type of stream they exist in.
        \item Growing evidence supports that the size of the deflection angle of switchbacks increases with Alfvén Mach number, and that it is rare to observe polarity reversals in sub-Alfvénic wind. 
    \end{itemize}
      
\section{Summary and Open Questions}\label{sec: 6_open_questions}

We have summarized the main properties of ``magnetic switchbacks'', solar wind fluctuations which have been shown to be ubiquitous close to the Sun by Parker Solar Probe observations. In Sect.~\ref{sec: 2_SB_definition}, we discussed how switchbacks' essential features identify them as large amplitude Alfv\'enic fluctuations in which the magnetic field vector goes through a large, magnitude-conserving deflection while the radial velocity is simultaneously enhanced. The transition to the deflected state is sharp, meaning it occurs more rapidly than the time over which the deflection persists. Throughout this transition, field lines remain connected to the Sun as demonstrated by the fact that the electron heat flux remains aligned with the corresponding field line following the field line folds. At closest approach to the Sun, switchbacks are arranged into patches whose angular scales match those of coronal structures (supergranules).

In Sect.~\ref{sec: 3_methodology}, we discussed the different methodologies used to detect switchbacks with \textit{in situ} data. In particular, we highlight an overall dichotomy in the literature: while some studies simply examine the deflection angle of the field, others use all the defining features of switchbacks but require human intervention and are thus less statistically rich. The use of the rapidly growing field of machine learning or other AI tools may be a useful route to improve this limitation.  Further, we highlight that deflection-based thresholding for switchback detection requires some choice of background field orientation, and choices vary throughout the literature. 

Next, in Sect.~\ref{sec: 4_SB_properties} we listed some of the more detailed properties inferred in relation to switchbacks. These were sub-categorized into:
\begin{enumerate}
    \item \textbf{The steady plasma properties of switchbacks as compared to their surroundings.} In line with their identification as Alfv\'enic structures, it was noted that the magnetic field magnitude and density of switchbacks changes only modestly compared to their surroundings. The inside-outside ratios form distributions which peak at the ideal incompressible (i.e., Alfv\'enic) case but with a finite population which exhibits some compression. Additionally, there is some evidence for increased proton temperatures inside individual switchbacks. This may suggest that switchbacks contain plasma distinct from their surroundings and indicative of an injection process at the source. However, these results need to be bolstered with additional evidence, both in terms of the plasma temperatures as well as other compositional properties.
    \item \textbf{The properties of switchback boundaries, including boundary geometry, discontinuity classification, and boundary-associated electromagnetic wave activity.} We summarize how the processes at these boundaries offer clues on switchback generation and decay. The nature of their boundary discontinuity is important for establishing the extent to which switchbacks can mix with their surroundings and therefore lose memory of their origin and merge slowly, rather than rapidly via instabilities and rapid dynamical processes such as reconnection. 
    \item \textbf{The interaction between switchbacks, turbulence, and ion scale waves.} We summarize a general consensus of a more developed turbulence in the presence of switchbacks. As an energy source at large wavelengths, switchbacks are likely to be an important actor within the full spectrum of solar wind turbulence. Nevertheless, results differ in their details and a consensus picture of cause and effect between turbulence and switchbacks, and implications for whether switchbacks are carrying coronal information, is yet to be established. 
\end{enumerate}

Finally, in Sect.~\ref{sec: 5_sw_structure} we reviewed results on the collective behavior of switchbacks, most prominently their organization into patches, but also the extent to which their presence varies in different wind types.
Specifically, we recounted how switchbacks patch spatial extent (with variation across a range of solar wind parameters) and solar supergranular structure are well associated. However, questions remain about the radial extent of patches, with some suggestion some are time-dependent structures. Comparing different solar wind types, a consensus emerges that switchbacks are generally most prominent, high amplitude, and energetically relevant in faster wind streams. More systematic approaches that bin wind, especially slow types, into different levels of Alfv\'enicity are needed to complete this. Lastly, we touched on the connection between switchbacks and low-Alfv\'en Mach number wind, where several authors report that polarity reversal switchbacks are rare in such wind streams. However, counter-examples are possible to find, and additionally, further work is needed to distinguish the effects of transient wakes from steady sub-Alfvénic wind streams to fully understand the behavior of switchbacks at this transition point.

We close this summary with a list of some significant remaining gaps in our knowledge of the properties of magnetic switchbacks, whose resolution may provide evidence for the origin of these structures and their subsequent interaction with the solar wind. 

\subsection{Knowledge Gaps}

\subsubsection*{Are Switchbacks Meaningfully Distinct from Alfv\'enic Turbulence?}
    
    Switchbacks are large amplitude Alfv\'enic fluctuations embedded in solar wind dominated by Alfv\'enic turbulence. Viewing the spectrum of solar wind fluctuations in terms of just Alfv\'enic fluctuation power or rotational increments \citep{Larosa2024}, switchbacks appear continuous with the background. However, switchbacks are also defined (Sect.~\ref{sec: 2_SB_definition}) by geometrical constraints: sharp boundaries and continuous electron strahl. These aspects are not conditions generally applied to turbulent fluctuations, and also are not enforced for switchback detection methods (Sect.~\ref{sec: 3_methodology}), which are based solely on magnetic deflection angle. 
    
    It remains to be robustly established how important applying these geometric criteria to switchbacks is in determining their properties and whether they are truly special fluctuation types with respect to background solar wind turbulence.

\subsubsection*{Do Switchbacks Contain a Different Plasma Population from the Surrounding Solar Wind?}
    
    Switchbacks are characterized by a number of properties that do not change much compared to their surroundings. Most notably, this is true of the magnetic field magnitude and density. This property supports the characterization of switchbacks as nearly incompressible Alfv\'enic fluctuations. 
    
    However, their thermal and compositional properties are not firmly established with respect to their surroundings. While the best indications to date suggest a potential enhancement in proton temperature and potential modulation of anisotropy \citep{Woodham2021,HuangJ2023b,Laker2024}, measurement and statistical limitations mean further work is needed to more concretely establish this.  The presence of proton (and alpha) beams is also something that may be systematically affected by observational effects and therefore needs careful further attention. Moreover, it is not excluded that part of the observed modulation of the plasma pressure within switchbacks may occur locally, due to the coupling to velocity shears within the structure \citep{DelSarto_Pegoraro_2018}.
    
    Direct \textit{in situ} thermal compositional data is also limited by instrumentation at Parker to studying proton and alpha particles, and these results are also currently ambiguous \citep{McManus2022}. Possible inroads may also be made by the consideration of energetic particles by the IS$\odot $IS instrument on Parker \citep{McDougall2025}, or by combining measurements farther out by the Solar Orbiter Heavy Ion Sensor \citep{Livi2023} for well-aligned streams. This latter strategy was recently applied by \citet{Rivera2024b} to examine the heavy ion composition of patches and did not reveal distinct differences inside patches to outside, but did find a range of compositional signatures indicating a solar wind source with open and closed field in close proximity. These additional insights are vital given that there are \textit{in situ} instability mechanisms which may also lead to simple temperature differences occurring in the solar wind rather than being created at the source.
    
    Answering these questions is of direct importance to the formation of switchbacks in that distinct thermal and compositional characteristics would imply they are created from distinct coronal sources. It should be cautioned, though, that the inverse is not necessarily true if there is significant mixing between switchbacks and their neighboring plasma, then such thermal or compositional information may be lost. This highlights the importance of establishing the discontinuity type of switchbacks since tangential discontinuities may inhibit such mixing.

\subsubsection*{What do Switchback Boundaries and Associated Wave Activity Tell us About How Switchbacks Form and Decay?}
    
    In this review, we highlighted the importance of switchback boundaries, which are unique features of the fluctuations. Investigations to date have focused on classifying the boundary type, looking for signatures of reconnection, and examining the zoo of different wave types produced there.
    
    All of these aspects have important implications for how switchbacks will evolve, determining their stability and ability to mix with the surrounding plasma. As yet, there is no consensus nor unifying model for the dynamics at the boundaries of switchbacks.
    Thus we still do not fully understand how they evolve and decay (and thereby lose their energy to the overall solar wind).
    
    For example, certain electromagnetic waves could indicate switchback compression or merging. Rotational or tangential discontinuity types could indicate the extent to which switchback material can leak or mix with its neighboring plasma. Observations of reconnection could imply switchbacks impart energy impulsively to their surroundings. Departures from exact incompressibility are also of great importance to understanding this stability and evolution.

\subsubsection*{Do All Reported Switchback Properties Continue to Hold in Later Orbits and Closer Distances?}
    
    By the nature of the Parker Solar Probe mission, whose orbit has shrunk slowly over time to closer and closer perihelia, and the fact that switchbacks were prominently observed from the very first orbits, many results have been derived from these early encounters. As seen in various places in this review and discussed at length in \citet{Mallet_this_issue}, switchbacks do evolve with radial distance. For example, polarity reversals become much less common closer to the Sun.
    
    Therefore, it is important that observational results concerning switchbacks obtained from data in the earlier encounters be validated using observations over the full mission both to test if inferred properties remain robust with radial distance and to obtain larger statistical samples.
    
    For example, conclusions about the limit to the velocity perturbations in switchbacks used measurements from the first encounter only \citep{Agapitov2023ApJ}. In another case, it has been observed that switchbacks whose boundaries show signs of reconnection have a smaller velocity enhancement relative to the Alfv\'en speed \citep{Suen2023}, but the number statistics where this is observed could be much improved using all available data. Lastly, it has been shown that Single Value Decomposition gives different results for classifying the discontinuity type at switchback boundaries \citep{Bizien2023} as compared to prior Minimum Variance Analysis methods \citep{AkhavanTafti2021,Larosa2021}, and this finding and its implications for switchback physics should be applied to a much wider statistical population.

\subsubsection*{How Do Switchback Patches Evolve in both Time and Space?}
    
    The relationship between switchback patches and solar supergranulation scale size is an important linkage between these structures and the corona. However, there are some reports of switchback patch structure appearing in time while the spacecraft does not cross far in longitude \citep{Shi_2022_patches}. This is an important phenomenon to investigate and determine statistically the relevant temporal and spatial scales.

\subsection{Final Remarks}

Magnetic switchbacks have provoked intense interest in heliophysics since their ubiquity was revealed by the launch of Parker Solar Probe. As discussed in Sect. \ref{sec: 1_Intro}, there are numerous reasons motivating this interest -- both in how they interact and affect the dynamics of the solar wind and also the way in which they reveal its underlying physics and coronal connections. In this review, we have presented the major properties and areas of study of these complex phenomena. We highlight a consensus that magnetic switchbacks are of central importance to our understanding of the young solar wind, but also that their study is complicated by the need for consistent definitions and detection methodologies, as well as the encounter-based and time-evolving nature of the Parker mission. Future work, as more orbits and statistical datasets are built, is essential to fully understanding their role in shaping the inner Heliosphere.

\vspace{\baselineskip}
\noindent
{\large\bf Acknowledgments}

STB and SBD were supported by the Parker Solar Probe project through the SAO/SWEAP subcontract 975569. MV acknowledges support from PSP through the FIELDS experiment on the Parker Solar Probe spacecraft was designed and developed under NASA contract NNN06AA01C.
Parker Solar Probe was designed, built, and is now operated by the Johns Hopkins Applied Physics Laboratory as part of NASA's Living with a Star (LWS) program (contract NNN06AA01C). Support from the LWS management and technical team has played a critical role in the success of the Parker Solar Probe mission.  We acknowledge the extraordinary contributions of the PSP mission operations and spacecraft engineering team at the Johns Hopkins University Applied Physics Laboratory
NF was supported by UKRI/STFC grant ST/W001071/1 and by the Centre National d'Etudes Spatiales (CNES). 
DR was supported by the National Science and Technology Development Agency (NSTDA) and National Research Council of Thailand (NRCT): High-Potential Research Team Grant Program (N42A650868), and from the NSRF via the Program Management Unit for Human Resources \& Institutional Development, Research and Innovation (B37G660015). 
N.B., T.D. and C.F. acknowledge funding from CNES. 
C.F. acknowledges funding from the CEFIPRA Research Project No. 6904-2. 
C.F, E.P \& TDdW acknowledge the support of the French Agence Nationale de la Recherche (ANR) for the JET2SB project under grant ANR-25-CE31-7416. E.P. and A.P.R. acknowledge support by the Action Thématique Soleil-Terre (ATST) of CNRS/INSU PN Astro as well as from the APR program of CNES
C.F., A.P.R., TDdW and N.B. acknowledge support from CNES (APR SCM).
O.P. is supported by NSF SHINE grant \# 2229566.
J.H. is supported by NASA grants 80NSSC23K0737 and 80NSSC23K0450.  M.M. acknowledges DFG grant WI 3211/8-2, project number 452856778, and the support of the Brain Pool program funded by the Ministry of Science and ICT through the National Research Foundation of Korea (RS-2024-00408396).
A.L. is supported by STFC Consolidated Grant ST/T00018X/1. 
T.A.B is supported by NASA grant 80NSSC24K0272.
L.S.-V. is supported by the Swedish Research Council (VR) Research Grant N. 2022-03352, by the projects “2022KL38BK– The ULtimate fate of TuRbulence from space to laboratory plAsmas (ULTRA)” (B53D23004850006) and ‘Data-based predictions of solar energetic particle arrival to the Earth: ensuring space data and technology integrity from hazardous solar activity events’ (H53D23011020001), funded by the Italian Ministry of University and Research through PRIN/NRRP-NextGenerationEU. 
VKJ acknowledges support from the Parker Solar Probe mission as part of NASA's Living with a Star (LWS) program under contract NNN06AA01C. 
LM, MV, PO, and LSV are supported by the International Space Science Institute (ISSI) in Bern, through the ISSI International Team project \#23-591 (Evolution of Turbulence in the Expanding Solar Wind). 
G.H.H.S. is partially supported by the Royal Society (RS) and the Consiglio Nazionale delle Ricerche (CNR) through the project 'Multi-scale electrostatic energisation of plasma: comparison of colllective processes in laboratory and space" (IEC\textbackslash\textbackslash R2\textbackslash\textbackslash222050 and SAC.AD002.043.021).
O.V.A. is supported by NASA grants 80NSSC21K1770, 80NSSC20K0218, and 80NSSC22K0433.
The work done at the University of Michigan was supported by NASA contract Nos. NNN06AA01C, 80NSSC20K1847, 80NSSC20K1014, and 80NSSC21K1662. STB acknowledges Y. Rivera and M. Terres for helpful discussions on various sections of this work.  Finally, all authors would like to thank ISSI for hosting the workshop that led to this review article. The work at AIP carried out by A.P.R. was supported by the Alexander von Humboldt foundation.

\vspace{\baselineskip}
\noindent
{\large\bf Statements and Declarations}

The authors have no competing interests to declare that are relevant to the content of this article.

\bibliography{sn-bibliography}


\end{document}

%% file: authors.tex
\author[1]{\fnm{Samuel T.} \sur{Badman} \orcidlink{0000-0002-6145-436X}}
\equalcont{These authors contributed equally to this work.}

\author[2, 3]{\fnm{Naïs} \sur{Fargette}
\orcidlink{0000-0001-6308-1715}
}
\equalcont{These authors contributed equally to this work.}

\author*[3]{\fnm{Lorenzo} \sur{Matteini}
\orcidlink{0000-0002-6276-7771}
}
\email{l.matteini@imperial.ac.uk }
\equalcont{These authors contributed equally to this work.}


\author[4, 5]{\fnm{Oleksiy V.} \sur{Agapitov}
\orcidlink{0000-0001-6427-1596}
}

\author[6]{\fnm{Mojtaba} \sur{Akhavan-Tafti}
\orcidlink{0000-0003-3721-2114}
}

\author[4]{\fnm{Stuart D.} \sur{Bale}
\orcidlink{0000-0002-1989-3596}
}

\author[1]{\fnm{Srijan} \sur{Bharati Das}
\orcidlink{0000-0003-0896-7972}
}

\author[7]{\fnm{Nina} \sur{Bizien}
 \orcidlink{0000-0001-6767-0672}
}

\author[4]{\fnm{Trevor A.} \sur{Bowen}
\orcidlink{0000-0002-4625-3332}
}

\author[7,8]{\fnm{Thierry} \sur{Dudok de Wit}
\orcidlink{0000-0002-4401-0943}
}

\author[7]{\fnm{Clara} \sur{Froment}
\orcidlink{0000-0001-5315-2890}
}

\author[3]{\fnm{Timothy} \sur{Horbury}
\orcidlink{0000-0002-7572-4690}
}

\author[4]{\fnm{Jia} \sur{Huang}
\orcidlink{0000-0002-9954-4707}
}

\author[9]{\fnm{Vamsee Krishna} \sur{Jagarlamudi}
\orcidlink{0000-0001-6287-6479}
}

\author[10, 11]{\fnm{Andrea} \sur{Larosa}
\orcidlink{0000-0002-7653-9147}
}

\author[12,13,14]{\fnm{Maria S.} \sur{Madjarska}
\orcidlink{0000-0001-9806-2485}}

\author[15]{\fnm{Olga} \sur{Panasenco}
\orcidlink{0000-0002-4440-7166}
}

\author[16,17]{\fnm{Etienne} \sur{Pariat}
\orcidlink{0000-0002-2900-0608}
}

\author[9]{\fnm{Nour E.} \sur{Raouafi}
\orcidlink{0000-0003-2409-3742}
}

\author[2,18]{\fnm{Alexis P.} \sur{Rouillard}
\orcidlink{0000-0003-4039-5767}
}

\author[19]{\fnm{David} \sur{Ruffolo}
\orcidlink{0000-0003-3414-9666}
}

\author[20,4]{\fnm{Nikos} \sur{Sioulas}
\orcidlink{0000-0002-1128-9685}
}

\author[21]{\fnm{Shirsh Lata} \sur{Soni}
\orcidlink{0000-0002-5550-739X}
}

\author[11,22]{\fnm{Luca} \sur{Sorriso-Valvo}
\orcidlink{0000-0002-5981-7758}
}

\author[23,7]{\fnm{Gabriel Ho Hin} \sur{Suen}
\orcidlink{0000-0002-9387-5847}
}

\author[20]{\fnm{Marco} \sur{Velli}
\orcidlink{0000-0002-2381-3106}
}

\author[24]{\fnm{Jaye} \sur{Verniero}
\orcidlink{0000-0003-1138-652X}
}



\affil[1]{\orgdiv{Center for Astrophysics}, \orgname{Harvard \& Smithsonian}, \orgaddress{\city{Cambridge}, \state{Massachusetts}, \postcode{02138}, \country{USA}}}

\affil[2]{\orgdiv{Institut de Recherche en Astrophysique et Planétologie}, \orgname{CNES, CNRS, Université de Toulouse}, \orgaddress{\city{Toulouse}, \postcode{31400}, \country France}}

\affil[3]{\orgdiv{Department of Physics}, \orgname{Imperial College London}, \orgaddress{\city{London}, \country{UK}}}

\affil[4]{\orgdiv{Space Sciences Laboratory}, \orgname{University of California}, \orgaddress{ \city{Berkeley}, \state{California}, \postcode{94720}, \country{USA}}}

\affil[5]{\orgname{Astronomy and Space Physics Department, National Taras Shevchenko University of Kyiv}, \orgaddress{\city{Kyiv}, \postcode{01601}, \country{Ukraine}}}
\affil[6]{\orgdiv{Department of Climate and Space Sciences and Engineering}, \orgname{University of Michigan}, \orgaddress{\city{Ann Arbor}, \state{Michigan}, \postcode{48109}, \country{USA}}}

\affil[7]{\orgdiv{LPC2E, OSUC}, \orgname{Univ Orleans, CNRS, CNES}, \orgaddress{\postcode{F-45071}, \city{Orleans}, \country{France}}}

\affil[8]{\orgname{International Space Science Institute}, \orgaddress{\city{Bern}, \postcode{3012}, \country{Switzerland}}}

\affil[9]{\orgname{John Hopkins University Applied Physics Laboratory}, \orgaddress{\city{Laurel}, \state{Maryland}, \postcode{20723}, \country{USA}}}

\affil[10]{\orgdiv{School of Physical and Chemical Sciences}, \orgname{Queen Mary University of London}, \orgaddress{\city{London}, \country{UK}}}

\affil[11]{\orgdiv{Institute for Plasma Science and Technology (ISTP)}, \orgname{CNR},  \postcode{70126}, \orgaddress{\city{Bari}, \country{Italy}}}

\affil[12]{ \orgname{Max Planck Institute for Solar System Research}, \orgaddress{\city{Gottingen}, \postcode{37077}, \country{Germany}}}

\affil[13]{ \orgname{Korea Astronomy and Space Science Institute}, \orgaddress{\city{Daejeon}, \postcode{34055}, \country{South Korea}}}

\affil[14]{ \orgname{Space Research and Technology Institute, Bulgarian Academy of Sciences}, \orgaddress{\city{Sofia}, \postcode{1113}, \country{Bulgaria}}}

\affil[15]{\orgname{Advanced Heliophysics}, \orgaddress{\city{Pasadena}, \state{California}, \postcode{91106}, \country{USA}}}


\affil[16]{\orgdiv{French-Spanish Laboratory for Astrophysics in Canarias}, \orgname{CNRS, Instituto de Astrofísica de Canarias}, \orgaddress{\postcode{38205}, \city{La Laguna},  \state{Tenerife}, \country{Spain}}}

\affil[17]{\orgdiv{Sorbonne Universit\'e, \'Ecole polytechnique, Institut Polytechnique de Paris, Universit\'e Paris Saclay, Observatoire de Paris-PSL, CNRS, Laboratoire de Physique des Plasmas}, \orgaddress{\postcode{75005}, \city{Paris}, \country{France}}}

\affil[18]{\orgname{Leibniz-Institut für Astrophysik Potsdam (AIP)}, \orgaddress{\city{Potsdam}, \postcode{11482}, \country{Germany}}}

\affil[19]{\orgdiv{Department of Physics, Faculty of Science}, \orgname{Mahidol University}, \orgaddress{\city{Bangkok}, \postcode{10400}, \country{Thailand}}}

\affil[20]{\orgdiv{Department of Earth, Planetary, and Space Sciences}, \orgname{University of California}, \orgaddress{\city{Los Angeles}, \state{California}, \postcode{90095}, \country{USA}}}

\affil[21]{\orgdiv{Department of Physics and Astronomy}, \orgname{University of Iowa}, \postcode{IA 52242-1479}, \orgaddress{\city{Iowa City}, \country{USA}}}

\affil[22]{\orgdiv{School of Electrical Engineering and Computer Science, Department of Electromagnetics and Plasma Physics}, \orgname{KTH Royal Institute of Technology}, \postcode{SE-11428}, \orgaddress{\city{Stockholm}, \country{Sweden}}}

\affil[23]{\orgdiv{Department of Space and Climate Physics}, \orgname{University College London}, \orgaddress{\city{London}, \country{UK}}}

\affil[24]{\orgdiv{Heliophysics Science Division}, \orgname{NASA Goddard Space Flight Center}, \orgaddress{\city{Greenbelt}, \state{Maryland}, \postcode{20771}, \country{USA}}}